\documentclass{PoS}
\usepackage{latexsym,amsmath,amsfonts,amsbsy,amssymb,fontenc,times,mathptmx,graphicx,stmaryrd}

\title{On boundary conditions and spacetime/fibre duality  \\
 in Vasiliev's higher-spin gravity} %unfolded description of soln spaces?  fibre description of solns spaces? 

\ShortTitle{On boundary conditions and spacetime/fibre duality in Vasiliev's HS gravity}

\author{\speaker{Carlo Iazeolla}\thanks{The review part of this paper is based on \cite{Iazeolla:2017vng,DeFilippi:2019jqq}.}\\
        NSR Physics Department, G. Marconi University \\
	via Plinio 44, Rome, Italy\\
        E-mail: \email{c.iazeolla@gmail.com}}

\abstract{This paper discusses some aspects of the Vasiliev system, beginning with a review of a recent proposal for an alternative perturbative scheme: solutions are built by means of a convenient choice of homotopy-contraction operator and subjected to asymptotically anti-de Sitter boundary conditions by perturbatively adjusting a gauge function and integration constants. At linear level the latter are fibre elements that encode, via unfolded equations, propagating massless fields of any spin. Therefore, linearized solution spaces, distinguished by their spacetime properties (regularity and boundary conditions), have parallels in the fibre. The traditional separation of different branches of linearized solutions via their spacetime features is reviewed, and their dual fibre characterization, as well as the arrangement of the corresponding fibre elements into AdS irreps, is illustrated. This construction is first reviewed for regular and singular solutions in compact basis, thereby capturing massless particles and static higher-spin black holes, and then extended to solutions in conformal basis, capturing bulk-to-boundary propagators and certain singular solutions with vanishing scaling dimension, related to boundary Green's functions. The non-unitary transformation between the two bases is recalled at the level of their fibre representatives.\ }

\FullConference{Corfu Summer Institute 2019 "School and Workshops on Elementary Particle Physics and Gravity" (CORFU2019)\\
		31 August - 25 September 2019\\
		Corfù, Greece}

 \csname
@addtoreset\endcsname{equation}{section}

\def\mso{\mathfrak{so}}

\def\msp{\mathfrak{sp}}

\def\mhs{\mathfrak{hs}}

\def\mD{\mathfrak{D}}

\def\Real{{\mathbb R}}
\def\Comp{{\mathbb C}}

\def\bec{\begin{center}}
\def\ec{\end{center}}
\def\a{\alpha} \def\ad{\dot{\a}} \def\ua{{\underline \a}}

\def\b{\beta}  \def\bd{\dot{\b}} 
\def\c{\gamma} \def\cd{\dot{\c}}
\def\C{\Gamma}
\def\d{\delta} \def\dd{\dot{\d}}
\def\D{\Delta}
\def\e{\epsilon} 
\def\ve{\varepsilon}

\def\k{\kappa}
\def\vark{\varkappa}
\def\l{\lambda}
\def\L{\Lambda}
\def\m{\mu}
\def\n{\nu}
\def\r{\rho}
\def\s{\sigma}

\def\t{\tau}

\def\y{\eta}

\def\O{\Omega}
\def\o{\omega}

\def\sb{{\bar\s}}

\def\cP{{\cal P}}

\def\bfx{{\bf{x}}}

\def\kb{{\bar \kappa}}
\def\yb{{\bar y}}
\def\zb{{\bar z}}

\def\jb{{\bar j}}

\def\tn{\widetilde{\n}}

\def\tP{\widetilde{P}}
\def\tcP{\widetilde{\cP}}

\def\nn{\nonumber}
\newcommand{\eq}[1]{(\ref{#1})}

\def\be{\begin{equation}}
\def\ee{\end{equation}}
\def\bea{\begin{eqnarray}}
\def\eea{\end{eqnarray}}
\def\ba{\begin{array}}
\def\ea{\end{array}}

\def\ft#1#2{{\textstyle{{\scriptstyle #1}
\over {\scriptstyle #2}}}}

\def\ket#1{|#1\rangle}
\def\bra#1{\langle#1|}
\def\scs#1{\section{\bf{\sc \large #1}}}
\def\scss#1{\subsection{\bf{\sc  #1}}}
\def\scsss#1{\subsubsection{\bf{\sc \small #1}}}

\def\ad{\dot\alpha}
\def\bd{\dot\beta}
\def\sb{\bar\sigma}

\def\acd{D^{(0)}_{\rm ad}}
\def\tcd{D^{(0)}_{\rm tw}}

\def\au{{\underline{\alpha}}}
\def\bu{{\underline{\beta}}}

\newcommand{\starcomm}[2]{\left[#1\,,#2\right]_\star}

\newcommand{\picomm}[2]{\left[#1\,,#2\right]_{\pi}}

\begin{document}

%%%%%%%%%%%%%%%%%%%%%%%%%%%%%%%%%%%%%%%%%%%%%%%%%%%%%%%%%%%%%%%%%%%%%

\scs{Introduction}\label{sec:intro}

%%%%%%%%%%%%%%%%%%%%%%%%%%%%%%%%%%%%%%%%%%%%%%%%%%%%%%%%%%%%%%%%%%%

Higher-spin gravity theories are extensions of ordinary gravity by massless fields with spin greater than two, based on the gauge principle. As such, they offer great opportunities and pose important challenges. While it is somehow natural to speculate that higher-spin symmetries capture the deep rationale behind the good properties of String Theory \cite{Gross,sundborg,perjohan,minwalla,Gaberdiel:2014cha,Gaberdiel:2015wpo}, and offer a particularly favourable window in which to study and, possibly, prove holographic dualities \cite{Sezgin:2002rt,Klebanov:2002ja,Douglas:2010rc,Gaberdiel:2010pz,Eberhardt}, many aspects of higher-spin gravity are difficult to grasp. This is largely due to the fact that the theory involves infinite-dimensional local symmetry algebras that  mix fields of different spins and different numbers of derivatives. While such invariance places very powerful constraints on the interactions, it also makes the theory quite different from ordinary field theories. For instance, as the order of derivatives in a vertex grows linearly with the spins involved, the necessary presence of fields of arbitrarily high spin means that theory is expected to contain non-local interaction terms. Moreover, the ordinary concepts of Riemannian geometry have no invariant meaning, and need to be replaced by some higher-spin-extended notion thereof --- which in turn may have some lesson in store for a deeper understanding of String Theory.

In this sense, it is remarkable that many features of the non-linear theory \cite{vasiliev,Vasiliev:1990vu,more,Vasiliev:2003ev} (see \cite{Vasiliev:1999ba,Bekaert:2005vh,Iazeolla:2008bp,Didenko:2014dwa} for reviews) can be controlled, largely due to the highly constraining symmetry and to the first-order unfolded formalism in which they are manifest. By reformulating the equations in terms of zero-curvature and covariant constancy conditions for master fields, transforming covariantly under the infinite-dimensional local symmetry algebra, it is possible to encode a highly complicated dynamics into a compact generating system for the non-linear spacetime equations. The master fields are functions of fibre coordinates $Y$, and the base manifold is an extension of spacetime with additional coordinates $Z$, with $Y$ and $Z$ each making up a non-commutative twistor space. The evolution along $Z$ generates the interaction vertices among physical fields, with gauge and field-redefinition ambiguities encoded into the choice of resolution operator for the $Z$-dependence.  It is by studying the consequences of this choice of homotopy-contraction operator on the resulting spacetime interactions that progress has been recently made \cite{Vasiliev:2017cae,Gelfond:2018vmi,Didenko:2018fgx,Didenko:2019xzz,Gelfond:2019tac} in addressing the problem of non-locality  \cite{Boulanger:2015ova,Bekaert:2015tva,Skvortsov:2015lja,Sleight:2017pcz} for some vertices. It will be interesting to see whether a generalized notion of locality can emerge from this analysis, powerful enough to fully constrain the ambiguity in field redefinitions and extract physical information at the level of vertices among spacetime fields. 

An alternative and, possibly, complementary approach is to try and extract all possible physical information at the level of master fields in the extended $(x,Y,Z)$ space and of the gauge-invariant observables of the full theory. This has the advantage of making use of the simplicity of the equations in the extended space to push the perturbative resolution as far as to get exact solutions, provided certain algebraic conditions are satisfied (that put constraints on the allowed class of functions of $(Y,Z)$). In practice, dropping one assumption %necessary to see non-linear terms as interaction vertices among physical fields 
(i.e., analyticity in twistor space for \emph{all} master fields) in the intermediate steps of the perturbative analysis, it is possible to choose a particularly convenient homotopy contraction. Star products among certain non-polynomial elements and distributions in twistor-space variables are handled via a regularization prescription based on specific integral transforms. In the resulting choice of field variables, all fluctuations are included in the spacetime zero-forms of the theory.  In fact, due to the unfolded equations, that locally reconstruct spacetime solutions from functions in twistor space \cite{Vasiliev:1999ba,Bekaert:2005vh}, to a large extent the spacetime features of the resulting master-field configurations are entirely stored in the twistor-space dependence of the zero-forms --- in a sort of spacetime/twistor-space duality much akin to a Penrose transform (see \cite{Vasiliev:2012vf} and references therein). Then, in order to interpret a solution in term of Fronsdal fields, it is possible to switch on a specific gauge function that glues the gauge fields to the local data encoded in twistor space. Moreover, any non-analyticity in the fibre coordinates $Y$ that may come from the first part of the perturbative procedure turns out to be cohomologically trivial, thus having no impact on the gauge fields. In \cite{DeFilippi:2019jqq} it was proved that this approach indeed provides the expected result at first order, and a scheme was proposed that imposes asymptotically anti-de Sitter boundary conditions at master-field level, by perturbatively adjusting the gauge function and the integration constants (i.e., the twistor space local data) at higher orders. This perturbative scheme, if proved successful, would have the advantage of giving a way of controlling boundary conditions at the level of master fields, thereby helping in setting up a well-posed boundary value problem for the Vasiliev system in the extended space. The latter may help in the analysis of locality, allowed field redefinitions, allowed class of functions in twistor space, and in general in giving a better understanding of the geometry behind higher-spin fields.

The infinite-dimensional fibre allows the encoding of the propagating degrees of freedom into the $Y$-dependence. The Weyl zero-form equation contains the Klein-Gordon equation and the generalized Bargmann-Wigner equations for all linearized Weyl tensors of every spin $s$. Its solution at a spacetime point $x_0$ is a fibre element with the coefficients of the $Y$-expansion corresponding to all spacetime derivatives of the physical fields at $x_0$. %Fibre elements dual to linearized solutions correspond to the Weyl 0-form master field at a spacetime point $x_0$, the rigid coefficients of the $Y$-expansion corresponding to all spacetime derivatives of physical fields at $x_0$. 
Hence, solutions in spacetime can be reconstructed from appropriate $Y$-space elements, and one expects that the classification of linearized solution spaces according to their spacetime properties (regularity and boundary conditions) can be mapped to the fibre. Moreover, fibre elements corresponding to different modes of a propagating field should arrange into irreps under the action of the $AdS$ rigid isometry algebra generators. Indeed, one can show that this happens and that the whole linearized Weyl 0-form master field admits a slicing in terms of $\mso(2,3)$-modules \cite{Iazeolla:2008ix,Iazeolla:2017vng,Aros:2019pgj,DeFilippi:2019jqq}. This extends not only to regular and normalizable solutions, but more generally to non-normalizable and singular solutions as well. In \cite{Iazeolla:2008ix}, indeed, working in any $D\geq 4$ and in \emph{compact basis} (i.e., slicing the $\mso(2,D-1)$-modules with respect to the compact  $\mso(2)\oplus\mso(D-1)$ subalgebra) elements corresponding to massless particle and anti-particle modes of any spin were explicitly constructed and shown to span respectively lowest- and highest-weight modules --- with the additional benefit of recognizing, facilitated by the formalism, that such modules are in fact ideals in a wider, indecomposable module where non-normalizable solutions also appear\footnote{Solutions appearing in the complement of (anti-)particle modules are referred to as \emph{wedge modes} in this paper, and strictly speaking are non-normalizable only for $D>4$, with $D$ the dimension of the $AdS$ bulk. However, even in $D=4$ they do not fall off fast enough at the boundary as to lead to conservation of the Killing energy, see Section \ref{ptandbh}.}. Moreover, in \cite{Iazeolla:2017vng,DeFilippi:2019jqq}, it was shown  that the singular $D=4$ static higher-spin black hole solutions of \cite{Didenko:2009td,us,us2} also can be mapped to fibre elements, but singular ones: delta functions and their derivatives (at least in Weyl-ordering). The latter, however, have good star-product properties, and can thus be considered elements of a well-defined star-product algebra. We shall comment more on this fact, discussing its possible interpretation as a smoothening of gravitational curvature singularities via their embedding into higher-spin gravity\footnote{See also \cite{Aros:2019pgj} for similar results on singularity-resolution mechanisms for fluctuations over a BTZ black-hole and for degenerate metrics via the unfolded formalism.}.

A similar study of linearized solution spaces via unfolding in \emph{conformal basis} (i.e., slicing the $\mso(2,3)$-modules according to a non-compact  $\mso(1,1)\oplus\mso(1,2)$ subalgebra) is then initiated. In this basis, the regular\footnote{This is strictly true in euclidean $AdS$, not in minkowskian signature (and without a specific prescription to avoid poles). Bulk-to-boundary propagators and their descendants are however referred to as ``regular'' here in the sense that, as we shall see, their fibre representatives are.} and normalizable solutions are bulk-to-boundary propagators, with corresponding fibre element also studied in \cite{Giombi:2010vg,DidenkoSkvortsov}. Together with their descendants, they form lowest-weight modules that (in minkowskian signature) can be thought of as resulting from a ``Wick rotation'' of those of particles. Whereas the latter are representations bounded from below in energy, the bulk-to-boundary propagator module is bounded from below in the eigenvalue of the operator that extracts, in the adapted Poincar\'e coordinates, %minus the homogeneity degree, i.e., 
the (classical) scaling dimension. Correspondingly, the anti-particle module is rotated into bulk-to-boundary propagators in inverted variables $(z',\bfx^{\prime m})=(\frac{z}{z^2+\bfx^2},\frac{\bfx^m}{z^2+\bfx^2})$. Finally, we also build, via the fibre/spacetime duality, the counterpart of static black-hole solutions: a tower of singular solutions with vanishing scaling dimension of the form $z^{s+1}\phi_{m_1\ldots m_s}(\bfx)$, where $\phi_{m_1\ldots m_s}(\bfx)$ are solutions to the boundary wave equation that are singular on the light-cone, related to boundary Green's functions\footnote{For these solutions, as well as for the bulk-to-boundary propagators, in this paper we do not consider the appropriate $i\e$ prescription that would enable a precise association with a specific boundary Green's function in minkowskian signature.}. We defer a more systematic and detailed study of these solution branches to future work. 

This paper is organized as follows. In Section \ref{sec:in} we collect the basic material on Vasiliev's four-dimensional bosonic equations. Section \ref{Sec:pertth} contains a short review of general features of their perturbative expansion around $AdS_4$, and then reviews the scheme recently proposed in \cite{DeFilippi:2019jqq}. In Section \ref{Sec:linsol} we study the linearized solution space of the twisted-adjoint equation, emphasizing their fibre/spacetime dual descriptions. To this purpose, we start by reviewing in some detail the classification of solutions of the Klein-Gordon equation at the spacetime level and then present some of their fibre counterparts, in order to compare the characterization of different sectors in the two pictures. This is done both in compact and conformal basis. We finally unfold the fibre representatives in spacetime by virtue of vacuum gauge functions adapted to spherical and Poincar\'e coordinates, in order to recover explicitly the spacetime solutions treated before.

%%%%%%%%%%%%%%%%%%%%%%%%%%%%%%%%%
\scs{Vasiliev's Four-Dimensional Bosonic Equations}\label{sec:in}
%%%%%%%%%%%%%%%%%%%%%%%%%%%%%%%%%%

%%%%%%%%%%%%%%%%%%
\scss{Correspondence space}\label{Sec:corr}
%%%%%%%%%%%%%%%%%%

Vasiliev's higher spin gravity is formulated in terms of a finite set 
of master fields living on a total bundle space ${\cal C}$, locally equivalent to  
\be {\cal C} \ \cong \ {\cal B}_8\times {\cal Y}_4 \ \cong \ {\cal X}_4\times {\cal Z}_4\times {\cal Y}_4\ ,\ee
where the base manifold ${\cal B}_8$ is an extension of the spacetime manifold ${\cal X}_4$ via the non-commutative manifold ${\cal Z}_4$, while ${\cal Y}_4$ plays the role of a fibre. ${\cal C}$ is referred to as \emph{correspondence space}, as reductions from the total space to either ${\cal X}_4\times  {\cal Y}_4$ or ${\cal Z}_4\times {\cal Y}_4$ yield dual formulations of the full dynamics. 
 The master fields are horizontal forms belonging to the differential associative algebra $\Omega({\cal C})$ of differential forms on ${\cal C}$, which is equipped with  a differential 
$d_{\cal C}:\O({\cal C})\rightarrow \O({\cal C})$ and 
a compatible associative binary composition rule $(\cdot)\star(\cdot):
\O({\cal C})\otimes \O({\cal C})\rightarrow \O({\cal C})$,
that are (finite) deformations of the de Rham differential and the
wedge product, respectively, such that if $f,g,h\in\O({\cal C})$
then%\footnote{The hat on $d$ indicates that it is in general
%a nontrivial deformation of the de Rham differential, whereas the
%hats on the elements in $\O({\cal C})$ are used to distinguish them
%from elements that are independent of the non-commutative coordinates $Z$.}
%
\be d^{\,\,2}_{\cal C} f~=~0\ ,\qquad 
d^{\,\,2}_{\cal C}\left(f\star  g\right)~=~
\left(d_{\cal C} f\right)\star g+
(-1)^{{\rm deg}( f)} f\star\left(d^{\,\,2}_{\cal C}  g\right)\ ,\ee
\be  f\star ( g\star  h)~=~(f\star  g)\star  h\ .\ee
These operations are in addition assumed to be compatible with an 
hermitian conjugation operation $\dagger$, \emph{viz.}
\be \left(f\star g\right)^\dagger~=~
(-1)^{{\rm deg}( f) {\rm deg}(g)}\big(g\big)^\dagger\star \big(f\,\big)^\dagger\ ,
\qquad \left(d_{\cal C}\,f\right)^\dagger~=~d_{\cal C}\left(\big(f\,\big)^\dagger\right)\ ,\qquad ((f)^\dagger)^\dagger=f\ ,\ee
for all $f,g\in\O({\cal C})$.

We coordinatize ${\cal C}$ using local coordinates $(x^\mu; Z^{\underline\a}; Y^{\underline\a})$ ($\underline\a=(\a,\ad)$; $\a,\ad=1,2$)\footnote{Our spinor conventions
are collected in Appendix \ref{App:conv}.},
\be %\Xi^{\underline M}~=~ (X^M;Y^{\underline\a})\ ,\qquad X^M~=~ (x^\mu; Z^{\underline\a})\ ,\qquad 
(Y^{\underline\a};Z^{\underline\a})~=~(y^\a,\yb^{\ad};z^\a,-\zb^{\ad})\ ,\ee
with reality properties $(x^\mu)^\dagger= x^\mu$, 
$(y^\a)^\dagger=\yb^{\ad}$,
$(z^\a)^\dagger=\zb^{\ad}$, and canonical commutation rules
\begin{align}
\starcomm{Y_\au}{Y_\bu}&=2iC_{\au\bu} 
\,,&
\starcomm{Z_\au}{Z_\bu}&=-2iC_{\au\bu} 
\,,&
\starcomm{Y_\au}{Z_\bu}&=0
\,,
\end{align}
where $C_{\au\bu}$ is the $Sp(4)$-invariant tensor.
We shall also refer to ${\cal Z}_4 \times {\cal Y}_4$ as the \emph{twistor space} $\cal T$. 
The local representatives of horizontal forms and the differential acting on them are then given, respectively, by 
\bea &  f|_{\rm hor} \stackrel{\rm loc}{=} f|_{dY^{\underline\a}=0}~=~f(z,Z,dx,dZ;Y)\ , & \\
&\hat d := d_{{\cal C}}|_{\rm hor}  \stackrel{\rm loc}{=}  d_{{\cal C}}|_{dY^{\underline\a}=0}~=~d+q ~:=~d x^\m \partial_\m+dZ^{\ua}\partial_{\ua}^{(Z)}& \ .\eea
The master fields can thus be represented as sets of
locally defined forms on ${\cal X}_4\times {\cal Z}_4$ valued 
in oscillator algebras ${\cal A}({\cal Y}_4)$ generated by the fibre 
coordinates. %glued together by transition functions, that we shall
%assume are defined locally on ${\cal X}_4$, resulting in a bundle 
%over ${\cal X}$ with fibers given by 
%$\Omega({\cal Z})\otimes {\cal A}({\cal Y})$.
%

As studied in \cite{us,us2,Aros:2017ror,Aros:2019pgj,DeFilippi:2019jqq,Didenko:2009td} and as we shall review in this paper, master fields subject to specific boundary conditions belong to oscillator algebras that are larger than the class of polynomials in $Y$ and $Z$, and in general contain distributions as well as regular non-polynomial functions. The star product on such an extended class of symbols can be realized as a twisted convolution formula, 
\begin{equation}
\label{eq:NOprod}
(f\star g)(Y,Z)
=
\int_{\mathbb{R}^8}\frac{d^4U d^4V}{(2\pi)^4}
\,e^{iV^{\au}U_{\au}}\,f(Y+U,Z+U)\,g(Y+V,Z-V)
\,,
\end{equation}
where $f,g\in \Omega_{[0]}({\cal Y}_4\times {\cal Z}_4)$,  and the integration domain is
\be  \mathbb{R}^8\ =\ \{~(u_\a,\bar u_{\ad};v_\a,\bar v_{\ad})~:~ 
(u_{\a})^\dagger=u_\alpha\ , \quad 
(\bar u_{\ad})^\dagger=\bar u_{\ad}\ , \quad
(v_{\a})^\dagger=v_\alpha\ , \quad 
(\bar v_{\ad})^\dagger=\bar v_{\ad}\ \}\ ,\label{realint}\ee
that is, all auxiliary variables are integrated over the real line.
 The relevant oscillator algebras can be given in various bases and functional presentations. The star product formula \eq{eq:NOprod} provides the representation of an operator algebra in terms of symbols given in \emph{normal order} with respect to the creation (+) and annihilation (-) operators
\begin{equation}
a_{\alpha }^{+ }\ := \ \frac{1}{2i}\left( y_{\alpha }- z_{\alpha }\right) \text{
,} \qquad a_{\alpha }^{- }\ := \ \frac{1}{2}\left( y_{\alpha }+z_{\alpha }\right) \text{
,}
\end{equation}
since $f(Y-Z)\star g(Y+Z)=f(Y-Z)g(Y+Z)$.  The normal order reduces to Weyl order for elements that are independent
of either $Y$ or $Z$. Moreover, as $Y$ and $Z$ commute, the normal-ordered star product of symbols $f(Y,Z)=f_1(Y)\star f_2(Z)$ that admit factorization with respect to $Y$ and $Z$  corresponds to (total) \emph{Weyl order}. In other words, keeping the $Y$ and $Z$ dependence always factorized is equivalent to working in a basis where contractions between $Y$ and $Z$ vanish, and the symbol of  $f(Y,Z)=f_1(Y)\star f_2(Z)$ defined using Weyl order is $f_1(Y) f_2(Z)$.

\scss{Master fields}

Vasiliev's equations are formulated in terms of :
\begin{enumerate}
\item an \emph{adjoint} master one-form connection $A$ on ${\cal C}$,  with curvature
\be F:= \hat d A+A\star A\ ,\ee
and obeying the Bianchi identity
\be \hat D F:=\hat dF+[A, F]_\star \equiv 0\ ;\ee
\item a \emph{twisted-adjoint} \cite{Vasiliev:1990vu,more,Vasiliev:1999ba} zero-form master field $\Phi$, also referred to a \emph{Weyl zero-form}, with covariant covariant derivative 
\be \hat D\Phi:=\hat d\,\Phi+[A,\Phi]_\pi\ ,\ee
obeying the Bianchi identity
\be \hat D^2\Phi\equiv [F,\Phi]_\pi\ ,\ee
where 
\be [f,\,g\,]_\pi:=
f\star g-(-1)^{{\rm deg}(f){\rm deg}(g)}g\star\pi(f)\ ,\qquad 
f,g\in\O({\cal C})\ ,\ee
with $\pi$ and $\bar\pi$ involutive automorphisms of the star product algebra defined by 
$d\,\pi=\pi\,d$, $d\,\bar\pi=\bar\pi\,d$ and
\be \pi(x^\mu;y^\a,\yb^{\ad};z^\a,\zb^{\ad})~=~
(x^\mu;-y^\a,\yb^{\ad};-z^\a,\zb^{\ad})\ ,\qquad \pi(f\star g)~=~
\pi(f)\star \pi(g)\ ,\ee
\be \bar\pi(x^\mu;y^\a,\yb^{\ad};z^\a,\zb^{\ad})~=~
(x^\mu;y^\a,-\yb^{\ad};z^\a,-\zb^{\ad})\ ,\qquad \bar\pi(f\star g)~=~
\bar\pi(f)\star \bar\pi(g)\ ;\ee

\item a non-dynamic two-form $J$, 
\begin{equation}\label{J}
J=  j+\bar j := -\frac{ib}4 dz^\a\wedge dz_\a \k- \frac{i\bar b}4 d\zb^{\ad}\wedge d\zb_{\ad} \kb\,,\qquad 
b=e^{i\theta}\,,\qquad \bar b=e^{-i\theta}
\end{equation}
where $\k$ and $\bar \k$ are the inner Klein operators
\begin{align}
\label{eq:k^2=1}
\k:&=e^{iy^\a z_\a}
\,,&
\kb:&=\k^\dagger=e^{-i\yb^{\ad}\zb_{\ad}}
\,,&
\k\star\k&=\kb\star\kb=1
\,,
\end{align}
that make the $\pi,\bar \pi$ automorphisms inner
\begin{align}
\label{eq:pifromk}
\pi(f_{[p;q,\bar q]})
&=
(-1)^{q}\k\star f_{[p;q,\bar q]}\star\k 
\,,&
\bar\pi(f_{[p;q,\bar q]})
&=
(-1)^{\bar q}\kb\star f_{[p;q,\bar q]}\star\kb
\,,
\end{align}
for horizontal forms $f_{[p;q,\bar q]}$ of degree $p$ on ${\cal X}_4$ and mixed holomorphic and anti-holomorphic degree $q$ and $\bar q$ on ${\cal Z}_4$, respectively; and $b,\bar b$ are phases that can be fixed by requiring parity invariance \cite{Sezgin:2003pt}. The two-form $J$ is closed, $\hat d J=0$, and twisted-central, $[J,f]_\pi=0$, for all $f\in \Omega({\cal C})$.  

\end{enumerate}

The inner Klein operators can be factorized in inner Klein operators $\k_y$ and $\k_z$ (\emph{idem} their hermitian conjugates) that anticommute only with $y$ and $z$ variables \cite{Didenko:2009td,berezin}, respectively,
\begin{align}
\label{eq:k fact}
\k&=\k_y\star\k_z
\,,&
\k_y:&=2\pi\delta^2(y)
\,,&
\k_z:&=2\pi\delta^2(z)
\,,&
\k_y\star\k_y&=
\k_z\star\k_z=1
\,,\\
\bar \k &=\kb_{\yb}\star\kb_{\zb}
\,,&
\kb_{\yb}:&=2\pi\delta^2(\yb)
\,,&
\kb_{\zb}:&=2\pi\delta^2(\zb)
\,,&
\kb_{\yb}\star\kb_{\yb}&=
\kb_{\zb}\star\kb_{\zb}=1
\,,
\end{align}
hence one can write
\be
\label{eq:def J}
J=\k_y\star j_z+\kb_{\yb}\star\jb_{\zb}
\,,\qquad
j_z:= -\frac{ib}{4} dz^\a\wedge dz_\a \k_z
\,,\qquad
\jb_{\zb}:= -\frac{i\bar b}{4} d\zb^{\ad}\wedge d\zb_{\ad} \kb_{\zb}\ . 
\ee
The star product of a function on ${\cal Y}_4$ with $\k_y$ amounts to a Fourier transform on the holomorphic variables \cite{Didenko:2009td},
\be f(y,\yb)\star \k_y \ = \ \int\frac{d^2\xi}{2\pi}\,e^{-i y^\a \xi_\a}\,f(\xi,\yb) \ ,\label{kyFT}\ee
and similarly $\bar \k_{\yb}$ for $\yb$, as well as $\k_z,\bar \k_{\zb}$ on functions that depend on $Z$.
It is important to observe that while the normal-ordering symbol for $\k,\bar \k$ \eq{eq:k^2=1} is a regular function of the oscillators, the one for each separate factor \eq{eq:k fact} -- or, equivalently, the (total) Weyl-ordered symbol $(\k)_{\rm Weyl} = (2\pi)^2\d^2(y)\d^2(z)$ -- are delta functions\footnote{The delta functions are assumed to be real-analytic
(in the sense that they preserve real-analyticity of the test function), i. e.  $\delta^2(My)=(\det M)^{-1} \delta^2(y)$ where $(My)^\alpha\equiv M^{\alpha\beta} y_\beta$, idem $\delta^2(z)$ \cite{us,DeFilippi:2019jqq}.}.  As a consequence, one can expect that solving the equations in Weyl ordering leads to admitting distributions (and their derivatives) in the oscillator algebra, which one must be able to deal with. We shall see that this is indeed the case in Section \ref{Sec:pert}.

The master fields obey the reality conditions
\be (\Phi,A)^\dagger~=~
(\pi(\Phi),-A)\ ,\label{reality}\ee
and the truncation to bosonic fields is implemented via the condition
\be \pi\bar\pi(\Phi,A)~=~(\Phi,A)\ ,\label{piofj}\ee
together leading to real 
Fronsdal fields with integer rank, each occurring once.
A further truncation to the minimal bosonic model is possible, in which the odd-spin Fronsdal fields are
removed by the stronger projection
\be \tau(\Phi,A,J,\,\bar J)~=~(\pi(\Phi),-A,-J,-\bar J)\ ,\label{min}\ee
where $\tau$ is the graded anti-automorphism defined by $d\,\tau ~=~\tau\, d$ and
\be \tau(x^\mu;Y^{\underline\a};Z^{\underline\a})~=~(x^\m;iY^{\underline\a};-iZ^{\underline\a})\ ,\qquad \tau(f\star g)~=~(-1)^{fg} \tau(g)\star \tau(f)\ ;\ee
from $\tau^2=\pi\bar\pi$ it follows that \eq{min} implies \eq{piofj}.
%

%%%%%%%%%%%%%%%%%%%%%%%%%%%%%
\scss{Field equations}\label{fieldequations}
%%%%%%%%%%%%%%%%%%%%%%%%%%%%%%

%\paragraph{Master field equations.}
Vasiliev's equations are given by
\be
F+\Phi\star J \ = \ 0
\ , \qquad
\hat D\Phi \ = \ 0
\ ,
\ee
The equations are Cartan-integrable, and hence
admit the following on-shell gauge transformations:
\be \delta_{\e}\Phi~=~-[\e,\Phi]_\pi\ ,\qquad \delta_{\e}A~=~ \hat d\e+[A,\e]_\star\ ,\ee
where the local parameters obey the same kinematic conditions as $A$,
and the two-form is treated as a background in the sense that $\delta_{\e}J~=~0$. 

In order to make contact with the free Fronsdal equations on $AdS_4$ as well as to construct exact solutions, it is useful to decompose 
\be A~=~U+V\ ,\ee
where $U$ and $V$ are respectively a spacetime and $Z$-space one-form connection, 
\be U~:=~ dx^\mu U_\mu(x;Z,Y)\ ,\ee\be  V~:=~ dZ^{\underline\a}V_{\underline\a}(x;Z,Y)~=~dz^\a V_\a(x;Z,Y)+d\zb^{\ad} \bar V_{\ad}(x;Z,Y)\ ,\ee
%
%and introduce the deformed oscillators
%
%\be S_{\underline\a}~:=~( S_{\a}, -{\bar S}_{\ad})~:=~Z_{\underline\a} -2iV_{\underline\a}~=~(z_\a-2i V_\a,-\zb_{\ad}+2i {\bar V}_{\ad})\ .\label{SV}\ee
%
Letting $d=dx^\mu\partial_\mu$ %, the master field equations can be rewritten as
%
%\be d U+U\star U~=~0\ ,\label{MC}\ee 
%\be d\Phi+U\star\Phi-\Phi\star \pi(U)~=~0\ ,\label{MC2}\ee
%\be dS_{\underline\a}+[U,S_{\underline\a}]_\star~=~0\ ,\label{dSa}\ee
%\be S_\a\star\Phi+\Phi\star\pi(S_\a)~=~0\ ,\quad
%{\bar S}_{\ad}\star\Phi+\Phi\star\bar\pi({\bar S}_{\ad})~=~ 0  \ ,\label{INT1}\ee\be
%[S_\a,S_\b]_\star~=~ -2i\e_{\a\b}(1-b\,\Phi\star\kappa)\ ,\quad [{\bar S}_{\ad},{\bar S}_{\bd}]_\star~=~ -2i\e_{\ad\bd}(1- \bar b \, \Phi\star{\bar \kappa}) \ ,\label{INT2}\ee\be
%[S_\a,{\bar S}_{\ad}]_\star~=~ 0\ .\label{INT3}\ee
%
%Alternatively, using the exterior derivative 
and $q=dZ^{\ua}\partial^{(Z)}_{\ua}$, the master field equations can be rewritten as 
\be dU+U\star U \ = \ 0 \ , \label{MC}\ee
\be  d\Phi+\picomm{U}{\Phi}  \ = \ 0 \ ,\label{MC2}\ee
\be   qU+dV+\starcomm{U}{V} \ = \ 0 \  ,\label{dSa} \ee
\be q\Phi+\picomm{V}{\Phi} \ = \ 0 \ ,\label{INT1}\ \ee
\be  qV+V\star V+\Phi\star J  \ = \ 0 \label{INT2} \ee
The gauge transformations now read
\be \delta_{\e} \,\Phi~=~-[\e,\Phi]_\pi\ ,\qquad 
\delta_{\e}\,V \ =\ q\e+[V,\e]_\star\ ,\qquad 
\delta_{\e}\, U~=~ d\e+[U,\e\,]_\star\ .\ee
%

%Thus, the master field equations describe a flat connection on ${\cal X}$ and
%a covariantly constant deformed oscillator algebra 
%
%\be  {\cal A}:= \left\{ (\Phi\star \kappa)^{\star k} \star (\Phi\star{\bar\kappa} )^{\star \bar k}\star S_{(\underline \a_1}\star\cdots \star S_{\underline \a_m)}\right\}\ ,\ee
%
%generated by the internal master fields $(\Phi,S_{\underline\a})$.

%%%%%%%%%%%%%%%%%%%%%%%%%%%%%%%%%%%%%%%%%%%%%%%%

\scs{Perturbative schemes, homotopy contractions and boundary conditions}\label{Sec:pertth}

%%%%%%%%%%%%%%%%%%%%%%%%%%%%%%%%%%%%%%%%%%%%%%5

%%%%%%%%%%%%%%%%%%%%%%%%%%%%%%%%%

\scss{$AdS_4$ vacuum solution and gauge functions}

%%%%%%%%%%%%%%%%%%%%%%%%%%%%%%%%%%

The simplest vacuum solution of the equations gives the four-dimensional Anti-de Sitter spacetime, which is obtained as
\bea & \Phi \ = \ \Phi^{(0)} \ = \ 0\ , & \\
& V \ = \ V^{(0)}\ = \ 0  % \qquad {\rm i.e.} \qquad S_{\ua} \ =  S^{(0)}_{\ua} \ = \ Z_{\ua}
\ , &\\
& \displaystyle U \ = \ U^{(0)} \ = \ \Omega \ = \ \frac{1}{4i}\left(\o^{\a\b}y_\a y_\b+\bar \o^{\ad\bd}\yb_{\ad }\yb_{\bd}+2e^{\a\bd} y_\a\yb_{\bd}\right)\ .  & \label{AdSconn} \eea
and can be written in terms of a gauge function 
\be
\label{eq:AdS L}
\Omega \ = \ L^{-1}\star dL \  , \qquad L: \, U_8\to SO(2,3)
\,,\qquad
qL=0
\,,
\ee
where $U_8$ is a chart in ${\cal X}_4\times {\cal Z}_4$. Different choices of coordinate systems and charts correspond to different, adapted gauge functions. In this paper (see Appendix \ref{App:conv} for our conventions and more details on the coordinate systems), we shall consider :
\begin{itemize}

\item Stereographic coordinates $x^a$, $ a=0,1,2,3$, in which the metric is given on manifestly Lorentz covariant form as (setting the $AdS$ radius to $1$)
\be ds^2 \ = \ \frac{4dx^2}{(1-x^2)^{2}}\ ,\qquad x^a\in \Real^4\ ,\qquad  x^2\neq 1\ , \label{stereo}\ee
where $x^2:=x^a x^b\eta_{ab}$ and $dx^2:=dx^a dx^b\eta_{ab}$.  The corresponding gauge function is \cite{Bolotin:1999fa,Sezgin:2005pv} 
%
%\be e^{(0)}_{\a\ad}= - h^{-2}(\s^a)_{\a\ad}dx_a\ ,\qquad 
%\o^{(0)}_{\a\b}=- h^{-2} (\s^{ab})_{\a\b} dx_a x_b\ ,\qquad h~:=~\sqrt{1-x^2}\ .\label{Vbetc}\ee
%
%Defining $ x^a=-\frac12 (\s^a)^{\a\ad}x_{\a\ad}$, $x^{\a\ad}=(\s_a)^{\a\ad} x^a$, and $\xi~:=~(1-h^2)^{-\ft12}\tanh^{-1}\sqrt{\ft{1-h}{1+h}}$, the gauge function $L$ can be written as 
%
%\be  L_{\rm stereo}~=~\exp_\star (4i\xi(x^2)  x^a  P_a) \ =\ {2h\over 1+h} \exp {4ix^a P_a\over 1+h}\ .\label{Lgf}\ee
%
%where $P_a$ are $AdS_4$ transvections.
%
%\be %e^{(0)\a\bd}  \ =  \	  -\frac{dx^{\a\bd}}{1-x^2} \ , \qquad \o^{(0)\a\b} \ = \  -\frac{dx^{\a\bd}}{1-x^2}  ...  \qquad 
%ds^2 \ = \ \frac{4dx^2}{(1-x^2)} \ , \quad x^2<1 \ , \ee
%
%corresponding to 
%
\be L_{\rm stereo} \ =  \ \exp_\star (4i\xi(x^2)x^a P_a) \ = \ {2h\over 1+h} \exp {4ix^a P_a\over 1+h}\label{Lgf}\ , \ee
where $\xi(x^2):=  \frac{{\rm arctanh}(\sqrt{\frac{1-h}{1+h}})}{1-h^2}$, $h:=\sqrt{1-x^2}$ and $P_a$ are $AdS_4$ transvections\footnote{$\exp_\star A$ denotes the star-power expansion $\exp_\star A=1+A+\ft12 A\star A+...$ .}. The spacetime point at which the gauge function becomes trivial, which we shall refer to as the \emph{unfolding point}, is $x^a=0$ for $L_{\rm stereo}$ . %Two sets of stereographic coordinates are needed to cover $AdS_4$ (see \cite{} for more details); 

\item Global spherical coordinates $(t,r,\theta,\phi)$, $t\in [0,2\pi)$, $r\in[0,\infty)$, $\theta\in[0,\pi]$ and $\phi\in[0,2\pi)$, in which the $AdS_4$ metric is
\be ds^2 \ = \ -(1+r^2)dt^2+\frac{dr^2}{1+r^2}+r^2d\O^2_{S^2} \ , \ee
and is obtained from the gauge function \cite{Aros:2019pgj}
\be  L_{\rm spherical} \ =  \ \exp_\star (-iEt)\star \exp_\star (i\,{\rm arcsinh} (r)\, n^k P_k) \ ,\label{Lsph}\ee
where $k=1,2,3$, $n^k n_k=1$ and $E\equiv P_0$ is the energy generator. Evidently the spacetime origin is the unfolding point, as $L_{\rm spherical}(t=0=r)=1$. 

\item  Poincar\'e coordinates $(\bfx^m,z)$, $m=0,1,2$, where $\bfx^m\in \Real^3$ are coordinates on the three-dimensional flat Minkowski boundary and $z\in(0,\infty)$, in which the metric is
\be ds^2 \ = \ \frac{d\bfx^m d\bfx_m + dz^2}{z^2} \ ,\ee
and is obtained from 
\be L_{\rm Poincar\acute{e}} \ = \ \exp\left(i\,\bfx^m T_m\right) \star \exp_\star\left(\frac{i}4\log(z)P\right) \label{LP}\ee
(see also \cite{Giombi:2010vg,DidenkoSkvortsov,Rahman:2015pzl}), where $P$ is a spatial $AdS_4$ transvection and $T_m$ are the (commuting) boundary translations (see Appendix \ref{App:conv} for their realization in terms of the $\mso(2,3)$ generators). The unfolding point for $L_{\rm Poincar\acute{e}}$  is at $z=1$, $\bfx^m=0$. 

\end{itemize}

The Cartan gauge transformations with rigid group elements $\e^{(0)}$ that preserve the vacuum obey
\be D^{(0)} \e^{(0)}=0\,,\qquad q\e^{(0)}=0\,,\label{Keq}\ee
that is,
\be \e^{(0)}=L^{-1}\star \e^{\prime(0)}\star L\,,\qquad d\e^{\prime(0)}=0=q\e^{\prime(0)}\,,\ee
correspond to Killing symmetries (indeed Eq. \eq{Keq} contains, in the bilinears in $Y$, the Killing equation and its consistency condition \cite{Didenko:2009tc}).  $\e'(Y)$ belongs to the bosonic higher spin algebra $\mhs(2,3)$ 
or its minimal bosonic subalgebra $\mhs_0(2,3)$ \cite{Fradkin:1986ka,Vasiliev:1986qx,Konstein:1989ij}. The action of $L$ on functions of $Y$ amounts to a rotation of the oscillators,
\be f^L(Y)~:=~ L^{-1}(x,Y)\star f(Y)\star L(x,Y) \ = \ f(Y^L)\ ,\qquad  Y^{L}_{\underline\a}~:=~L^{-1}\star Y_{\underline\a}\star L~=~ L_{\underline\a}{}^{\underline\b} Y_{\underline\b}\ ,\label{YLrotn}\ee
with the matrix representation $L_{\underline{\a\b}}(x)$ of $L$ given by, e.g., \eq{3.20} in stereographic coordinates.  As a consequence of \eq{YLrotn}, taking $\e^{\prime(0)}$ to be an $\mso(2,3)$ isometry generator $M_{AB}=-\frac18 Y^{\au}(\Gamma_{AB})_{\au\bu}Y^{\bu}$, $L$-rotation gives 
\be  M_{AB}^L \ = \ -\frac18 Y^{\au}(\Gamma_{AB}^L)_{\au\bu}Y^{\bu} \ , \qquad  (\Gamma_{AB}^L)_{\au\bu} \ = \ -(L^T\Gamma_{AB} L)_{\au\bu} \ ,\ee
where, exhibiting the $2\times 2$-blocks,
\be   (\Gamma_{AB}^L)_{\au\bu}  \ = \ \left(\ba{cc} \vark^L_{\a\b} & v^L_{\a\bd} \\[5pt] \bar v^L_{\ad \b} & \bar \vark^L_{\ad\bd}\ea\right)_{AB} \ .\label{Gammakv}\ee
The off-diagonal blocks are components of the Killing vector $\vec v^L_{AB}$, and the diagonal ones are the (anti-)selfdual components of the corresponding Killing two-form. For instance, $L$-rotating $\Gamma_{0'0}$%=\left(\ba{cc} 0 & (\s_0)_{\a\bd} \\[5pt] (\bar \s_0)_{\ad\b}  & 0\ea\right)$ 
via $L_{\rm stereo} $, one obtains 
\be  \vec v^L_{0'0} \ = \ (v^L_{0'0})^\m\frac{\partial}{\partial x^\m} \ = \ \frac12\left[(1+x_0^2+x^k x_k)\frac{\partial}{\partial x^0}-2x_0 x^k\frac{\partial}{\partial x^k} \right] \ , \ee
or, in spherical coordinates via $L_{\rm spherical}$, 
\be \vec v^L_{0'0} \ = \ \frac{\partial}{\partial t} \ .  \ee
Similarly, in Poincar\'e coordinates (see \eq{yLP}-\eq{ybLP})
\be \vec v^L_{0'3} \ = \ -z\frac{\partial}{\partial z} -\bfx^m\frac{\partial}{\partial \bfx^m}  \ .  \label{KVD}\ee
%

%%%%%%%%%%%%%%%%%%%%%%%%%%%%%%%%%%%%%

\scss{Perturbative expansion around $AdS_4$}\label{Sec:pert}

%%%%%%%%%%%%%%%%%%%%%%%%%%%%%%%%%%%%%

\paragraph{Expansion in curvatures and homotopy contractions.} The ${\cal Z}_4$-space components of the master-field equations \eq{MC}-\eq{INT2} can be integrated iteratively in an expansion in curvatures \cite{Vasiliev:1990vu,more,Vasiliev:1999ba,Sezgin:2002ru}, with initial data
\be  \Phi|_{Z=0} \ = \ C(x,Y)\ ,\qquad U|_{Z=0} \ = \ W(x,Y) \ , \label{intconst}\ee
i.e.,
\bea
q\Phi+[V,\Phi]_\pi &=&0 \qquad \qquad \longrightarrow \qquad \qquad \Phi = C(x,Y)+q^\ast \left(-[V,\Phi]_\pi \right) \ ,
\\
qV+V\star V+\Phi\star J&=&0  \qquad \qquad \longrightarrow \qquad \qquad V = q\epsilon +q^\ast \left(-V\star V - \Phi\star J  \right)  \ , \\
qU+dV+[U,V]_\star &=&0  \qquad \qquad \longrightarrow \qquad \qquad U= W(x,Y)+q^\ast \left(-[U,V]_\star -dV \right)\ ,
\eea
where $\e=\e(x, Z, Y)$ is a gauge parameter and $q^\ast$ is a resolution operator, a formal inverse of $q$ giving rise to a particular solution. $\Phi$ and $V$ are assumed to be of at least first order in the curvatures contained in $C$, while the zeroth order term in $U$ corresponds to the background $AdS$ connection \eq{AdSconn}-\eq{eq:AdS L}, 
\begin{eqnarray}
\Phi& = & \sum_{n\geq 1} \Phi^{(n)}\ , \qquad \Phi^{(1)}=C^{(1)}(x,Y) \nn\\
V &= & \sum_{n\geq 1}V^{(n)}\nn \\
U & = & \sum_{n\geq 0} U^{(n)}\ , \qquad U^{(0)} \ = \ \O \ = \ L^{-1}\star dL\nn \ .
\end{eqnarray}
Note that there are in principle initial data $C^{(n)}(x,Y) $ emerging at every perturbative step, as solutions to the homogenous equation at order $n$. 

At every perturbative order, the $Z$-space equations have the typical form $q f= g$, with $f$ a form field, and their general solution can thus be written as 
\be f \ = \ q^\ast Jg+ qh+ c \ , \ee
where $h$ is a gauge function (or form), $c$ is an element of the $q$-cohomology\footnote{Actually $H_{[0]}(q)$ is the only cohomology that is relevant for the (duality unextended) Vasiliev system, see \cite{DeFilippi:2019jqq} and references therein.} $H(q)\subset \Omega({\cal Z}_4)$ valued in $\Omega({\cal X}_4)\otimes {\cal A}({\cal Y}_4)$, that is, $c \ = \ {\cal P}f\,,\ {\cal P}:\Omega({\cal Z}_4)\to H(q)\,, \ {\cal P}^2={\cal P}$, and $q^\ast $ is some homotopy-contraction operator that gives a resolution of the identity \cite{Didenko:2015cwv}
\be q q^\ast + q^\ast q \ = \ 1 - {\cal P} \ . \ee
The initial dta \eq{intconst}, encoding the local degrees of freedom of the system, are in fact such $q$-cohomology projections onto $H_{[0]}(q)$.  A particular form of $q^\ast$ can be obtained by contracting along the Euler vector field $\vec E:=Z^\au\vec\partial^Z_{\au}$ shifted by a vector $\vec V$,
\begin{equation}
\label{eq:rhoV}
q^{(E+V)\ast} g = \imath_{\vec E+\vec V} 
\int_0^1 \frac{d t}{t} g(x,tZ+(t-1)V;dx,tdZ;Y)
\,,
\end{equation}
where, as usual, $\imath_{\vec E+\vec V} := (Z^\au+V^\au)\frac{\partial}{\partial dZ^{\au}}$ and $V^\au$ is $Z$-independent. Solutions obtained via two different contracting homotopies differ by gauge choices and field redefinitions \cite{Vasiliev:2017cae,Didenko:2018fgx}.  Given a decomposition using a specific choice $(q^{(A)\ast},{\cal P}^{(A)})$, we shall refer to the projection 
\begin{equation}
f^{(A)} := q^{(A)\ast} g +  c^{(A)}\equiv \left(q^{(A)\ast}q+{\cal P}^{(A)}\right)f\,, \label{decompf}
\end{equation}
of $f$ obtained by setting $h^{(A)}$ to zero, as the \emph{twistor space $A$-gauge}. Note that two such gauges may be physically inequivalent, as the gauge function may carry physical degrees of freedom (arising via boundaries or other topological defects) \cite{DeFilippi:2019jqq}.

\paragraph{Linearized analysis.}

Armed with these notions, let us look more in detail at the linearization of the system \eq{MC}-\eq{INT2} around the $AdS_4$ vacuum: 
\begin{align}
\label{eq:qPhi1}
q {\Phi}^{(1)}
&= 0\,,\\
\label{eq:DPhi1}
\tcd {\Phi}^{(1)}
&= 0\,,\\
\label{eq:qV1}
q {V}^{(1)}+
 {\Phi}^{(1)}\star {J}
&=0\,,\\
\label{eq:qU1+DV1}
q {U}^{(1)}
+\acd {V}^{(1)}
&=0\,,\\
\label{eq:DU1}
\acd {U}^{(1)}
&=0\,,
\end{align}
where the adjoint and twisted-adjoint background covariant derivatives of a master field $f=f(x,Z,dx,dZ;Y)$ are defined by 
\begin{align}
\label{eq:D0 from L}
\acd{f}:&=
L^{-1}\star d\left(L\star f\star L^{-1}\right)\star L=
df+\starcomm{ {U}^{(0)}}{{f}}\,,\\
\label{eq:D0t from L}
\tcd f:&=L^{-1}\star d\left(L\star f\star \pi(L^{-1})\right)\star \pi(L)=
d f+\picomm{ {U}^{(0)}}{f}
\,,
\end{align}
respectively.
It follows that 
\begin{align}
\label{eq:Dad as diffop}
\acd=&d+\Omega^{\au\bu}Y_\au\partial^{(Y)}_\bu -i\Omega^{\au\bu}\partial^{(Y)}_\au\partial^{(Z)}_\bu
\,,\\
\tcd=&
d
+\omega^{\au\bu}Y_\au\partial^Y_\bu-i\omega^{\au\bu}\partial^Y_\au\partial^Z_\bu
\nonumber\\&
-\frac{i}{2}e^{\au\bu}Y_\au Y_\bu-e^{\au\bu}Y_\au\partial^Z_\bu
+\frac{i}{2}e^{\au\bu}\partial^Y_\au\partial^Y_\bu+\frac{i}{2}e^{\au\bu}\partial^Z_\au\partial^Z_\bu
\,.
\end{align}

From Eq. \eqref{eq:qPhi1}, it follows that $\Phi^{(1)}$ is purely given by its cohomological part, that is,
\be\Phi^{(1)}={\cal P}\Phi^{(1)} \ \equiv \ C^{(1)}(x,Y)\,,\ee
independently of the choice of ${\cal P}$.
Decomposing $V^{(1)}$ and $U^{(1)}$, respectively, using $(q^{(A)\ast},{\cal P}^{(A)})$ and $(q^{(B)\ast},{\cal P}^{(B)})$, it follows from Eqs. \eqref{eq:qV1} and \eqref{eq:qU1+DV1} that
\begin{align} {V}^{(1)}&=-q^{(A)\ast}(\Phi^{(1)}\star {J})+qh^{(1,A)}\,,\\
{U}^{(1)}&=q^{(B)\ast}\acd \left(q^{(A)\ast}(\Phi^{(1)}\star{J})-qh^{(1,A)}\right)+W^{(1,A,B)}\,,\end{align}
where the cohomological part 
\be W^{(1,B)}={\cal P}^{(B)}{U}^{(1)}\,,\ee
is a one-form field on ${\cal X}_4$ that does not depend on the $Z$ variables.
Thus, in the $A$-gauge, 
\begin{align}
\label{eq:VA}
{V}^{(1,A)}&=-q^{(A)\ast}(\Phi^{(1)}\star {J})
\,,\\
\label{eq:WAB}
{U}^{(1,A)}&=q^{(B)\ast}\acd q^{(A)\ast}(\Phi^{(1)}\star{J})+W^{(1,A,B)}
\,;
\end{align}
as the notation indicates, the choice of $(q^{(B)\ast},{\cal P}^{(B)})$ affects $W^{(1,A,B)}$ but not ${U}^{(1,A)}$.

The remaining linearized field equations, that is, Eqs. \eqref{eq:DPhi1} and \eqref{eq:DU1}, now read 
\begin{align}
\label{eq:COMST AB}
\acd W^{(1,A,B)}=&-(\acd q^{(B)\ast})(\acd q^{(A)\ast})(\Phi^{(1)}\star{J})\,,\\ \label{D0C} D^{(0)}_{\rm tw} \Phi^{(1)}&=0 
\,,
\end{align}
which constitute a free differential algebra on ${\cal X}_4$. 
%
%The Cartan integrability of the original system on ${\cal X}_4\times {\cal Z}_4$ implies that, as the left-hand side of Eq. \eqref{eq:COMST AB} is $Z$-independent, so is its right-hand side, whose normal and Weyl ordered forms are hence equal; for further details, see for example \cite{Didenko:2015cwv}  .

The simplest choice $V^\au=0$ --- that is, $q^{(A)\ast}=q^{(B)\ast}=q^{(E)\ast}$ --- was originally used for the linearized analysis of the equations \cite{more,Vasiliev:1999ba}. This corresponds to assuming a twistor gauge condition (in normal order)
\be {\cal O}_E V^{(1,E)}:= \imath_{\vec E} V^{(1,E)} \ = \ Z^{\au}V_{\au}^{(1,E)} \ = \ 0 \ . \label{Vasg}\ee
This procedure gives rise, from \eq{eq:COMST AB}, to the condition 
\begin{equation}
\label{eq:COMST Vas}
\acd W^{(1,E,E)}=
-\frac{ib}{4}e^{\a\ad}e_{\a}{}^{\bd}\partial^\yb_{\ad} \partial^\yb_{\bd} \Phi^{(1)}(x;0,\yb)
-\frac{i\bar b}{4}e^{\a\ad}e^{\b}{}_{\ad}\partial^y_\a \partial^y_\b \Phi^{(1)}(x;y,0)
\,,
\end{equation}
which provides a cocycle $\Sigma(e,e;\Phi^{(1)})$ that glues the twisted-adjoint zero-form module to the adjoint one-form module in a manifestly Lorentz-covariant fashion. Equations \eq{eq:COMST Vas} and  \eq{D0C} decompose under $Sp(4)$ into unfolded equations of motion for a set of Fronsdal fields of spins $s\in \{0,1,2,3,\dots\}$ under the bosonic projection \eq{piofj} ($s\in \{0,2,4,\dots\}$ under the minimal bosonic projection \eq{min}), a result known as Central On Mass Shell Theorem (COMST) \cite{Vasiliev:1990vu,Vasiliev:1999ba,Vasiliev:1988sa}. 
%
%In other words, the Vasiliev system can be subjected to boundary and gauge conditions in twistor space such that its linearization around anti-de Sitter spacetime describes the gluing of a Weyl zero-form $\Phi^{(1)}$ to an adjoint one-form $W^{(1)}$ via the two-form cocycle appearing in the constraint on the linearized curvature two-form $\acd W^{(1)}$ in \eqref{eq:COMST}.
%
The spin-$s$ Fronsdal fields are identified as the components 
\be
\label{eq:comp Fronsdal}
\phi_{\mu(s)}
:=\left.
e_{(\mu_1}{}^{(0)\a_1\ad_1}\cdots
e_{\mu_{s-1}}{}^{(0)\a_{s-1}\ad_{s-1}}
(\partial^y_\a)^{s-1}(\partial^\yb_{\ad})^{s-1}W^{(1)}_{\mu_s)}
\right|_{Y=0}\ee
of the expansion of the generating function $W^{(1)}$ in monomials of all degrees in $Y$, while the remaining components of each spin-$s$ sector (degree $2(s-1)$ in $Y$) are auxiliary connections; and the gluing \eq{eq:COMST Vas} identifies the components 
\be 
\label{eq:comp Weyl}
C_{\a(2s)}:= \left.(\partial^y_\a)^{2s} C^{(1)}\right|_{Y=0}
\,,\qquad
\bar C_{\ad(2s)}:= \left.(\partial^\yb_{\ad})^{2s} C^{(1)}\right|_{Y=0}
\ee
of the twisted adjoint field $\Phi^{(1)}\equiv C^{(1)}$ as generalized spin-$s$ Weyl tensor, while the higher components $C_{\a(m)\ad(n)}$  of each spin-$s$ sector $|m-n|=2s$ are all their on-shell non-trivial derivatives (for more details see, e.g., \cite{Vasiliev:1999ba,Iazeolla:2008bp,Didenko:2014dwa}). 

Note that, for this interpretation to make sense, it is important that the initial data  $W^{(1)}$ and $C^{(1)}$ are real-analytic functions of $Y$ around $Y=0$, at least in any spacetime region in which the interpretation of their physical content in terms of Fronsdal fields holds. In the standard perturbative analysis, in fact, this is a condition that is satisfied, since it is assumed that \emph{all} master fields are formal polynomials in the oscillators \emph{everywhere} on ${\cal X}_4$ and the total inner Klein operator is real-analytic on $\cal T$ in normal order, $\k=e^{iy^\a z_\a}$ and \emph{idem} for ${\bar{\k}}$.  %the source term on the r.h.s. of the field equations is a regular function on $\cal T$ provided $\Phi$ is.

\paragraph{Higher orders and shifted homotopy contractions.} Sticking to these assumption, it is possible in principle to proceed to higher orders and obtain a formal perturbative expansion in terms of self-interacting Fronsdal fields, locally Lorentz covariant and diffeomorphism invariant on ${\cal X}_4$. Solving the higher order $Z$-space equations and plugging back into the pure spacetime equations \eq{MC} and \eq{MC2} gives rise to non-linear constraints of the form
\bea  dW& =& {\cal V}_2(W,C) \ := \ W\star W +{\cal V}_{[2]}^{(1)}(W,W,C)+{\cal V}_{[2]}^{(2)}(W,W,C,C)+...\\
dC & = & {\cal V}_1(W,C)  \ := \ [W,C]_\pi+{\cal V}_1^{(2)}(W,C,C)+... \ , \eea 
where ${\cal V}_{[p]}^{(n)}$ schematically denote $n$th-order interaction terms of form-degree $p$. However, the standard homotopy contraction $q^{(E)\ast}$ is known to lead to non-local interaction vertices \cite{Boulanger:2015ova,Skvortsov:2015lja}. Thus, certain shifted homotopies  $q^{(E+V)\ast}$ have been proposed and studied in \cite{Vasiliev:2017cae,Gelfond:2018vmi,Didenko:2018fgx,Didenko:2019xzz,Gelfond:2019tac} that give rise to interaction vertices exhibiting \emph{spin-locality}, i.e. locality of the vertices in $Y$-space --- which coincides with spacetime locality at the lowest order only, and may provide criteria to restrict the large freedom in gauge transformations and field redefinitions. In these modified resolution operators, the vector $Z^{\ua}$ in $q^{(E)\ast}$ is shifted with combination of $Y$-derivatives acting on the master fields entering a given vertex, plus an extra shift to include explicit differentiation over the $y$ variable $\b \frac{\partial}{\partial y}$, with $-\infty< \b< 1$. Certain type of vertices up to quartic order (at the level of the field equations) have been shown to be spin-local in the limit $\b\to -\infty$ \cite{Didenko:2019xzz}. It will be interesting to see what can be said about other vertices, such as those of type ${\cal V}^{(4)}_{[1]}(W,C,C,C)$, suspected to be problematic from other types of analysis \cite{Bekaert:2015tva,Sleight:2017pcz}, and whether such scheme can be extended to all orders. Besides, it would also be important to get a deeper understanding of the meaning of spin-locality from a more geometric perspective. 

\paragraph{An alternative perturbative scheme.} An alternative perturbative scheme was independently proposed in \cite{DeFilippi:2019jqq}, based on ideas developed in a series of works \cite{us,us2,Iazeolla:2015tca,Iazeolla:2017vng,review,Aros:2017ror,Aros:2019pgj} that had a different point of view, i.e., finding and systematizing a convenient method to build exact solution spaces.  This scheme  is articulated in two perturbative reconstructions : 

\begin{enumerate}

\item first, it makes use of a specific decomposition $(q^{\ast},{\cal P})$ to facilitate the perturbative analysis of a large solution space --- in fact, enabling one to push the resolution to all orders under certain conditions, thereby obtaining full solutions in closed form; 

\item then, it perturbatively reconstructs a gauge function that connects to a different decomposition $(q^{\ast\prime},{\cal P}')$ in which Fronsdal fields can be extracted at first order (glued to unmodified linearized Weyl curvatures by means of the COMST \eq{eq:COMST Vas}). Beyond the first order, it was proposed that the gauge function can be determined by imposing boundary conditions at the level of master fields.

\end{enumerate}

Step 1 consists of singling out a choice of homotopy contraction that greatly simplifies the perturbative analysis. The need for step 2 arises from the fact that in the frame determined by $(q^{\ast},{\cal P})$ there is no cocycle gluing the spacetime one-form connection and the twisted-adjoint zero-form. In other words, there is no dualization of the local degrees of freedom, contained in $\Phi^{(1)}$, into $U^{(1)}$, the two modules remain disconnected and therefore it is possible to gauge away all fluctuations from the spacetime one-form. Step 2 thus amounts to a procedure to find a family of gauge functions that map the propagating degrees of freedom, encoded in the zero-form integration constants, to Fronsdal fields. More precisely, the gauge function provides the Chevalley-Eilenberg cocycle required by the COMST, that glues the on-shell curvatures of the Fronsdal fields, i.e. their Weyl tensors, to the corresponding components of the Weyl zero-form already built in step 1 (and which remain untouched by the transformation at first order).  The gauge function is determined at first order by solving a condition which is a relaxed version of the gauge condition \eq{Vasg} used in the ``standard'' perturbative analysis. As proposed in \cite{DeFilippi:2019jqq}, the remaining ambiguities can be fixed order by order by imposing asymptotically AdS (AAdS) boundary conditions on the solutions, which can in principle determine the choice $(q^{\ast\prime},{\cal P}')$ at higher orders while directly connecting it to the construction of a specific superselection sector (as well as to selecting a specific class of functions of the oscillators).

\paragraph{Perturbative expansion in factorized gauge.} Step 1 exploits two features of the Vasiliev system: 

\begin{itemize}

\item[a)] First, the linearized equations \eq{eq:qPhi1} and \eq{eq:DPhi1} always imply that the first-order Weyl zero-form is $Z$-independent and $AdS$-covariantly constant, i.e.,
\be \Phi^{(1)} \ = \ {\cal P}\Phi^{(1)} \ = \ C^{(1)}(x,Y) \ , \qquad   C^{(1)}(x,Y) \ = \ L^{-1}\star C^{\prime(1)}(Y)\star \pi(L) \ , \label{LCpiL}\ee
where $C^{\prime(1)}(Y)$ is an $x$-independent fibre element, and $L$ is the $AdS_4$ gauge function \eq{eq:AdS L} adapted to a chosen coordinate system. The last equation is the statement that the integration constants $C^{\prime(1)}(Y)$ contain, as coefficients of their expansion in $Y$, all spacetime derivatives of the physical fields that are invariant under linearized higher-spin gauge transformations evaluated at the unfolding point. The gauge function $L$ then ``spreads'', or ``unfolds'' this local datum on the spacetime chart where $L$ is defined \cite{Vasiliev:1999ba,Bekaert:2005vh,Vasiliev:1990bu}.  In fact, due to the structure of the equations, any solution can (locally) be written by means of a gauge function, even at the fully nonlinear level --- a solution method frequently referred to as \emph{gauge function method} \cite{us,us2,review,Aros:2017ror,Sezgin:2005pv,Vasiliev:1990bu,Iazeolla:2007wt,Sundell:2016mxc}. The vacuum gauge function $L$ was also used in the same way, to locally gauge away the spacetime dependence, for the calculation of the holographic correlation function in \cite{Giombi:2010vg}. At first order one can write the solutions to \eq{eq:qPhi1}-\eq{eq:DU1} in terms of the vacuum gauge function $L$ and a (linearized) gauge function $H^{(1)}$,
\begin{equation}
{U}^{(1)}=\acd{H}^{(1)}\,,\qquad  \Phi^{(1)}=C^{(1)}=L^{-1}\star C^{\prime(1)}\star \pi(L)\,, \label{UCfirst}\end{equation}
\begin{equation}  V^{(1)}=L^{-1}\star  V^{\prime(1)}\star L+q{H}^{(1)}\,,\label{Vfirst}
\end{equation}
where $V^{\prime(1)}$ is a particular solution to the pure twistor space conditions
\be  q V^{\prime(1)}+C^{\prime(1)}\star J=0\,,\qquad dV^{\prime(1)}=0\,.\ee 

Eq. \eq{LCpiL} can equivalently be rewritten in terms of the adjoint element $\Psi$,
\be \Psi \ := \ \Phi\star \k_y \ = \ L^{-1}\star\Psi'\star L\ . \  \label{psipr}\ee
The notion of integration constants, encoding the propagating degrees of freedom, can therefore equivalently be referred to $\Phi^{\prime(1)}\equiv C^{\prime(1)}$ and to $\Psi^{\prime(1)}:=\Phi^{\prime(1)}\star\k_y$. 

\item[b)]  By virtue of the identity \eq{eq:k fact}, the non-trivial source terms of the Vasiliev system featuring on the r.h.s. of \eq{INT2}, which trigger all the non-linear corrections, can be rewritten as
\be \Phi \star \k \ = \ \Psi \star \k_z  \ ,\qquad  \Phi \star \bar \k \ = \ \bar \Psi \star \bar \k_{\zb} \ , \ee
where $\bar \Psi := \Phi\star\bar \k_{\yb}$. Therefore, taking into account the remark made in a), the $Z$-dependence of the source comes exclusively from $\k_z$ ($\bar \k_{\bar z}$ in the anti-holomorphic sector). This suggests that, if we organize the perturbative expansion of all fields in terms of star-powers of $\Psi$ and \emph{keep the $Y$ and $Z$ dependence of all the master fields factorized}, we can actually solve \emph{universally} for the $Z$-dependence. 
 
\end{itemize}

Indeed, $\Psi$ can be kept first-order exact, $\Psi=\Psi^{(1)}$, and inserting
\be V \ = \ \sum_{n=1}^{\infty}V^{(n)} \ = \ \sum_{n=1}^\infty \left((\Psi^{(1)})^{\star n}\star v_n(z)+(\bar \Psi^{(1)})^{\star n}\star \bar v_n(z)\right) \ , \label{VPsi}\ee
into Eq. \eq{INT2}, which, in components and for the holomorphic sector (with the obvious notation $\partial_\a=\partial/\partial z^\a$), reads
\bea \partial_{[\a} V_{\b]}+V_{[\a}\star V_{\b]}  & = & -\frac{ib}{4}\,\e_{\a\b} \, \Psi^{(1)}\star\k_z \ , 
\eea
the latter separates into the conditions for the $Z$-dependent coefficients in \eq{VPsi},
\bea 
& \displaystyle \partial_{[\a} v_{1\b]} \ = \ -\frac{ib}{4}\,\e_{\a\b} \,\k_z \ , \qquad n=1 \ ,& \label{curlVd}\\
& \displaystyle  \partial_{[\a} v_{n\b]} + \frac{1}{2} \sum_{p+q=n}\left[v_{p\a}, v_{q\b}\right]_\star \ = \ 0 \ ,\qquad n\geq 2 \ ,&\label{Hordcurl}
\eea
which can be solved by the standard homotopy contraction. The first equation \eq{curlVd} essentially says that the curl of the first-order coefficient $v_{1\a}$ is a delta function, i.e., admits a solution in terms of distributions in $z$.  We shall soon recall how such distributions can be handled as star-product elements. The sourceless, higher-order equations \eq{Hordcurl} can be solved by known techniques, developed in \cite{us,Sezgin:2005pv,Iazeolla:2007wt,Prokushkin:1998bq}, see Appendix C in \cite{DeFilippi:2019jqq}.  Note that we have chosen a \emph{holomorphic factorized gauge} for $V$ \eq{VPsi}, in which the $v$ ($\bar v$) coefficients are only dependent on $z$ ($\bar z$), see \cite{Iazeolla:2017vng,DeFilippi:2019jqq} for details. 

Recalling the observations at the end of Section \ref{Sec:corr} on the Weyl-ordering prescription, it is clear that this entire procedure of solving with factorized $Y$ and $Z$ dependence  can equivalently be described \cite{DeFilippi:2019jqq} by solving the $Z$-space equations \emph{in Weyl ordering}. More precisely, the above solution method with its gauge choices can be rephrased by using a shifted homotopy contraction: in fact, homotopy contracting in normal order along $\vec E+i\vec \partial_Y$ with $\vec \partial_Y:=\partial^{\au}_Y \vec\partial^Z_{\au}$,
\begin{equation}
\label{eq:rhoF}
q^{(E+i\partial_Y)\ast} g = \imath_{\vec E+i\vec \partial_Y} 
\int_0^1 \frac{d t}{t}\int d^4Y' g(x,tZ+i(t-1)\partial_Y; dx,tdZ;Y')
\delta^4\left(Y-Y'\right)
\,,
\end{equation}
is equivalent to homotopy contracting in Weyl order using the standard homotopy $q^{(E)\ast}$, viz.
\begin{equation}
q^{(E+i\partial_Y)\ast} = \hat\tau^{-1}q^{(E)\ast}\hat\tau 
\,,
\end{equation}
where 
\begin{equation}
\hat\tau f(Y,Z)
:=
\int\frac{d^4Y' d^4Z'}{(2\pi)^4}
\exp\left(-i(Y-Y')(Z-Z')\right)
f(Y',Z')
\,,
\end{equation}
maps symbols from normal to Weyl order (that is, if $f_N$ and $f_W$, respectively, are
the Weyl and normal ordered symbol of an operator, then $f_W=\hat\tau f_N$).

The advantage of homotopy contracting in Weyl order is the factorization property
\begin{equation}
\label{eq:rhoF fact}
q^{(E+i\partial_Y )\ast}(f(Y)\star g(Z; dZ))
=
f(Y)\star q^{(E)\ast} g(Z; dZ)
\,,
\end{equation}
which is what enables solving universally for the $Z$ dependence, as sketched above in more elementary terms, provided that $q^{(E)\ast} j_z$ can be somehow regularized. It is interesting to note that $q^{(E+i\partial_Y)\ast} $ is equivalent to the case $\b=1$ of the shifted contracting homotopies studied in \cite{Didenko:2019xzz}\footnote{In fact, $\b\to1$ is the opposite limit of the $\b\to -\infty $ one that was preferred there, as it guarantees spin-locality of the vertices examined in that work. In \cite{Didenko:2019xzz} it was also shown how all $\b$-shifts in $q^{\ast}$, with $-\infty<\b<1$ can be related to reorderings, parameterized by $\b$. The main reason why the $\b=1$ case was not used in \cite{Didenko:2019xzz} is precisely that, in Weyl ordering, the source term of the Vasiliev becomes a distribution, which leads to divergencies in the solution for $V$. As we shall recall later on in this work, however, in \cite{DeFilippi:2019jqq} said divergencies are interpreted as reconstructing distributions in twistor space, and the latter are handled, as elements of an enlarged star-product algebra, by means of a regularization scheme. Then, in performing Step 2. of our perturbative scheme, the singularities are shown to be harmless, at least at first order, in the gauge in which one reads off the Fronsdal fields, as they are cohomologically trivial \cite{DeFilippi:2019jqq}.}.  In terms of $q^{(E+i\partial_Y)\ast} $, then, the solution to Eq.\eqref{eq:qV1},
\begin{equation}
q V^{(1)}=-\Psi^{(1)}\star j_z-\bar\Psi^{(1)}\star\jb_z
\,,\label{qV1}
\end{equation}
is given by
\begin{equation}
\label{eq:V1 fact}
V^{(1)}=
-\Psi^{(1)}\star q^{(E)\ast} j_z-\bar\Psi^{(1)}\star q^{(E)\ast} \jb_z
\,.
\end{equation}
where property \eq{eq:rhoF fact} was used. Since $\acd$ annihilates $\Psi^{(1)}$ and $\bar\Psi^{(1)}$,
we can rewrite Eq.\eqref{eq:qU1+DV1} as
\begin{equation}
qU^{(1)}=\Psi^{(1)}\star d q^{(E)\ast} j_z+\bar\Psi^{(1)}\star d q^{(E)\ast} \jb_z\,.
\end{equation}
Assuming that $q^{(E)\ast} j_z$ and $q^{(E)\ast} \jb_z$ are $x$-independent (which is possible, see \eq{kzreg}),
we then have
\begin{equation}
\label{eq:U1=0}
U^{(1)}=W^{(1)}\,,
\end{equation}
which we can gauge-fix to zero by virtue of Eq.\eqref{eq:DU1}.

Fixing gauges and proceeding to higher orders, one can recursively prove that the solution
\begin{align}
\label{eq:Phi^m fact}
\Phi \ = \ \sum_{n=1}^\infty \Phi^{(n)}\ , \qquad \Phi^{(n)}&=\delta_{n,1}\left(
\Psi^{(1)}\star\k_y+\bar\Psi^{(1)}\star\kb_y
\right)
\,,\\
\label{eq:V^m fact}
V \ = \ \sum_{n=1}^\infty V^{(n)}\ , \qquad V^{(n)}&=
(\Psi^{(1)})^{\star n}\star v_n(z)+({\bar\Psi^{(1)}})^{\star n}\star\bar{v}_n(\zb)
\,,\\
\label{eq:U^m fact}
U\ = \ \sum_{n=0}^\infty U^{(n)}\ , \qquad U^{(n)}&=\delta_{n,0}\,\Omega
\,,
\end{align}
is actually a formal exact solution to the Vasiliev equations. Note that, with this choice of homotopy contraction and gauges, $\Phi$ (equivalently, $\Psi$) is first-order exact, $U$ remains uncorrected (up to pure gauge terms), identical to its vacuum value \eq{AdSconn}, and the $Z$-dependence is solved in a universal manner. 

For \eq{eq:Phi^m fact}-\eq{eq:U^m fact} to be considered an actual solution, for a given initial datum $\Psi'(Y)$, we should be able to make sense of the resulting distributions in twistor space (of $q^{(E)\ast} j_z$, in particular) and the star products that build up the solutions, $(\Psi^{(1)})^{\star n}$ as well as $(\Psi^{(1)})^{\star n}\star v_n$ must be finite. Note that, in the standard perturbative scheme, this is not a problem because one assumes that all master fields are formal polynomials, $\k=e^{iy^\a z_\a}$ is regular in normal order, and one can prove \cite{more,Prokushkin:1998bq} that all functions obtained by solving in such perturbative scheme are also regular. In solving easily with $q^{(E+i\partial_Y)\ast} $, instead, the resulting delta function source term induces a distributional behaviour of the $v_n(z)$. Moreover, as found in \cite{Iazeolla:2008ix,us,us2,Iazeolla:2017vng,Aros:2017ror,Aros:2019pgj,Didenko:2009td,Giombi:2009wh,Giombi:2010vg,Kraus:2012uf}, many interesting solutions to the free and/or fully non-linear equations are actually in correspondence, via unfolding, with non-polynomial fibre elements that, in certain cases, may give rise to star-product divergencies. Both problems can actually be handled by means of \emph{regular presentations} --- i.e., typically, integral transforms of non-polynomial elements with gaussian kernels in $Y$ and $Z$, ensuring finiteness and closure of the resulting star-product algebra. 

More precisely, in this \emph{regular computational scheme} (see \cite{Aros:2019pgj} and \cite{DeFilippi:2019jqq} for details) the $Z$-dependent coefficients of the Weyl-ordered solution \eq{eq:V^m fact}, starting with $v_1$ being a potential for a delta-function source, are represented via open-contour integrals similar to a Schwinger parametrization. Let us introduce a \emph{spin-frame} $(u^+_\a,u^-_\a)$, that is, a holomorphic metric on ${\cal Z}_2$,
\be ds^2_z:= {\cal D}_{\alpha\beta}dz^\alpha dz^\beta\,,\qquad {\cal D}_{\alpha\beta}:=2u^+_{(\alpha} u^-_{\beta)}\,,\qquad u^{+\a} u^-_\a=1\ee
(analogously for their anti-holomorphic counterparts), to get a representation of the delta function in gaussian basis \cite{us,Iazeolla:2017vng,DeFilippi:2019jqq}, 
\be  \k_z \ = \ \lim_{\e\to 0^+} \frac{1}{\e}\,e^{\frac{i}{2\e}z{\cal D}z} \ = \ 2
\int_{-1}^{1}\frac{d s}{1+s}\,\delta(1+s)
\exp\left(
\tfrac{i}{2}\tfrac{1-s}{1+s}\, z{\cal D}z
\right) \label{kzreg}
\ee
where spinor products with omitted indices are meant in terms of matrix notation, with the NW-SE contraction rule (see Appendix \ref{App:conv}). We shall use this notation from now on, whenever unambiguous. Then, a parametric integral presentation of $v_1(z)$ satisfying \eq{curlVd} is 
\be   v_{1\a} \ = \ -\frac{ib}{2}\, z_\a \int_{-1}^1 \frac{dt}{(t+1)^2}\,e^{-\ft{i}2\ft{t-1}{t+1}z{\cal D}z} \ = \ -\frac{ib}{4}\,z_\a\,\int_0^{+\infty} d\t \,e^{-i\t w_z} \,,\label{intpresv1}\ee
that is
\be v_{1}^\pm \ = \ -\frac{b}{4}\frac{1}{z^\mp}\left(1-\lim_{\e\to 0}e^{-\ft{i}{\e}z^+ z^-}   \right) \,, \label{eq:V1bdry}\ee
where $w_z := -\frac12 z{\cal D}z \equiv z^+ z^-$, $z^\pm:=u^{\pm}z$. One can show that, as expected, such integral is indeed a representation of the distribution $\theta(z^\pm)\delta(z^\mp)$ \cite{DeFilippi:2019jqq}. The divergencies of the integrand at the boundary of the integration domain are therefore interpreted as encoding distributions. Integral presentation for all the $v_n(z)$ can be obtained from this one \cite{Iazeolla:2017vng,Aros:2017ror,DeFilippi:2019jqq}, leading to the general solution for the $Z$-space connection 
\bea  V_\a \ = \ \sum_{k\geq 1}v_{n\a} \star( \Psi^{(1)})^{\star n} \ = \ \int_{-1}^1 \frac{dt}{(t+1)^2} {}_1F_{1\star}\left(\frac12; 2; b\log t^2\,\Psi^{(1)}\right)\star z_\a\,e^{i\frac{t-1}{t+1}w_z}  \ .\eea

Essential part of the regular computational scheme is, of course, the prescription that any operation involving oscillators --- derivatives, star products, traces, homotopy integrals --- be performed on the integrand, before evaluating the auxiliary integrals (see \cite{DeFilippi:2019jqq} for details). As for the basis elements on which to expand $\Psi'(Y)$, which define the fibre algebra ${\cal A}({\cal Y}_4)$, their regular presentation will be recalled in Section \ref{Sec:genproj}.

Thus, this perturbative scheme is, so far, is entirely driven by the fibre elements $\Psi^{\prime (1)}(Y)$ (the gauge function $L$ is the $AdS_4$ vacuum one, and the $Z$-dependent coefficient are solved for universally), and it is important to stress that the requirement of finiteness of the star products in \eq{eq:V^m fact} puts non-trivial restrictions on the class of functions to which the elements of the fibre algebras  ${\cal A}({\cal Y}_4)$ can belong. While in this frame we are not directly able to extract Fronsdal fields, some gauge-invariant quantities of the Vasiliev system can be computed\footnote{See \cite{Sundell:2016mxc} and \cite{DeFilippi:2019jqq} for a definition of trace adapted to ${\cal A}({\cal Y}_4)$.} \cite{review,DeFilippi:2019jqq,Sezgin:2005pv,Sezgin:2011hq,FCS,Bonezzi:2016ttk}, and requiring that they be finite gives more constraints on  ${\cal A}({\cal Y}_4)$ (including restrictions on the admissible regular presentations). 

Finally, it is worth recalling that, for many interesting exact solutions (such as the higher-spin black holes \cite{us,us2,Iazeolla:2017vng,Didenko:2009td}), the solution given by \eq{eq:V^m fact}  is regular on twistor space at generic spacetime points (only irregular at $r=0$ for spherically symmetric black holes, for example) in normal order (that is, after the star products are computed), even though the $v_n(z)$ coefficients are singular.

%%%%%%%%%%%%%%%%%%%%%%%%%%%%
\paragraph{Fronsdal fields and AAdS boundary conditions.}
%%%%%%%%%%%%%%%%%%%%%%%%%%%%
 
Interpreting the solution found in Step 1 in terms of interacting Fronsdal fields requires a change of resolution frame $(q^{\ast},{\cal P})$ which can be induced by a modification of the gauge function 
\be L  \ \longrightarrow \  G \ := \ L\star \left(1+\sum_{n=1}^{\infty} H^{(n)} \right)\ ,  \ee
 where $H^{(1)}$ can be determined via imposing a gauge condition, and higher orders from imposing AAdS boundary conditions at the level of master fields. 
 
Starting from a gauge $A$ and transforming to $G$, $\Phi$, encoding the propagating degrees of freedom, will remain untouched in the first order, while $H^{(1)}$ will modify $V$ and $U$ as
\begin{align}
\label{eq:HAB}
{V}^{(1,G)}&={V}^{(1,A)} + q {H}^{(1,A\to G)}
\,,\\
\label{eq:U1gt}
{U}^{(1,G)}
&=
{U}^{(1,A)}+\acd {H}^{(1,A\to G)}
\,.
\end{align}

 The generating function of Fronsdal fields will be given by 
\be W^{(1,G,G)}\equiv 
{\cal P}^{(G)} U^{(1,G)} \ , \label{Gproj}
\ee
with the cohomological projection
\be {\cal P}^{(G)}f:= f|_{Z=0}
\label{eq:proj Vgauge}
\,,\ee
where $Z$ is set to zero in normal order (prior to performing all auxiliary integrals). Clearly, it is required that $W^{(1,G,G)}$ be real analytic on ${\cal Y}_4$ at $Y=0$.

Then, a gauge function that maps solutions $(\Phi^{(1,E+i\partial_Y)},V^{(1,E+i\partial_Y)},U^{(n,E+i\partial_Y)})$ found in the Weyl-ordered scheme to a different decomposition $(q^{\ast G},{\cal P}^{G})$ in which the COMST holds --- and in which, therefore, a non-trivial $W^{(1)}$ is activated and connected to the propagating degrees of freedom in $\Phi^{(1)}$ --- can be obtained, in the first order, by solving a gauge condition that relaxes \eq{Vasg}. In fact, it only sets to zero the interior product of $V$ along $\vec E$ up to terms of at least second order in $Z$, schematically $O(Z^2)$.  More precisely, the gauge condition can be written as
\be 
\label{eq:ZV=O(Z2)}
{\cal O}_G V^{(1,G)}:= \imath_{\vec E} V^{(1,G)}- \vec E H_2^{(1,G)}=0\,,
\ee
where $H_2^{(1,G)}$ is an arbitrary symbol such that
\begin{equation}\label{eq:H2atZ=0}
{\cal P}^{(G)}\acd H_2^{(1,G)}=0
\,,
\end{equation}
 which implies that $H_2^{(1,G)}$ does not affect the definition of the gauge field generating function. Indeed, as shown in \cite{DeFilippi:2019jqq}, this gauge can be reached starting from any gauge $A$ in which $V^{(1,A)}$ and $U^{(1,A)}$ are real-analytic in $Z$ at $Z=0$ in normal order (which gives another condition on ${\cal A}({\cal Y}_4)$, that is satisfied by the solutions found in Step 1 for a set of elements $\Psi'(Y)$ that we shall examine later on). As \eq{Gproj} itself requires $U^{(1,G)}$ to be analytic in $Z$ at $Z=0$, \eq{eq:H2atZ=0} implies that $H_2^{(1,G)}$ is a $O(Z^2)$ function. 
 
${H}^{(1,A\to G)}$ can then be explicitly constructed by inserting \eq{eq:HAB} into \eq{eq:ZV=O(Z2)} and solving, thereby obtaining
\begin{align}
{H}^{(1,A\to G)}
&=
-\frac{1}{{\cal L}_{\vec E}}\imath_{\vec E} V^{(1,A)}
+h^{(1,A\to G)}
+H_2^{(1,A\to G)}
\nonumber\\&=
-\int_0^1 d t\,Z^{\au} V^{(1,A)}_{\au}(x,tZ,Y) +h^{(1,A\to G)} +H_2^{(1,A\to G)}
\,,
\label{eq:H = int VL + h}
\end{align}
 where $H_2^{(1,A\to G)}$ is a function satisfying Eq.\eqref{eq:H2atZ=0}
and where
$h^{(1,A\to G)}\in {\rm Ker}({\cal O}_G) $, i.e., due to the analyticity requirements on ${\cal Z}_4$, is an arbitrary $Z$-independent function,
\be qh^{(1,A\to G)}=0\,.\ee

 Thus, the generating function for the spacetime gauge fields is given by
\begin{align}
W^{(1,G,G)}
=& 
{\cal P}^{(G)}U^{(1,A)}
+
{\cal P}^{(G)}\acd H^{(1,A\to G)}
\nonumber\\=&
{\cal P}^{(G)}U^{(1,A)}
+i\Omega^{\bu\underline{\gamma}}\partial^Y_{\bu}{\cal P}^{(G)}\partial^Z_{\underline{\gamma}} \int_0^1 dt\,Z^{\au} V^{(1,A)}_{\au}(x,tZ,Y)+\acd h^{(1,A\to G)}
\nonumber\\=&
\label{eq:def omega_phys}
{\cal P}^{(G)}U^{(1,A)}
+
i\Omega^{\au\bu}\partial^Y_{\bu}{\cal P}^{(G)} V^{(1,A)}_{\au}
+\acd h^{(1,A\to G)}
\,.
\end{align}
Now, due to the distributional nature of the source term in the Weyl-ordered scheme, for some linearized field configuration (such as $AdS$ massless particles, see Section \ref{fibresol} and \cite{Iazeolla:2017vng,DeFilippi:2019jqq}) $V^{(1,A)}=V^{(1,E+i\partial_Y)}$ inherits the singular behaviour of the $v_n(z)$ coefficients, only exhibiting it in $Y$ instead of $Z$. As a consequence of \eq{eq:H = int VL + h}, ${H}^{(1,A\to G)}$ will inherit such singularities. The $Z\to Y$ trade-off of the latter is all-important, however, since singularities in $Y$ can entirely be absorbed by $h^{(1,A\to G)}$, leaving a regular gauge field generating function. In other words, as \eq{eq:def omega_phys} suggests, the singularities produced by the Weyl-ordered perturbative solution are cohomologically trivial. Requiring real-analyticity on ${\cal Y}_4$ at $Y=0$ fixes $h^{(1)}$ modulo its real-analytic part. In \cite{DeFilippi:2019jqq} the resulting Fronsdal field generating functions were given explicitly for both massless particles and higher-spin black holes.  Indeed, by applying $D^{(0)}$ to the r.h.s. of  \eq{eq:def omega_phys}, it is possible to check that $W^{(1,G,G)} $ satisfies the COMST \cite{DeFilippi:2019jqq},
\begin{align}
\acd W^{(1,G,G)}
&=
%\acd {\cal P}^{(G)}U^{(1,A)}
%+
%i\left(\left\{\Omega^{\au\bu}\partial^Y_{\bu},\acd\right\}-\Omega^{\au\bu}\partial^Y_{\bu}\acd\right){\cal P}^{(G)} V^{(1,A)}_{\au}
%\nonumber\\&=
%\acd{\cal P}^{(G)}U^{(1,A)}
%-i\Omega^{\au\bu}\partial^Y_{\bu}\acd{\cal P}^{(G)}V^{(1,A)}_{\au}
%\nonumber\\&=
%i\Omega^{\au\bu}\partial^Y_{\au}{\cal P}^{(G)}\left(\partial^Z_{\bu} U^{(1,A)}-\acd V^{(1,A)}_{\bu}\right)
%+\Omega^{\au\bu}\partial^Y_{\bu}\Omega^{\cu\bu}\partial^Y_{\bu}{\cal P}^{(G)}\partial^Z_{[\cu} V^{(1,A)}_{\au]}
%\label{eq:COMST eom step}\\
%&=
%-\Omega^{\au\bu}\partial^Y_{\bu}\Omega^{\cu\bu}\partial^Y_{\bu} {\cal P}^{(G)}(\Phi^{(1)}\star J)_{\cu\au}
%\nonumber\\&=
%-\frac{ib}{4}\Omega^{\a\bu}\partial^Y_{\bu}\Omega^{\c\bu}\partial^Y_{\bu}{\cal P}^{(G)}\left(\epsilon_{\c\a}\Phi^{(1)}(x;-z,\yb)e^{iyz}\right)
%-{\rm h.c.}
%\nonumber\\&=
\label{eq:COMST H1}
-\frac{ib}{4}e^{\a\ad}e_{\a}^{\phantom\a\bd}\partial^\yb_{\ad} \partial^\yb_{\bd} \Phi^{(1)}(x;0,\yb)
-\frac{i\bar b}{4}e^{\a\ad}e^{\b}_{\phantom\a\ad}\partial^y_\a \partial^y_\b \Phi^{(1)}(x;y,0)
\,.
\end{align}

As we pointed out, this scheme determines ${H}^{(1,A\to G)}$ only up to some $O(Z^2)$ construct $H_2^{(1,A\to G)}$. This freedom can actually be used to impose AAdS bondary conditions order by order on the solutions (while at the same time keeping analyticity in the gauge field generating function), in an iterative procedure that can in principle determine the higher orders ${H}^{(n,A\to G)}$. Field configurations for which this is possible would indeed admit an interpretation in terms of interacting Fronsdal fields. 

Letting $r$ denote a coordinate such that the limit $r\to\infty$ defines the boundary of $AdS_4$, the idea is to set up a Fefferman-Graham-like scheme at the level of master fields. Therefore, on a fixed chosen basis ${\cal B}$ for ${\cal A}({\cal Y}_4)$, we expand the master fields $(\Phi,U,V)$ in powers of $1/r$, and fix AAdS boundary conditions by demanding that 
\begin{align}
\label{eq:AAdS bc}
\Phi^{(G)}&=\widetilde{\Phi}^{(G)}+O_{\cal B}(1/r)
\,,&
V^{(G)}&=\widetilde{V}^{(G)}+O_{\cal B}(1/r)
\,,&
U^{(G)}&=\widetilde{U}^{(G)}+O_{\cal B}(1/r)
\,,
\end{align}
where $O_{\cal B}(1/r)$ stands for forms that are sub-leading in the $1/r$ expansion; and $\widetilde{\Phi}^{(G)}$, $\widetilde{V}^{(G)}$ and $\widetilde{U}^{(G)}$ form a solution to the linearized Vasiliev equations (\ref{eq:qPhi1}--\ref{eq:DU1}),
\bea
\label{eq:AsVE Phi}
&
q\widetilde{\Phi}^{(G)}
= 0
\,, \qquad \quad
\tcd\widetilde{\Phi}^{(G)}
= 0  
\,,&\\
\label{eq:AsVE UV}
& q\widetilde{V}^{(G)}+\widetilde{\Phi}^{(G)}\star {J}
=0
\,,\qquad
q\widetilde{U}^{(G)}+\acd\widetilde{V}^{(G)}
=0
\,,\qquad 
\acd\widetilde{U}^{(G)}
=0\,, &
\eea
that encode free unfolded Fronsdal fields. The master fields admit expansions in perturbation theory. Thus, expanding both sides of the AAdS conditions \eq{eq:AAdS bc}, at every order $n$ we get, schematically,
\begin{align}
\label{eq:AsGlu Phi}
\widetilde\Phi^{(n,G)}
&=
C^{(n,A)}+\Phi^{(n,G)}_{\rm l.o.}
+O_{\cal B}(1/r)
\,,\\
\widetilde{V}^{(n,G)}
&=
{\cal V}^{(n,A)}[C^{(n,A)}]+qH^{(n,A\to G)}+V^{(n,G)}_{\rm l.o.}
+O_{\cal B}(1/r)
\,,\\
\label{eq:AsGlu U}
\widetilde{U}^{(n,G)}
&=
{\cal U}^{(n,A)}[C^{(n,A)}]+\acd H^{(n,A\to G)}+U^{(n,G)}_{\rm l.o.}
+O_{\cal B}(1/r)
\,.
\end{align}
where %$(\widetilde\Phi^{(n,G)},\widetilde{V}^{(n,G)},
%\widetilde{U}^{(n,G)})$ obey the free master field equations (\ref{eq:AsVE Phi},\,\ref{eq:AsVE UV}) for each $n$,  
$\Phi^{(n,G)}_{\rm l.o.}$, $V^{(n,G)}_{\rm l.o.}$ and $U^{(n,G)}_{\rm l.o.}$ are particular solutions for the bulk master fields that only depend on the lower order moduli contained in $C^{(n'<n,A)}$ and $H^{(n'<n,A\to G)}$, $C^{(n,A)}$ are the homogeneous solution, and ${\cal V}^{(n,A)}[C^{(n,A)}]$ and ${\cal U}^{(n,A)}[C^{(n,A)}]$ solve 
\begin{align}
q{\cal V}^{(n,A)}[C^{(n,A)}]+C^{(n,A)}\star {J}
&=0
\,,&
q{\cal U}^{(n,A)}[C^{(n,A)}]+\acd{\cal V}^{(n,A)}[C^{(n,A)}]
&=0
\,,&
\acd{\cal U}^{(n,A)}[C^{(n,A)}]
&=0\,.
\end{align}
At every order, the perturbative corrections to the bulk master fields involve star-product interactions that may affect their leading order in the asymptotic expansion. The idea is then to use the freedom in $H_2^{(m,A\to G)}$ and $C^{(m+1,A)}$ to adapt them to the lower-order constructs and impose that the r.h.s. of \eq{eq:AsGlu Phi}-\eq{eq:AsGlu U} indeed obey \eq{eq:AsVE Phi}-\eq{eq:AsVE UV} \cite{DeFilippi:2019jqq}.

In fact, it is in principle possible to use them to impose the stronger constraint that the full master fields linearize asymptotically (\emph{maximal subtraction scheme}), i.e., for $n\geq 2$
\begin{align}
\label{eq:maxsub bc}
\Phi^{(n,G)}&=O_{\cal B}(1/r)
\,,&
V^{(n,G)}&=O_{\cal B}(1/r)
\,,&
U^{(n,G)}&=O_{\cal B}(1/r)
\,.
\end{align}
Thus, in general, imposing boundary conditions will generate corrections to the zero-form initial data as well as to the gauge function. Note in particular that upon adapting every $C^{(n,A)}$ in terms of $\Phi^{(n,G)}_{\rm l.o.}$, which are $n$-linear functionals of $\Phi^{(1)}$, observables like the on-shell action functional $K =\oint_{{\cal Z}_4} {\rm Tr}_{{\cal A}({\cal Y}_4)}\left(
\Phi\star\Phi^{\dagger}\star J^{\star2}\right)$ \cite{DeFilippi:2019jqq,FCS,Bonezzi:2016ttk} turn into an infinite expansion in $\Phi^{(1)}$ (see \cite{DeFilippi:2019jqq} for more comments and details).

An important question to be addressed in the future is whether or not the perturbative approaches described in this Section are equivalent. A proper comparison can only be carried out at the level of the gauge-invariant observables of the theory, that coordinatize the solution space. In this respect note that, at Step 1 of our perturbative procedure, we have a large moduli space which essentially consists of the integration constants $\Phi'^{(n)}$ and of the local data $H'^{(n)}$ for the gauge function (modulo small gauge transformations).  As described above, Step 2 relates these two data: imposing boundary conditions on given bulk solutions should indeed restrict the moduli space accessible to them, fixing a superselection sector. Then, within a given sector, constructs based on asymptotic spacetime data (such as asymptotic charges) also become relevant observables. In summary, a well-defined notion of equivalence of two solution schemes first of all relies on being able to control boundary conditions, and on singling out the relevant set of observables that coordinatize the corresponding accessible sector of the moduli space.

\scs{Linearized solution spaces and spacetime/fibre duality.}\label{Sec:linsol}

The perturbative scheme explored in the previous Section makes use of a convenient choice of homotopy contraction to construct a solution space in which $\Phi$ is first-order exact and all non-linear correction are shifted into $V$. If the star products $\Psi^{\star n}$ make sense (i.e., are not divergent or are at least suitably regularizable) and close over a common basis of functions for all orders $n$, one can actually write down a full solution in closed form immediately. In such cases, the Weyl-ordered perturbative expansion \eq{eq:Phi^m fact}-\eq{eq:U^m fact}  can be seen as a way of 
dressing solutions $\Phi(x,Y)=\Phi^{(1)}(x,Y)$ to the linearized twisted adjoint equation into solutions to the full Vasiliev equations. This is indeed the case for the solutions first studied in \cite{Didenko:2009td,us,us2,Iazeolla:2017vng,review,Aros:2017ror,DeFilippi:2019jqq}, where the initial datum $\Phi'$  (equivalently, $\Psi'$) is expanded on generalized projector bases: while their elements are non-polynomial, and in fact include distributions, they have well-defined star-products, and form a star-product subalgebra which will be taken as our ${\cal A}({\cal Y}_4)$. 

In this Section we show how different linearized solutions can be encoded into fibre elements, the integration constants $\Phi'(Y)$.  We shall therefore first of all review in some detail the ordinary classification of solutions to the Klein-Gordon equation on $AdS_4$ that will be of relevance in the following, in spherical and Poincar\'e coordinates. The focus will be on how the choice of manifest symmetries, spacetime regularity and boundary conditions distinguish the various solution sectors. We shall then show how the solutions are mapped, via the unfolded equations, to non-polynomial functions on ${\cal Y}_4$ spanning modules of $\mso(2,3)$ and of the higher-spin algebra. The focus will be on massless scalar particle modes and spherically symmetric black hole modes \cite{Iazeolla:2008ix,us,Iazeolla:2017vng} in compact basis, and on bulk-to-boundary propagators and singular solutions with vanishing conformal dimension, related to boundary Green's functions, in conformal basis. We shall  also recall how the fibre realization of these representations is flexible enough to capture the full indecomposable structure of the module of regular solutions --- in which the familiar particle modes form an ideal that is complemented by less studied  ``wedge'' modes \cite{Iazeolla:2008bp,Iazeolla:2008ix}.

%%%%%%%%%%%%%%%%%%%%%%%%%%%%%%%%%%%
\scss{Spacetime analysis in global coordinates}\label{ptandbh}
%%%%%%%%%%%%%%%%%%%%%%%%%%%%%%%%%%%

The linearized twisted adjoint equation contains the Klein-Gordon equation and the Bargmann-Wigner equations in $AdS_4$ for all spin-$s$ linearized Weyl tensors \cite{Vasiliev:1999ba,Bekaert:2005vh,Didenko:2014dwa,Boulanger:2008up}) encoded in $\Phi^{(1)}$. Each solution space is mapped to specific fibre elements $\Phi'^{(1)}(Y)$ via \eq{LCpiL}.  Let us then first of all recall how various sectors of solutions of the free field equations in $AdS_4$ are distinguished by different regularity and boundary conditions. First, let us perform this analysis in \emph{global spherical coordinates} $(t,r,\theta,\varphi)$ \eq{metricglob} (equivalently, $(t,\xi,\theta,\varphi)$ \eq{globxi}). 

The most familiar solution sector is the one spanned by \emph{particle modes}. They correspond to solutions that are regular everywhere in spacetime and satisfy additional boundary conditions which ensure that they are normalizable and have finite, conserved Killing energy \cite{Breitenlohner:1982jf}. They fill unitary irreducible $\mso(2,3)$ lowest-weight modules $\mD(e_0,(s_0))$, organized in terms of the eigenvalues under the \emph{compact} generators, the energy $E$ and the spatial rotations $M_{rs}$. Correspondingly, they are denoted via the energy and $\mso(3)$-spin eigenvalues of their lowest weight state $\ket{e_0=s_0+1,(s_0)}$ as $\mD(s_0+1,(s_0))$. Their negative-energy counterparts, the anti-particles, fill highest-weight modules $\mD(-s_0-1,(s_0))=\pi [\mD(s_0+1,(s_0))]$. In the present paper we shall mainly be interested in massless scalar particle modes in $AdS_4$, although there is neither conceptual nor technical limitation to repeating our construction with particles of arbitrary spin (see \cite{Iazeolla:2008ix,DeFilippi:2019jqq}). In $D=4$ there are two unitary scalars, $\mD(1,(0))$ and $\mD(2,(0))$, satisfying Neumann and Dirichlet boundary conditions, respectively. Their lowest-weight modes are
\be \phi_{1,(0)} \ = \ \frac{e^{-it}}{\sqrt{1+r^2}} \ , \qquad \phi_{2,(0)} \ = \ \frac{e^{-2it}}{1+r^2} \ . \label{lwspherical}\ee

Foregoing some of the restrictions mentioned above, the space of admissible solutions enlarges to include other sectors, such as \emph{wedge modes} and \emph{static black-hole modes}\footnote{The solutions that we refer to as higher spin black hole states, or modes, are so called essentially because they possess a tower of Weyl tensors of all integer spins that include and generalize the spin-2 Weyl tensor of an AdS Schwarzschild black hole. However, at present there is no known higher-spin invariant quantity ensuring that the singularity of the individual Weyl tensors is physical, and whether there exists any invariant notion of an event horizon --- as well as an entropy attached to it --- remains an open problem. On the other hand, the fact that each such solution has identical black-hole asymptotics but is possibly non-singular and horizon-free may suggest an interpretation in terms of black-hole microstates, similar to fuzzballs \cite{Lunin:2002qf,Mathur:2002ie,Mathur:2005zp,Skenderis}. See \cite{Iazeolla:2017vng} for more details on this proposal and on our usage of the terminology.}. 

Let us examine how all these modules arise as solutions of the massless Klein-Gordon equation in $AdS_4$, and how regularity and the choice of boundary conditions select the various solution spaces (mainly following the analysis in \cite{Breitenlohner:1982jf,Mezincescu:1984ev,Balasubramanian:1998sn}).  The free equation for a massless\footnote{Here we mean \emph{critically massless} \cite{Boulanger:2008up,Boulanger:2008kw}, also referred to as \emph{composite-massless}, in the sense that, as we shall recall, the massless scalar fields here treated are composites of two singleton UIRs.} scalar field in $AdS_4$ is
\be (\Box+2)\phi \ = \ 0 \ ,\label{KGads}\ee
where we are setting the AdS radius to $1$ and $\Box$ is the covariant D'Alembertian%, acting on a scalar field as $\Box\phi=\nabla^\m \partial_\m\phi$
. Note the mass-like term $m^2=-2$ coming from the coupling with the $AdS$ scalar curvature $(1/6)R\bar{\phi}\phi$ in the action. In the global coordinates \eq{globxi}, in terms of which the boundary is at $\xi\to\frac{\pi}2$, the Klein-Gordon equation reads
\be -\cos^2\xi\,\partial_t^2\phi+\cos^2\xi\,\partial_\xi^2\phi+\frac{2}{\tan\xi}\partial_\xi\phi+\frac{1}{\tan^2\xi}\left[\partial^2_\theta\phi+\frac{1}{\tan\theta}\partial_\theta\phi+\frac{1}{\sin^2\theta}\partial^2_\varphi\phi\right] +2\phi \ = \ 0 \ , \label{KGglob}\ee
where $\partial^2_\theta+\frac{1}{\tan\theta}\partial_\theta+\frac{1}{\sin^2\theta}\partial^2_\varphi=\nabla^2_{S^2}$. Separating the coordinates and inserting the factorized Ansatz
\be \phi \ = \ \phi_{\o\ell m} \ = \ e^{-i\o t}\,Y_{\ell,m}(\theta,\varphi)\,\chi(\xi) \ee
(where clearly $\ell=0,1,2...$ and $m= -\ell, -\ell+1,...\ell-1, \ell$, and $\nabla^2_{S^2}Y_{\ell,m}=-\ell(\ell+1)Y_{\ell,m}$) in \eq{KGglob} we are left with the radial equation
\be  \cos^2\xi\,\partial_\xi^2\chi+\frac{2}{\tan\xi}\partial_\xi\chi+ \left(\o^2\cos^2\xi-\frac{\ell(\ell+1)}{\tan^2\xi}+2\right)\chi \ = \ 0\ .\ee
Substituting $\chi(\xi) = \cos\xi(\sin\xi)^{2b} f_{\o\ell}(\xi)$, the latter becomes a hypergeometric equation for $f_{\o\ell}$, with two independent solutions corresponding to the two roots $b_\pm$ of the indicial equation,
\be 2b(2b+1) -\ell(\ell+1)\ = \ 0 \ ,\qquad b_\pm \ = \ \left\{\frac{\ell}{2},-\frac{\ell+1}{2}\right\}\ .\ee
The two solutions are distinguished by their behaviour at the origin $\xi=0$: one is regular, 
\be \phi_{\textrm{reg}} \ = \  e^{-i\o t}\,Y_{\ell,m}(\theta,\varphi)\,\cos\xi(\sin\xi)^{\ell}\, {}_2F_1\left(\frac{\ell+1+\o}{2},\frac{\ell+1-\o}{2},\ell+\frac{3}{2};\sin^2\xi\right) \ ,\label{phireg}\ee
whereas the second one is singular,
\be\phi_{\textrm{sing}} \ = \  e^{-i\o t}\,Y_{\ell,m}(\theta,\varphi)\,\cos\xi(\sin\xi)^{-\ell-1}\, {}_2F_1\left(\frac{\o-\ell}{2},-\frac{\o+\ell}{2},-\ell+\frac{1}{2};\sin^2\xi\right) \ . \label{phising}\ee
\paragraph{Particle modes.} Insisting on regularity in the interior one is then led to discard $\phi_{\textrm{sing}}$ and keep only $\phi_{\textrm{reg}}$\footnote{This is crucial, in the AdS/CFT context, to avoid any contribution from the interior to the boundary correlation function \cite{Balasubramanian:1998sn}.}. In order to study the boundary behaviour of the solutions \eq{phireg}-\eq{phising} it is useful to rewrite them in terms of functions of $\cos^2\xi$ by using the identity %(provided $c-a-b\notin \mathbb{Z}$)
\bea {}_2F_1(a,b,c;x) & = & \C(c)\left\{\frac{\C(c-a-b)}{\C(c-a)\C(c-b)}\, {}_2F_1(a,b,a+b-c+1;1-x)\nn\right.\\
&+& \left.\frac{\C(a+b-c)}{\C(a)\C(b)}\,(1-x)^{c-a-b}\, {}_2F_1(c-a,c-b,c-a-b+1;1-x)\right\}  \ .\eea
The regular solution can then be written as the linear combination
\bea \phi_{\textrm{reg}} \ = \ A\phi_{(-)}+B\phi_{(+)} \ , \label{phiregAB}\eea
where
\be A \ := \ \frac{\C\left(\ell+\frac{3}{2}\right)\C\left(\frac{1}{2}\right)}{\C\left(\frac{\ell + 2-\o}{2}\right)\C\left(\frac{\ell + 2+\o}{2}\right)} \ , \qquad B\ := \ \frac{\C\left(\ell+\frac{3}{2}\right)\C\left(-\frac{1}{2}\right)}{\C\left(\frac{\ell + 1-\o}{2}\right)\C\left(\frac{\ell + 1+\o}{2}\right)}\ ,\label{AandB}\ee
and
\bea  \phi_{(-)} & = &  e^{-i\o t}\,Y_{\ell,m}(\theta,\varphi)\,\cos\xi(\sin\xi)^{\ell}\, {}_2F_1\left(\frac{\ell+1+\o}{2},\frac{\ell+1-\o}{2},\frac{1}{2};\cos^2\xi\right) \ , \\
\phi_{(+)} & = &  e^{-i\o t}\,Y_{\ell,m}(\theta,\varphi)\,(\cos\xi)^2(\sin\xi)^{\ell}\, {}_2F_1\left(\frac{\ell+2+\o}{2},\frac{\ell+2-\o}{2},\frac{3}{2};\cos^2\xi\right) \ . \eea
It is then immediate to see that 
\be \phi_{\textrm{reg}} \ \xrightarrow[\xi \rightarrow\pi/2] \ e^{-i\o t}\,Y_{\ell m}(\theta,\varphi)\,\left\{A\cos\xi\left(1+O(\cos^2\xi)\right)+B\cos^2\xi\left(1+O(\cos^2\xi)\right)\right\} \ .\ee 
In this case (four spacetime dimensions,  $m^2=-2$) both $ \phi_{(-)}$ and $\phi_{(+)}$ are normalizable, and in order to select a single complete, normalizable set of field modes, it is necessary to impose extra boundary conditions\footnote{In higher dimensions (or for massive scalar fields) normalizability is all one needs to demand in order to select one complete set of field modes (the one that decays more rapidly at the boundary, $\phi_{(+)}$) satisfying the Klein-Gordon equation and leading to conserved energy. The precise condition can be stated in terms of the value mass term appearing in \eq{KGads} \cite{Balasubramanian:1998sn} as $m^2\geq 1-(D-1)^2/4$. So in general the regular solution \eq{phiregAB} is non-normalizable, unless one introduces the quantization condition on the frequencies that leads to $A=0$. On the other hand, for $-(D-1)^2/4<m^2<1-(D-1)^2/4$ (where we recall that $-(D-1)^2/4$ corresponds to the Breintenlohner-Freedman bound, the lowest value of $m^2$ compatible with stability in $AdS_D$.) there are two independent normalizable solutions, and one needs to impose further conditions in order to have conserved Killing energy.}. 

In fact, as found in \cite{Avis:1977yn,Breitenlohner:1982jf}, stronger boundary conditions are necessary in order to have a unique solution of the Cauchy problem, due to the well-known problem of the absence of a global Cauchy surface in $AdS$\footnote{This problem is caused by the fact that information can propagate along null geodesics from the origin  $\xi=0 $ to the boundary $\xi=\pi/2$, or vice versa, in finite time, thereby preventing any spacelike surface crossing the whole spacetime manifold from intercepting all null geodesics.}.  As a consequence, only specific boundary conditions at spatial infinity can lead to conservation of energy, and these lead to selecting a single normalizable set of modes. The authors of \cite{Avis:1977yn,Breitenlohner:1982jf,Mezincescu:1984ev} achieved this by imposing %conservation of the scalar product 
%
%\be (\phi_1,\phi_2) \ = \ i\int_\Sigma d^{3}x\,\sqrt{-g}\,g^{00}\,(\bar{\phi}_1\partial_0 \phi_2-\partial_0\bar{\phi}_1\phi_2 )\ ,   \label{inprod}\ee
%
%that is, that the flux of the formally conserved current $j^\m = ig^{\m\n}(\bar{\phi}_1\partial_\n \phi_2-\partial_\n\bar{\phi}_1\phi_2 )$ through the boundary vanishes; and 
conservation of the Killing energy 
\be  E \ = \  \int_\Sigma d^{3}x\,\sqrt{-g}\,T^0{}_0 \ ,\ee
that is, vanishing of the flux through the boundary 
\be F_E \ = \  \int_{\partial\Sigma} d\Sigma_i \,\sqrt{-g}\,T^i{}_0 \ . \label{FE}\ee
The latter condition translates to\footnote{Strictly speaking, this condition is related to the Killing energy flux calculated from the improved energy-momentum tensor $T_{\m\n}$, related to the canonical energy-momentum tensor $T^{\textrm{can}}_{\m\n}=2\partial_{(\m}\bar{\phi}\,\partial_{\n)}\phi-g_{\m\n}(\partial^\r\bar{\phi}\,\partial_\r\phi+(1/6)\,R\,\bar{\phi}\phi)$  by $T_{\m\n}=T^{\textrm{can}}_{\m\n}+(1/3)\Delta T_{\m\n}$, with $\Delta T_{\m\n}:=(g_{\m\n}\Box -D_\m \partial_\n+R_{\m\n})\bar{\phi}\phi$. The same calculation with $T^{\textrm{can}}_{\m\n}$ leads to the more restrictive conclusion that only the $\mD(2,0)$ scalar conserves the energy, and, in fact, has finite Killing energy. On the other hand, the improved energy-momentum tensor allows for weaker conditions, admitting also the $\mD(1,0)$ field modes. While on general curved backgrounds only the improved tensor is conserved, here the ambiguity arises because both $T_{\m\n}$ and $\Delta T_{\m\n}$ are separately conserved due to the constant curvature of $AdS$. Supersymmetry resolves the ambiguity, however, by including both the scalars in the same supermultiplet, which in particular means that only the improved energy-momentum tensor can be used to define a conserved energy functional \cite{Breitenlohner:1982jf}. }
\be \left.\tan^2\xi\left[\partial_\xi+2\tan\xi\right]\bar{\phi}\phi\,\,\right|_{\xi=\pi/2} \ = \ 0 \ , \label{noEflux}\ee
and in order for \eq{noEflux} to be satisfied one must require either $A$ or $B$ to vanish, which enforces the quantization condition $\o=\pm(\ell+2+2n)$ and $\o=\pm(\ell+1+2n)$, respectively, on the frequencies in order for the $\C$ functions at the denominators in \eq{AandB} to blow up. With these conditions on $\o$, the hypergeometric functions are truncated to Jacobi polynomials $P_n^{(\a,\b)}(x)=\frac{n!}{(1/2)_n}{}_2F_1(-n,n+\a+\b+1,\a+1;(1-x)/2)$, $P_n^{(\a,\b)}(-x)=(-1)^nP_n^{(\b,\a)}(x)$. Therefore, by insisting on conservation of energy one is left with only one of the two possible set of normalizable modes, each defining a complete set filling the $\mso(2,3)$ representation
\bea \mD(1,(0))\oplus\mD(-1,(0)) & : & \phi_{\o\ell m} \ = \  N_{n\ell}\,e^{-i\o t}\,Y_{\ell,m}(\theta,\varphi)\,\cos\xi(\sin\xi)^{\ell}\, P_n^{(\ell+1/2,-1/2)}(\cos 2\xi) \ ,\nn \\ 
& & \o \ = \ \pm(\ell+1+2n) \ , \quad n \ = \ 0,1,2,... \ , \quad \ell \ = \ 0,1,2, ... \ , \label{D10}\eea
and 
\bea \mD(2,(0))\oplus\mD(-2,(0)) & : & \quad \phi_{\o\ell m} \ = \  N_{n\ell}\,e^{-i\o t}\,Y_{\ell,m}(\theta,\varphi)\,\cos^2\xi(\sin\xi)^{\ell}\, P_n^{(\ell+1/2,1/2)}(\cos 2\xi) \ ,\nn \\ 
& & \o \ = \ \pm(\ell+2+2n) \ , \quad n \ = \ 0,1,2,... \ , \quad \ell \ = \ 0,1,2, ... \ ,\label{D20} \eea
respectively, where $N_{n\ell}$ is a normalization coefficient. Their lowest-weight modes $\phi_{1,(0)}$ and $\phi_{2,(0)}$, respectively, correspond to the expressions given in \eq{lwspherical} in terms of $r$ (see Appendix \ref{App:conv} for the relation between the two sets of coordinates). As it is well known, the selection of the above two $\mD(1,(0))$ and $\mD(2,(0))$ scalar fields correspond to imposing Neumann or Dirichlet boundary conditions, respectively, on the conformally transformed fields $\phi':=\phi/\cos\xi$ \cite{Avis:1977yn}. 

It will be useful for the following to have a single expression for all the modes in $\mD(1,(0))\oplus\mD(2,(0))\oplus\mD(-1,(0))\oplus\mD(-2,(0))$. Shifting the value of $n$ to $n=\pm 1, \pm2, ...$, one can write
\bea & (1+\pi)[\mD(1,(0))\oplus\mD(2,(0))] \ : \  \phi_{\o\ell m} \ = \  N_{n\ell}\,e^{-i\o t}\,Y_{\ell,m}(\theta,\varphi)\,\cos\xi(\sin\xi)^{\ell}\, P_n^{(\ell+1/2,\ell+1/2)}(\cos \xi) \ ,&\nn \\ 
& \o \ = \ n+\varepsilon_n\ell \ , \quad n \ = \ \pm1,\pm 2,... \ , \quad \varepsilon_n \ := \ \frac{n}{|n|} \ , \quad \ell \ = \ 0,1,2, ... \ .&\label{D120} \eea
In order to keep the usual notation for frequencies and spherical harmonics, in the above we have been using the symbols $\omega$ and $\ell$ for the energy eigenvalue $e$ and the $\mso(3)$-spin eigenvalue $s$ that label the states $\ket{e,s}$ in the $\mso(2,3)$ irreps $\mD(e_0,(s_0))$. 

\paragraph{Wedge modes and the indecomposable compact twisted-adjoint module.} Foregoing the conservation of energy --- that is, not imposing that the flux \eq{FE} through the boundary be zero --- one is allowed to retain a general combination of the two scalar irreps $\mD(1,0)$ and $\mD(2,0)$ (still normalizable, in $D=4$): in particular, combinations \eq{phiregAB} below the unitarity bound corresponding to states that fill a wedge $\mathfrak{W} $ in the compact-weight space  in between the (scalar) particle and anti-particle modules, hence the name \emph{wedge modes}. Such modes were first considered systematically --- and interpreted as representations of $\mso(2,3)$ and its higher-spin extension $\mhs(2,3)$ --- in \cite{Iazeolla:2008ix} from the unfolded point of view\footnote{By analogy with similar solutions in flat spacetime, they were there given the name of runaway modes. However, considering that they do not diverge at the $AdS$ boundary, and that in four dimensions they are even normalizable, in this work we use the more neutral denomination of wedge modes.}. More precisely, it was found in \cite{Iazeolla:2008ix} that massless (anti-)particle modules are invariant submodules of an indecomposable module $\mathfrak{M}$, and $\mathfrak{W}$ is their complement%\footnote{It is important to note in this respect that the fact that wedge modes are non-unitary is strictly true only with respect to the usual inner product $ (\phi_1,\phi_2) \ = \ i\int_\Sigma d^{3}x\,\sqrt{-g}\,g^{00}\,(\bar{\phi}_1\partial_0 \phi_2-\partial_0\bar{\phi}_1\phi_2 )$. In \cite{Iazeolla:2008ix}, on the other hand, a different, regularized inner product on the enlarged, indecomposable module was defined, with respect to which both particles and runaway modes are unitarizable. \framebox{what of the Breit-Freed bound?? If feels sus, eliminate this footnote and maybe keep it for v2.}}
, that is
\be \mathfrak{M} \ = \ \mD  \subsetplus \,\, \mathfrak{W}\ ,\label{MDW}\ee
where $\subsetplus$ denotes the semi-direct sum, and $\mD$ is further divided into $\mD=\mD^{(+)}\oplus \mD^{(-)}$, with the substructures
\bea  \mD^{(+)}& \simeq & (1+\pi)\bigoplus_{s=0}^\infty \mD(s+1;(s)) \ ,\label{mD}\\
 \mD^{(-)}& \simeq & (1+\pi)\bigoplus_{s=0}^\infty \mD(s+1+\d_{s,0};(s,1))\ .\label{mDprime}\eea
%
%where $\pi$ acts on the states by reversing the sign of the energy eigenvalue, i.e., sending particle to anti-particle states. 
Both the particle and the wedge submodules are divided into an even and odd submodule with respect to the conserved quantity $ (-1)^{e+s}$, which is preserved by the action of $\mso(2,3)$ and its higher-spin extension: the label $(\pm)$ denote the even/odd submodule (see \cite{Iazeolla:2008ix,Iazeolla:2017vng} for the even/odd submodules of $\mathfrak{W}$). 

The entire module $\mathfrak{M}$ (which is neither a lowest-weight nor a highest-weight module) can be generated by acting with the $\mhs(2,3)$ generators on the two static states $\phi_{0,(0)}$ and $\phi_{0,(1)}$. They can be obtained from \eq{phireg} via the static limit $\o\to 0$: the static rotationally-invariant mode is 
\be \phi_{0,(0)} \ := \ \left.\phi_{\textrm{reg}}\right|_{\o=0,\ell=0} \ = \  \cos\xi\, {}_2F_1\left(\frac{1}{2},\frac{1}{2};\frac{3}{2};\sin^2\xi\right) \ = \ \frac{\xi}{\tan\xi} \ ,\label{phi00} \ee
while the $\ell=1$ static element reads 
\bea \phi_{0,(1)} := \ \left.\phi_{\textrm{reg}}\right|_{\o=0,\ell=1} & = &  Y_{1,m}(\theta,\varphi)\,\cos\xi\,\sin\xi\, {}_2F_1\left(1,1,\frac{5}{2};\sin^2\xi\right)\nn\\ & = & 3Y_{1,m}(\theta,\varphi)\,\frac{1}{\tan\xi}\left(1-\frac{\xi}{\tan\xi} \right) \ .\label{phi01} \eea
Note that such field configurations, as it is obvious from their derivation, are regular at the origin. In fact, in terms of $r$ they read 
\be\phi_{0,(0)} \ = \ \frac{\arctan r}{r} \ , \qquad \textrm{and} \qquad \phi_{0,(1)} \ \sim  \ \frac{1}{r} - \frac{\arctan r}{r^2} \ . \ee

\paragraph{Static black-hole modes.} As anticipated, letting go also of the condition of regularity in the interior, one finds the solution branch \eq{phising}, which, for $\o=0$, contains more static solutions. Indeed, setting $\o=0$ and $\ell=0$ in \eq{phising}, one finds the simplest static, soliton-like solution of the Klein-Gordon equation (as well as fundamental solution of the three-dimensional laplacian), 
\be \phi_{\textrm{sing}(0,(0))} \ := \ \left.\phi_{\textrm{sing}}\right|_{\o=0,\ell=0} \ = \ \frac{1}{\tan\xi} \ = \  \frac{1}{r}  \ .\label{phiD00}\ee
This field configuration is singular in the origin and, in fact, represents the field generated by a delta-function source there centred. Note that this solution has the same quantum numbers as the regular solution \eq{phi00}, but belong to a separate weight space $\mathfrak{S}$ representing the singular branch, and has in fact a number of different properties. We shall be more precise about the relation between the two branches in the Section \ref{Sec:fibcomp}.

Singular, static and spherically-symmetric solutions of the free equations of any integer spin will be of interest in the following --- that is, we shall consider also the $s\geq 1$ counterparts of \eq{phiD00} that, as found in \cite{Didenko:2008va} (see also \cite{Didenko:2009tc,Didenko:2009td}), are distinguished by Weyl tensors of the form
\be C_{\a(2s)} \ \propto \ \frac{1}{r^{s+1}}\,u^+_{(\a_1}u^-_{\a_2}\ldots u^+_{\a_{2s-1}}u^-_{\a_{2s})} \ ,\label{WeylD}\ee
analogously for the anti-selfdual component, where $(u^+_{\a},u^-_{\a})$ is an $x$-dependent spin-frame (carrying the appropriate angular dependence)\footnote{Such fields correspond to static solutions of the spin-$s$ Bargmann-Wigner equations $\nabla^{(0)c}C_{ca(s-1),b(s)}=0 $,  $\nabla^{(0)}_{[c}C_{d|a(s-1),|f]b(s-1)} \ = \ 0$, with resulting mass-shell condition  $(\nabla^{(0)c}\nabla^{(0)}_c-m^2_s)C_{a(s),b(s)}  =  0$ , on generalized spin-$s$ Weyl tensors $C_{a(s),b(s)} $ with Young diagram of type $(s,s)$ of $\mso(1,3)$, with the critical masses $m^2_s = -2(s+1)$ (see for example their generalization to any dimension in \cite{Iazeolla:2008ix} and to mixed-symmetry fields and arbitrary mass in \cite{Boulanger:2008up,Boulanger:2008kw} ).}. These Weyl tensors are of Petrov-type D, and the expression \eq{WeylD} includes, for $s=0$, the static scalar \eq{phiD00}. For $s\geq 1$ they can be integrated and shown to correspond to a Kerr-Schild gauge field satisfying the Fronsdal equation,
%
%\be \nabla^\n\nabla_\n\phi_{\m(s)}-s\nabla_{\m}\nabla^\n\phi_{\n\m(s-1)}+2(s-1)(s+1)\phi_{\m(s)} \ = \ 0 \,\ee
%
of type
\be \phi_{\m(s)} \ = \ \frac{1}{r}\,k_{\m_1}\ldots k_{\m_s} \ , \ee
where $k_\m$ are Kerr-Schild vectors, with the properties $k^\m k_\m =0$, $k^\m\nabla_\m k_\n=0$ (see \cite{Didenko:2009td} for their explicit realization). The fact that they can be written in Kerr-Schild form is a manifestation of the fact that black-hole solutions solve both the linear and the nonlinear equations, and this is why they appear already in the linearized analysis. This is also why they are ideally suited to be elevated to solutions of the Vasiliev equations, as first noted in \cite{Didenko:2008va,Didenko:2009tc}. Indeed, we shall recall in Section \ref{Sec:genproj} that a superposition of black-hole modes of all spins (as required by higher-spin symmetry) of type \eq{WeylD} indeed solves the full equations.

%%%%%%%%%%%%%%%%%%%%%%%%%%%%%%%%%%%%%%%%%%%%%%%%%%%

\scss{Spacetime analysis in Poincar\'e coordinates}\label{sptP}

%%%%%%%%%%%%%%%%%%%%%%%%%%%%%%%%%%%%%%%%%%%%%%%%%%

Let us also examine some solutions of the $AdS_4$ massless Klein-Gordon equation in Poincar\'e coordinates \eq{Poincx} (not global, as is well-known, only covering the Poincar\'e patch). In this case, the highlighted symmetries are dilatation and Poincar\'e symmetry on the boundary coordinates ${\bfx}^m$. Accordingly, \eq{KGads} becomes
\be  z^2\left[z^2\partial_z\left(\frac1{z^2}\partial_z\phi\right)+\partial^m\partial_m\phi\right]+2\phi \ = \ 0 \ . \label{KGP} \ee
% 
%Given the translational symmetry of the (conformally rescaled) Minkowski boundary, a natural factorized Ansatz is
%
%\be \phi \ = \ e^{ik\cdot\bfx} z^{3/2}f_k(z) \ , \label{conffact}\ee
%
%and plugging into \eq{KGP} we find that $f_k(z)$ has to satisfy the condition
%
%\be k^2 z^2f''_k+kzf'_k-(\n^2+k^2z^2)f_k  \ = \ 0 \ .\ee
%
The natural factorized Ansatz this time is of the form
\be   \phi \ = \ f(z) g(\bfx) \ .\label{zyfact}\ee
We refrain from reviewing in detail the general solution of the Klein-Gordon equation in Poincar\'e coordinates, as it is covered in many papers and textbooks (see, e.g., \cite{Balasubramanian:1998sn,SonStarinets,Nastase:2015wjb}), limiting ourselves to pointing out a few facts that will be relevant in the following. This time solutions of the Klein-Gordon equation are organized in representations $\mD(\D_0,(s_0))$ where $\D$ is the eigenvalue under the dilatation operator $D\propto -z\partial_z -\bfx^m\partial_{\bfx^m}$ %\framebox{because it is preserved by the differential operator box in Poincare}
, generating an $\mso(1,1)$, and $(s_0)$ is the Lorentz-spin label of the boundary Lorentz symmetry algebra $\mso(1,2)$. The factor $g(\bfx)$ can in general be expanded in terms of plane waves $e^{i k\cdot \bfx}$ and for regular solutions $f(z)$ can be written in terms of Bessel functions. In $D=4$ there are again two normalizable scalar lowest-weight modes 
\be \phi_{1,(0)} \ = \ \frac{z}{z^2+\bfx^m\bfx_m} \ , \qquad \phi_{2,(0)} \ = \ \left(\frac{z}{z^2+\bfx^m\bfx_m}\right)^2 \ , \label{Btb}\ee
which are the well-known bulk-to-boundary propagators for (composite) massless scalars (for boundary reference point $\bfx^m=0$), and they form an $\mso(2,3)$-module together with their descendants, obtained via repeated action of the boundary translation operator $-i\partial_m$. To this module, via inversion $(z,\bfx^m)\to \frac{(z,\bfx^m)}{z^2+\bfx^m\bfx_m}$, correspond another module, with highest-weight modes
\be   \phi_{-1,(0)} \ = \  z\ , \qquad \phi_{-2,(0)} \ = \ z^2 \ ,\label{antiBtb}\ee
(irregular at the bulk point $z\to\infty$), which can be thought of as bulk-to-boundary propagators at the ``point at infinity''  $z=\infty$. There are also, in the singular branch, counterparts of the black-hole modes that we studied in global spherical coordinates. The scalar representative is 
\be \phi_{{\rm sing} (0,(0))}  \ = \ \frac{z}{\sqrt{\bfx^m\bfx_m}} \ ,\label{0confweight}\ee
which is irregular in the bulk and has evidently vanishing scaling dimension --- which is the counterpart, in the present coordinate split, of the time-independence of the black-hole modes.  This is an especially simple solution of the Klein-Gordon equation, that corresponds to choosing $g(\bfx)$ in \eq{zyfact} as a solution to the flat massless 3D Klein-Gordon equation and requiring that the overall bulk scaling dimension be zero. As a consequence, $g$ turns out to be related to the scalar propagator. In the limit $z\to 0$, \eq{0confweight} has only support on the boundary light-cone. One can build, in fact, a tower of such modes for any spin, solutions to the generalized Bargmann-Wigner equations just like the black-hole modes \eq{WeylD}, of the form $z^{s+1}\phi_{m_1\ldots m_s}(\bfx)$, where $\phi_{m_1\ldots m_s}(\bfx)$ are again solutions to the boundary wave equation that are singular on the light-cone. We shall explicitly construct them in Section \ref{Sec:singsol}, starting from their fibre duals and finding the spacetime solutions by virtue of \eq{LCpiL}. 

Note that the solutions here collected show an immediate similarity with those in found in spherical coordinates, which is especially evident in other coordinate systems, such as stereographic coordinates \eq{A.15}: indeed, the lowest-weight particle solutions are 
\be \phi_{e_0;(0)} \ = \ \left(\frac{1-x^2}{1+2ix^0+x^2}\right)^{e_0} \ , \qquad e_0=\{1,2\} \label{scalar12}\ee
whereas, identifying the $z$ coordinate as in \eq{Pemb} we get, for \eq{Btb}, 
\be \phi_{\D_0;(0)} \ = \ \left(\frac{1-x^2}{1-2x^3+x^2}\right)^{\D_0} \ , \qquad \D_0=\{1,2\} \ ,\ee
or, in embedding coordinates,
\be \phi_{e_0;(0)} \ = \ \left(\frac{1}{X^{0'}+iX^0}\right)^{e_0} \ , \qquad  \phi_{\D_0;(0)} \ = \ \left(\frac{1}{X^{0'}-X^3}\right)^{\D_0}\ .\ee
Similarly, the scalar solutions in the singular branches \eq{phiD00} and \eq{0confweight} are, in embedding coordinates,
\be  \phi_{{\rm sing}(e_0=0;(0))}  \ = \ \frac{1}{\sqrt{X^2_1+X^2_2+X^2_3}} \ , \qquad   \phi_{{\rm sing}(\D_0=0;(0))}  \ = \ \frac{1}{\sqrt{-X^2_0+X^2_1+X^2_2}} \ . \ee
The same ``Wick rotation'' of course connects also the anti-particle modules to the highest-weight modules generated from \eq{antiBtb}. We shall review precisely what this relation is and give more details on the structure of the solution spaces in Section \ref{Sec:fibcomp}, where we shall study them from the point of view of the unfolded formalism, and leave more details of the analysis to a future publication.

%For boundary momenta $k^2 < 0$ the solutions of the radial equation are Bessel functions, and for $D=4$, $m^2=-2$ one again finds two regular and normalizable solutions forse senza neanche mostrare q eq, subito dire che le sol naturali ora sono Jpmnu, che vanno come... e per nu<1 entr normalizzab. 

%%%%%%%%%%%%%%%%%%%%%%%%%%%%%%%%%%%

\scss{Fibre analysis: Cartan subalgebras of $\msp(4;\Comp)$ and linearized solution spaces} \label{fibresol}

%\scss{Operator realization of particle modes, wedge modes and black hole modes}\label{ptandbh2}
%%%%%%%%%%%%%%%%%%%%%%%%%%%%%%%%%%%

Having recalled the standard approach to finding and classifying the linearized solutions that will be considered in the following, we shall now recall how they can be encoded in functions on ${\cal Y}_4$, i.e., twisted-adjoint operators on which $\Phi'(Y)$  can be expanded. Just as solutions to the Klein-Gordon equation belonging to one representation are transformed into one another via the action of the $AdS_4$ Killing vector fields, so the corresponding fibre elements will fill $\mso(2,3)$-modules under the twisted adjoint action of the $AdS_4$ isometry generators. 

We shall focus on the construction of representations $\mD(e_0,(s_0))$ in \emph{compact slicing}, corresponding to the solution spaces of the Klein-Gordon equation in global spherical coordinates, and of representations $\mD(\D_0,(s_0))$ in \emph{conformal slicing} corresponding to the solutions of the same problem in Poincar\'e coordinates. %For massless particle and black-hole modes,  we can find a specific \emph{common} regular presentation such that repeated products of $\Psi$ are finite and meet all the conditions necessary to be deformed together to solutions of the full Vasiliev equations; while the same, simple regular presentation only allows the fibre Fourier duals of the bulk-to-boundary propagators to be elevated to solutions to the full equations. We leave the nonlinear dressing of \eq{Btb}, as well as of the wedge modes to future studies.

The general idea is to expand $\Phi'(Y)$ not in terms of monomials, but rather in terms of non-polynomial functions of $Y$ that realize the states of a given representation of $\mso(2,3)$: in other words, the (left) action of the algebra over  kets $\ket{m,n}$ should univocally correspond to the twisted-adjoint action on such fibre elements.  % In order to prepare for the non-linear completion of certain subspaces.
Let us briefly recall and extend the construction of such modules in the $Y$-fibre, following \cite{Iazeolla:2008bp,Iazeolla:2008ix} and \cite{us,us2,Iazeolla:2017vng,Aros:2019pgj,DeFilippi:2019jqq}\footnote{A similar change of basis for the $D=3$ case was performed and used in \cite{Raeymaekers:2016mmm,Kessel:2018zqm,Raeymaekers:2019dkc}.}.

Consider a pair of generators $(K_{(+)},K_{(-)})$ in the Cartan subalgebra of the complexified $AdS_4$ isometry algebra $\msp(4,\Comp)$, with oscillator realization 
\be K_{(\pm)}=\frac{1}{8} K^{(\pm)}_{\underline{\a\b}}Y^{\underline\a}\star Y^{\underline\b}\ ,\ee
where
\bea [K^{(q)},K^{(q')}]_{\au\bu}~=~0\ ,\qquad  K^{(q)}_{\au}{}^{\underline{\gamma}}\,K^{(q)}_{\underline{\gamma}}{}^{\bu}~=~-\delta_{\au}{}^{\bu}\ .\label{K2}
\eea
They can be written in terms of two number operators
\begin{equation}
w_{i} \ := \ a_{i}^+ a_{i}^- \ = \ a_{i}^+ \star a_{i}^- +\frac{1}{2}\text{ ,} \qquad \text{(no sum over \emph{i})}
\end{equation}
as
\be K_{(\pm)}=\frac{1}{2}(w_2\pm w_1)\ ,\ee
where the creation and annihilation operators can be extracted from linear combinations of the $Y$ oscillators, $a^{\pm}_i=(A^\pm_i)_{\underline\alpha}Y^{\underline\alpha}$, $i=1,2$, using projectors built from $K^{(q)}_{\underline{\a\b}}$ \cite{us,Aros:2019pgj,DeFilippi:2019jqq}.
It is then possible to build operators $P_{{\mathbf n}_L|{\mathbf n}_R}(Y)$ obeying 
\be P_{{\mathbf n}_L|{\mathbf n}_R}=\pi\bar\pi(P_{{\mathbf n}_L|{\mathbf n}_R})\ ,\ee
and 
\be P_{{\mathbf n}_L|{\mathbf n}_R} \star P_{{\mathbf m}_L|{\mathbf m}_R} \ = \ \delta_{{\mathbf n}_R,{\mathbf m}_L} P_{{\mathbf n}_L|{\mathbf m}_R} \ , \label{projalg}\ee
with ${\bf n}_{L,R}=(n_1,n_2)_{L,R}\in ({\mathbb Z}+1/2)\times ({\mathbb Z}+1/2)$, {\it idem} ${\bf m}_{L,R}$, being half-integer eigenvalues under the left or right star-product action of number operators $w_i$,
\be (w_i-n_{iL}) \star  P_{{\mathbf n}_L|{\mathbf n}_R} \ = \ 0 \ = \ P_{{\mathbf n}_L|{\mathbf n}_R} \star (w_i-n_{iR}) \ . \ee
Clearly, the $P_{{\mathbf n}_L,{\mathbf n}_R} $ also diagonalize the adjoint as well as twisted-adjoint actions of $K_{(\pm)}$, \emph{viz.} 
\be K_{(\pm)}\star P_{{\mathbf n}_L|{\mathbf n}_R} -P_{{\mathbf n}_L|{\mathbf n}_R}  \star K_{(\pm)} \ = \ \frac12\left(n_{2L}\pm n_{1L} -(n_{2R}\pm n_{1R}) \right)P_{{\mathbf n}_L|{\mathbf n}_R} \ ,\label{adjeig} \ee
\be K_{(\pm)}\star P_{{\mathbf n}_L|{\mathbf n}_R} -P_{{\mathbf n}_L|{\mathbf n}_R}  \star \pi(K_{(\pm)}) \ = \ \frac12 \left(n_{2L}\pm n_{1L} -(-1)^{\sigma_\pi(K_{(\pm)})} (n_{2R}\pm n_{1R}) \right)P_{{\mathbf n}_L|{\mathbf n}_R} \ ,\label{twadjeig} \ee
where $\pi(K_{(\pm)}) = \sigma_\pi(K_{(\pm)}) K_{(\pm)}$.

The diagonal elements $P_{\bf n|\bf n}\equiv P_{\bf n}=P_{n_1,n_2}$ are projectors and belong to the enveloping algebra of the number operators, and hence factorize as $P_{n_1,n_2}(w_1,w_2)=P_{n_1}(w_1)\star P_{n_2}(w_2)$. 
In particular, the projectors onto the lowest-weight state of the Fock space ($\e_2=+1$) and the highest-weight of the anti-Fock space ($\e_2=-1$) correspond to 
\begin{equation}
P_{\frac{\e_1}{2},\frac{\e_2}{2}} \ = \ 4e^{-2(\e_1 w_1+ \e_2 w_2)} \ = \ 4e^{-4\e_2K_{(\e_1\e_2)}}\text{ ,}\qquad \e_1,\e_2=\pm\ .
\label{Bgrndproj}
\end{equation}
In star-product form, the generic projector reads
\begin{equation}
P_{n_{1},n_{2}}=\frac{\left( a_{2}^{\e_2}\right) ^{\star
(|n_{2}|-1/2)}}{\sqrt{(|n_{2}|-1/2) !}}\star \frac{\left( a_{1}^{\e_1}\right)
^{\star (|n_1|-1/2)}}{\sqrt{(|n_1|-1/2) !}}\star P_{\frac{\e_1}{2},\frac{\e_2}{2}
}\star \frac{\left( a_{1}^{-\e_1}\right) ^{\star (|n_1|-1/2)}}{\sqrt{(|n_1|-1/2)!}
}\star \frac{\left( a_{2}^{-\e_2}\right) ^{\star (|n_2|-1/2)}}{\sqrt{(|n_2|-1/2) !}}\text{ ,}\label{projF}
\end{equation}
where $\e_i:={\rm sign} (n_i)$. 

Of course, adjoint and twisted-adjoint action \eq{adjeig}-\eq{twadjeig} are only different for $\pi$-odd Cartan generators (i.e., transvections). Since $K_{(\pm)}\star\k_y=\k_y\star\pi(K_{(\pm)})$, for any $\pi$-odd $K_{(\pm)}$ star-multiplication of $P_{{\mathbf n}_L|{\mathbf n}_R}$ by $\k_y$ exchanges adjoint and twisted-adjoint action, e.g. $K_{(\pm)}\star P_{{\mathbf n}_L|{\mathbf n}_R}\star\k_y-P_{{\mathbf n}_L|{\mathbf n}_R}\star\k_y\star\pi(K_{(\pm)}) = [K_{(\pm)},P_{{\mathbf n}_L|{\mathbf n}_R}]_\star \star\k_y$ .  This means, in particular, that in such cases one can define skew-diagonal, or \emph{twisted projectors} via star-multiplication by $\k_y$,
\be P_{\bf n|-\bf n} \ \equiv \ \tP_{\bf n} \ = \ \tP_{n_1,n_2} \ := \  P_{n_1,n_2}\star\k_y \ ,\ee
which will be distributions in $Y$. The addition of elements like the projectors and twisted projectors to the algebra of polynomials in oscillators (Weyl algebra) thus forms an extension of the latter which we refer to as \emph{extended Weyl algebra} ${\cal A}({\cal Y}_4)$ \cite{us,Aros:2019pgj,DeFilippi:2019jqq}.

There are \emph{three} distinct choices of $(K_{(+)},K_{(-)})$ modulo $Sp(4,\Real)$ rotations, given by \cite{us,us2}\footnote{We refer the reader to the Appendix \ref{App:conv} for our $AdS_4$ and spinor conventions.}
\be (E,J) \ , \qquad (J,iB) \ , \qquad (iB,iP) \ ,\ee
where $E:=P_0=M_{0'0}$ is the AdS energy, $J:=M_{23}$ is a spin, $B:= M_{01}$ is a boost and $P:=P_3=M_{0'3}$ is a transvection\footnote{Obviously, the concrete choices of embedding into $\mso(2,3)$ are purely conventional. What matters are only the possible inequivalent choices of $(K_{(+)},K_{(-)})$ as compact or non-compact  Cartan generators.}. As $E$ and $J$ are compact it follows that $\exp(\pm 4E)$ are projectors, while a factor of $i$ accompanies the non-compact generators $B$ and $P$ at the exponent in order for condition \eq{K2} to be satisfied, giving rise to projectors $\exp(\pm 4iB)$ and $\exp(\pm4i P)$.

Thus, starting from a pair of Cartan generators, one may form four lowest-weight ($\e=-$) or highest-weight ($\e=+$) projectors, namely $\exp(4\e K_{(\e')})$, where $\e,\e'=\pm$, and their twisted counterparts $\exp(4\e K_{(\e')})\star \kappa_y$, which are distinct elements iff $K_{(\e')}=E$ or $iP$ ($\exp(\pm 4J)\star \kappa_y=\exp(\pm 4J)$, \emph{idem} $iB$).
Once a pair is chosen, then the orbit of a chosen $\exp(4\e K_{(\e')})$ under the left and right actions of the extended Weyl algebra form an associative algebra ${\cal M}_\e(K_{(\e')};K_{(-\e')})$, with \emph{principal Cartan generator} $K_{(\e')}$; letting ${\cal M}(K_{(\e')};K_{(-\e')})={\cal M}_+(K_{(\e')};K_{(-\e')})\oplus {\cal M}_-(K_{(\e')};K_{(-\e')})$, we thus have \emph{six} possibilities, %\footnote{The orbits ${\cal M}(K_{(\e')};K_{(-\e')})\oplus {\cal M}_\e(K_{(-\e')};K_{(\e')})$ do not account for all stargenfunctions $P_{{\mathbf n}_L,{\mathbf n}_R}$ introduced above, see \cite{Aros:2019pgj}.} 
\be {\cal M}(E;J)\ ,\quad {\cal M}(J;E)\ ;\qquad {\cal M}(J;iB)\ ,\quad {\cal M}(iB;J)\ ;\qquad {\cal M}(iB;iP)\ ,\quad {\cal M}(iP;iB)\label{families} \ .\ee
Expanding $\Phi'$ over ${\cal M}(K_{(\e')};K_{(-\e')})$, we refer to the contributions from the Weyl algebra orbits of $\exp(\pm4K_{(\e')})$ and $\exp(\pm4K_{(\e')})\star \kappa_y$, respectively, as the \emph{regular} and \emph{twisted} sectors,% , since the former gives rise to a Weyl zero-form that is real-analytic in $Y$ at the unfolding point.%\footnote{The one-sided star multiplication by $\kappa_y$ exchanges a symbol by a dual symbol obtained by chiral Fourier transformation in $y$-space (but not $\yb$-space) followed by replacing the Fourier dual variable by $y$. 
%
%This duality transformation, that need not be a symmetry of the symbols of a generic quantum mechanical system, leaves the solution spaces found in \cite{Iazeolla:2017vng} invariant; whether it is a symmetry of higher spin gravity, possibly related to a GSO-like projection of an underlying topological open string, is an interesting open problem.}.
%
the latter being non-trivial iff the principal Killing vector is taken to be $E$ or $iP$. In such case we expand
\bea {\cal M}(E;J)\ ,\quad {\cal M}(iP;iB)\ :\qquad \Phi'(Y) \ = \ \sum_{{\mathbf n}_L,{\mathbf n}_R } \left( \tn_{{\mathbf n}_L|{\mathbf n}_R}P_{{\mathbf n}_L|{\mathbf n}_R}(Y) +\n_{{\mathbf n}_L|{\mathbf n}_R}P_{{\mathbf n}_L|{\mathbf n}_R}(Y)\star\kappa_y \right)\ , 
\eea
where $\tn_{{\mathbf n}_L|{\mathbf n}_R}$ and $\n_{{\mathbf n}_L|{\mathbf n}_R}$ are independent deformation parameters. %, while in the remaining families we set the $\mu$-parameters to zero, \emph{viz.}
%
%\bea {\cal M}(J;E)\ ,\quad {\cal M}(J;iB)\ ,\quad {\cal M}(iB;J)\ ,\quad {\cal M}(iB;iP)\ :\qquad \Psi(Y) \ = \ \sum_{{\mathbf n}_L,{\mathbf n}_R } \n_{{\mathbf n}_L,{\mathbf n}_R}P_{{\mathbf n}_L,{\mathbf n}_R}(Y) \ . \eea
%

In this paper we shall limit our considerations to the above two families, respectively with $E$ and $iP$ principal Cartan generators, and to expansions over projectors and twisted projectors only. In fact, for simplicity, we shall only consider \emph{symmetry-enhanced projectors}, obtained from summing all $P_{n_1,n_2}=P_{n_1,n_2}(K_{(+)},K_{(-)})$  with fixed eigenvalue of the principal Cartan generator, in such a way that the dependence on the other Cartan generator drops out and one is left with  ($n=\pm 1, \pm 2, ...$)
\bea {\cal P}_{n}(K_{(q)}) &=& \sum_{\tiny \ba {c}n_2+qn_1=n\\[-3pt] \e_1\e_2=q\ea}P_{n_1,n_2} %\ = \ 4(-1)^{n-1}\e_n \,e^{-4K_{(q)}}L^{(1)}_{n-1}(8K_{(q)})\ , \qquad \e_n=n/|n| \ ,\label{enhanced2} %\\[5pt]& = & 2(-1)^{n-\ft{1+\ve}2}\,\oint_{C(\ve)} \frac{d\eta}{2\pi i}\,\left(\frac{\eta+1}{\eta-1}\right)^{n}\,e^{-4\eta K_{(q)}}
\ ,\label{enhanced}\eea
that only depend on the principal Cartan generator.  The ${\cal P}_n$ admit the Weyl-ordered expressions
\bea {\cal P}_{n}(K_{(q)}) &=& 4(-1)^{n-1}\ve_n \,e^{-4K_{(q)}}L^{(1)}_{n-1}(8K_{(q)})\label{enhanced2} %\\[5pt]& = & 2(-1)^{n-\ft{1+\ve}2}\,\oint_{C(\ve_n)} \frac{d\eta}{2\pi i}\,\left(\frac{\eta+1}{\eta-1}\right)^{n}\,e^{-4\eta K_{(q)}}  
\ , \qquad \e_n=n/|n| \ ,
\ ,\label{enhanced2}\eea
where $L^{(1)}_{n-1}(x)$ are generalized Laguerre polynomials\footnote{It is easy to see, by using Kummer's transformation $L^{(\a)}_k(x) \ = \ (-1)^\a e^x L^{(\a)}_{-k-1-\a}(-x)$ ($k=-2,-3,...$ and $\a$ is a positive integer), that Eq. \eq{enhanced2} is a universal formula accommodating projectors with both signs of $n$.}.  %While the rank-$1$ projectors $P_{n_1,n_2}$ are only invariant under $\mso(2,\Comp)_{K_{(+)}}\oplus\mso(2,\Comp)_{K_{(-)}}$, the symmetry enhanced projectors share the same symmetry of the $\cP_{\pm1}$, i.e. under the whole centralizer  $ \mso(2,\Comp)\oplus\mso(3,\Comp) $ of the principal Cartan generator.  

Hence, we shall consider expansions
\bea {\cal M}(E;J)\ ,\quad {\cal M}(iP;iB)\ :\qquad \Phi'(Y) \ = \ \sum_{n} \left( \tn_n\cP_n(Y) +\n_n\cP_n(Y)\star\kappa_y \right)\ . \label{expphi}
\eea
For more general expansions and the physical meaning of other families, see \cite{us,us2,Aros:2017ror,Aros:2019pgj,DeFilippi:2019jqq}.

%%%%%%%%%%%%%%%%%%%%%%%%%%%%%%%%%%%%%%%%%%%%%%%%%%%%%%%%

\scss{Fibre representatives of linearized solutions in compact and conformal basis.}\label{Sec:fibcomp}

%%%%%%%%%%%%%%%%%%%%%%%%%%%%%%%%%%%%%%%%%%%%%%%%%%%%%%%%

Choosing a basis of eigenstates $P_{{\mathbf n}_L|{\mathbf n}_R}$ in the ${\cal M}(E;J)$ or the ${\cal M}(iP;iB)$ family essentially amounts to choosing between the compact or the conformal slicing of $\mso(2,3)$-modules. 

\paragraph{Compact basis.} Indeed, the first case corresponds to considering the infinite-dimensional orbits of the lowest-weight projector $P_{\ft12,\ft12}\equiv {\cal P}_1$ and of the highest-weight projector $P_{-\ft12,-\ft12}\equiv {\cal P}_{-1}$,
\be {\cal P}_{\pm1} \ = \ 4e^{\mp 4E} \ ,\label{eq:vac proj} \ee
both having a stability subalgebra given by the compact $\mso(3)_{M_{rs}}$ of spatial rotations, 
\be  [E,\cP_{\pm1}]_\pi \ = \ E\star \cP_{\pm1} + \cP_{\pm1} \star E \ = \ \pm \cP_{\pm1} \ , \qquad [M_{rs}.\cP_{\pm1}]_\pi \ = \ 0\ , \ee
$r,s=\{1,2,3\}$, and being annihilated respectively by the energy-lowering generators $L^-_r$, 
\bea  [L^-_r,\cP_1]_\pi \ = \ L^-_r\star\cP_1 - \cP_1\star L^+_r \ = 0 \ , \eea
and the energy-raising operator, 
\bea  [L^+_r,\cP_{-1}]_\pi \ = \ L^+_r\star\cP_1 - \cP_1\star L^-_r \ = 0 \ , \eea
of the compact basis \eq{ll}-\eq{ml}. In general, one can show \cite{Iazeolla:2008ix,DeFilippi:2019jqq} that linear combinations of the $P_{{\mathbf n}_L|{\mathbf n}_R}$ in compact basis realize all particle states $T_{e,(s)}$ that fill massless particle and anti-particle irreps $\mD(\pm(s+1),(s))$ of all spins, with  definite eigenvalues under the twisted-adjoint action of the compact Cartan generators $E$ and $J$ and of the quadratic Casimir $\frac12 M^{rs}\star M_{rs}$ of $\mso(3)$, 
\bea
&[E,T_{e;(s)}]_\pi=\{E,T_{e;(s)}\}_\star  =eT_{e;(s)}\,,&\\ 
&\frac12 [M^{rs},[M_{rs},T_{e;(s)}]_\pi]_\pi =\frac12 [M^{rs},[M_{rs},T_{e;(s)}]_\star]_\star =s(s+1)T_{e;(s)}\,,&
\eea
where each $T_{e;(s)}$ is a $(2s+1)$-plet with elements distinguished by the eigenvalue $j_s$ of $J$, $j_s = -s, -s+1,\ldots, s-1, s $,
and they span lowest-weight  modules (highest-weight modules for the anti-particle states) built via the action of energy-raising (lowering) operators $L^+_r$ ($L^-_r$) on a lowest-weight (highest-weight) state $T_{e_0;(s_0)}$ ($T_{-e_0;(s_0)}$), 
\be [L^-_r,  T_{e_0;(s_0)}]_\pi =L^-_r \star  T_{e_0;(s_0)}-T_{e_0;(s_0)}\star L^+_r =0
\,,\qquad{\rm for}\quad 
e_0 = s_0+1 
\,.
\label{eq:LWcond}\ee
One can thus establish a correspondence between each $T_{e,(s)}$ under twisted-adjoint action and each state $\ket{e,(s)}$ under left action. Massless particle and anti-particle states have all
\be |e|>s\,,\ee
and can in fact be built via the higher-spin algebra $\mhs(2,3)$ action on the $\mD(1,0)$ massless scalar particle lowest weight state \cite{Iazeolla:2008ix} $T_{1;(0)} \equiv {\cal P}_1$ and on the $\mD(-1,0)$ massless scalar anti-particle highest weight state $T_{-1;(0)}  \equiv {\cal P}_{-1}$. The particle and anti-particle module are exchanged by the action of $\pi$, that reverses the sign of $E$. 

The $P_{{\mathbf n}_L|{\mathbf n}_R}$ actually diagonalize the separate left and right action of the compact Cartan generators: in fact, from the point of view of the separate left/right action of $\mso(2,3)$, the $T_{e,(s)}$ correspond to enveloping algebra realizations of (anti-)singleton states \cite{Iazeolla:2008ix}. This is a reflection, at an operatorial level, of the compositeness of massless particle states in $AdS$. Thus, each element $T_{e;(s)}$ corresponds to a specific linear combination of operators on the (anti-)singleton Hilbert space, i.e., is an enveloping-algebra realization of the specific tensor product of singleton states corresponding to any massless particle state according to the Flato-Fronsdal theorem \cite{Flato:1978qz}. For example,
\be
E\star T_{1;(0)} 
=
\frac12\, T_{1;(0)} =T_{1;(0)}\star E 
\,,\qquad 
L^-_r\star T_{1;(0)} =0 =T_{1;(0)}\star L^+_r 
\,,
\ee
thus
\be
T_{1;(0)} =|\ft12;(0)\rangle \langle\ft12;(0)| 
\qquad\leftrightarrow\qquad  
|\ft12;(0)\rangle_1 |\ft12;(0)\rangle_2 = | 1;(0)\rangle  
\,,\ee
where $ |\ft12;(0)\rangle$ is the singleton lowest-weight state. 

In this correspondence,  the enhanced projectors \eq{enhanced2} ${\cal P}_n(E)$, $n=\pm 1,\pm 2,\dots$, correspond to the sum of the projectors onto all the (anti-)singleton states with energy $n/2$ and spin $(|n|-1)/2$, 
\be {\cal P}_n(E)\cong 
\ket{\frac{n}2;(\frac{|n|-1}2)}^{i(|n|-1)}{}_{i(|n|-1)}\bra{\frac{n}2;(\frac{|n|-1}2)}\ ,
\qquad n=\pm 1,\pm 2,\dots\ ,\label{calPnsum}\ee
where the notation $i(|n|-1)=i_1 i_2\ldots i_{|n|-1}$ denotes $|n|-1$ symmetrized doublet indices. These projectors obey
\be (E-\frac{n}2)\star {\cal P}_n(E)=0={\cal P}_n(E)\star (E-\frac{n}2)\ ,\qquad [M_{rs},\cP_n(E)]_\star  = 0\ ,\ee
from which it follows that they are rotationally invariant 
and that their twisted-adjoint energy eigenvalues are given 
by $n$, \emph{i.e.}
\be E\star\cP_n(E)-\cP_n(E)\star\pi(E) = \{E,\cP_n(E)\}_\star  = n\cP_n(E) \ .\label{twadj}\ee
One can thus recognize that the enhanced projectors correspond to rotationally-invariant massless scalar field states belonging to the $\mD(\pm 1,(0))$ ($n$ odd) and the $\mD(\pm 2,(0))$ ($n$ even) irreps. 

It also follows that the \emph{twisted projectors} $\tcP_n$, 
\be \widetilde{{\cal P}}_{n}:= {\cal P}_{n}(E)\star\k_y \cong 
\ket{\frac{n}2;(\frac{|n|-1}2)}^{i(|n|-1)}{}_{i(|n|-1)}\bra{-\frac{n}2;(\frac{|n|-1}2)}\ ,
\qquad n=\pm1,\pm2,\dots\ ,\label{twpr}\ee
i.e., can be thought of as endomorphisms on the (anti-)supersingleton module\footnote{Thus, the two-sided twisted-adjoint action of the algebra on the $\tcP_n$ is mapped to the left action on states that belong to the tensor product of a singleton and an anti-singleton \cite{Iazeolla:2017vng}, $\widetilde{{\cal P}}_{n} \ \leftrightarrow \ 
\ket{\frac{n}2;(\frac{|n|-1}2)}^{i(|n|-1)}_1\ket{-\frac{n}2;(\frac{|n|-1}2)}_{2,i(|n|-1)}$. It is therefore natural to expect that these black-hole-like solutions of the bulk theory correspond to singleton-anti-singleton composites in the boundary CFT \cite{Iazeolla:2017vng} (see also \cite{Iazeolla:2008ix,Basile:2018dzi} for results on the decomposition of this tensor product).} $(1+\pi)[\mD(1/2,(0))\oplus\mD(1,(1/2))]$. This is due to the fact that $\k_y$ changes the sign of the energy eigenvalue under the right-action, i.e.,
\be E\star\tcP_n \ = \ \frac{n}{2}\,\tcP_n \ = \ -\tcP_n\star E\ , \ee
and 
\be E\star\tcP_n-\tcP_n\star\pi(E) = [E,\cP_n(E)]_\star \star \k_y  = 0\ ,\ee
so these elements can be used as initial data for \emph{static}, \emph{spherically symmetric} solutions. Together with more general twisted states $P_{{\mathbf n}_L|{\mathbf n}_R}\star\k_y$, they fill the wedge in weight space in between particle and anti-particle modules, i.e. satisfy
\be |e| \leq s \,.\ee
However, states of this sort cannot be thought of as local data of wedge states, corresponding to solutions in $\mathfrak{W}$ \eq{MDW}, but rather as local data of singular black-hole-like solutions \eq{WeylD}. This will be checked later on, but evidently $\k_y$ is not an element of $\mso(2,3)$, while, as stated before, wedge modes and particle modes are elements of the same indecomposable module $\mathfrak{M}$ of $\mso(2,3)$ \eq{MDW}. Twisted projectors are instead singular (distributions) on ${\cal Y}_4$. For instance, the operators that maps the singleton and anti-singleton ground states into one another are represented by the Weyl ordered symbol (from \eq{eq:vac proj} and \eq{mab})
\begin{equation}
\tcP_{\pm1} \ \equiv \ {\cal P}_{\pm 1}\star\k_y \ = \ 8 \pi\delta^2(y\mp i\sigma_0\yb)
\,.
\end{equation}
Moreover, by its definition $L^\mp_r\star {\tcP}_{\pm1}=0=\tcP_{\pm1}\star L^\mp$, even though they are not lowest or highest-weight states in the twisted-adjoint sense.% whereas there are no lowest- or highest-weight elements within the wedge modes. 

An operator realization for the wedge modes was found in \cite{Sezgin:2005pv,Iazeolla:2008ix}\footnote{In the same reference, a composite origin for the wedge states (and, in fact, for the full compact twisted adjoint module) was suggested in terms of \emph{angletons}, more complicated than the one of particle states in terms of singletons as it involves a quotient over certain equivalence relations. For details, see \cite{Iazeolla:2008ix}.}. For instance, the elements corresponding to \eq{phi00} and \eq{phi01} are
\be T_{0,(0)} \ = \ \frac{\sinh 4E}{4E} \ , \qquad (T_{0,(1)})_r \ = \ \frac{3}{(4E)^2}P_r\left(\cosh 4E-\frac{\sinh 4E}{4E} \right) \ , \label{wedgestates} \ee
as we shall verify in Section \ref{Sec:sptreg}. Note that these are regular twistor-space elements, differently from the $\tcP_n$. 

An expansion \eq{expphi} in ${\cal M}(E,J)$ thus incorporates massless scalar particle and spherically-symmetric black-hole states, and the reality conditions \eq{reality} (together with the property that $\cP_n(E)\star\k_y\kb_{\yb}=(-1)^n\cP_n(E)$ \cite{Iazeolla:2017vng}) dictate that particles should be accompanied by anti-particles, $\tn^\ast_n=\tn_{-n}$, and that $\n^\ast_n=(-1)^n\n_n$  (i.e., $\n_n=i^n\m_n$, where $\m_n\in\Real$).

\paragraph{Conformal basis.} The choice ${\cal M}(iP,iB)$ leads to a completely parallel construction with $iP$ instead of $E$ as principal Cartan generator and the boost $iB$ in place of $J$. Identifying, for simplicity\footnote{This is of course purely conventional, and one could rather work more covariantly by introducing the embedding vectors $(L_i^a,L^a)$ with $L^a L_a=\e=\pm 1$, $L_i^a L_a =0 $, $L^a_i L_{ja}=\eta_{ij}=(+,+,-\e)$, with principal Cartan generator $L^aP_\a$, distinguishing between compact and conformal basis via the choice of timelike or spacelike embedding vector $L^a$. The embedding of the remaining generators in compact of conformal basis then involves defining $M_{ij}=L^a_i L^b_j M_{ab}$ and raising/lowering operators $(L^aM_{ab}+\b_\pm P_b)L^b_i$ (see also \cite{Sezgin:2005pv,Aros:2017ror}).}, $iP$ with $iP_3$, the lowest/highest vectors 
\be \cP_{\pm 1 }(iP) \ = \ 4e^{\mp iP} \ee
satisfy
\be  [iP,\cP_{\pm1}]_\pi \ = \ iP\star \cP_{\pm1} + \cP_{\pm1} \star iP \ = \ \pm \cP_{\pm1} \ , \qquad [M_{mn}.\cP_{\pm1}]_\pi \ = \ 0 \ ,\ee
$m,n=\{0,1,2\}$, and are respectively annihilated by the combinations $T_m:=M_{m3}-P_m$, 
\bea  [T_m,\cP_1]_\pi \ = \  0 \ , \eea
and $K_m:=M_{m3}+P_m$,
\bea  [K_m,\cP_{-1}]_\pi \  = \ 0 \ , \eea
where we note that $\pi(K_m)=T_m$. 

Choosing to embed the boundary conformal algebra in such a way that all its generators are hermitian, one can identify the dilatation generator $D=P_3$, with eigenvalue $\D$, with respect to which $K_m$ is the $\D$-lowering operator, generator of the special conformal transformations, and $T_m$ is the $\D$-raising operator, generator of the boundary commuting translations. The corresponding commutation relations read as in \eq{kt}-\eq{mt} and $ 4e^{\mp iP}$ are highest- ($\cP_{1}$) and lowest-weight ($\cP_{-1}$) states with respect to $D$. We shall stick to this definition from now on. 

All states created via the action of $(K_m,T_m)$ on $ 4e^{\mp iP}$ are organized in representations $\mD(\D_0,(s_0))$ bounded from below/above in the eigenvalue $\D$ of $D$, and with the $\mso(1,2)$-spin $s$ as second label, with $|\D|>s$. In other words, these $\mso(2,3)$ moduli are naturally sliced with respect to the so-called conformal basis $\mso(1,1)_{iP}\oplus\mso(1,2)_{M_{mn}}$ \cite{Gunaydin:1999jb}. Note that in ${\cal M}(iP,iB)$  the $\pi$-map exhanges highest- and lowest-weight submodules, i.e., reverses the sign of $\D$.  

Generic projectors $P_{n_1,n_2}$ depend on both $iP$ and $iB$, whereas symmetry-enhanced projectors $\cP_n(iP)$, defined as in \eq{enhanced2} with $K_{(q)}=iP$, have the same full boundary Lorentz-symmetry as $\cP_{\mp1}$. 
Twisted projectors $\tcP_n$ of ${\cal M}(iP,iB)$ satisfy
%
% \be D\star\tcP_n \ = \ \frac{n}{2}\,\tcP_n \ = \ -\tcP_n\star D\ , \ee lautoval di D in questo caso diventa IM
%
%and 
%
\be D\star\tcP_n(iP)-\tcP_n(iP)\star\pi(D) = [D,\cP_n(iP)]_\star \star \k_y  = 0\ ,\ee
i.e., have conformal weight $\D=0$, are 3D-Lorentz scalars, and are distributions in $Y$. In particular, 
\begin{equation}
\tcP_{\mp1} \ \equiv \ {\cal P}_{\mp 1}(iP)\star\k_y \ = \ 8 \pi\delta^2(y\mp \sigma_3\yb)
\,,
\end{equation}
and are annihilated from the left and from the right by $K_m$ ($\tcP_{-1}$) and $T_m$ ($\tcP_1$). Here, too, one can define analogues of the wedge states \eq{wedgestates}, the interpretation of which we leave for future work. 

The expansion \eq{expphi} on ${\cal M}(iP,iB)$ requires, via \eq{reality} and $\cP_n(iP)\star\k_y\kb_{\yb}= (-1)^{n-1}\cP_n(iP)$, $\tn^\ast_n=\tn_n$ and that every $\tcP_n$ be accompanied by a $\tcP_{-n}$, with $\n^\ast_n=(-1)^{n-1}\n_{-n}$. 

The ``Wick-rotation'' that connects solutions of the Klein-Gordon equation in spherical coordinates and Poincar\'e coordinates, that we examined in Section \ref{sptP}, can be here shown more precisely, at the level of fibre elements that generate those solutions via unfolding. Indeed, the compact $(E, M_{rs} , L^\pm_r)$ and non-compact split $(D,M_{mn},K_m,T_m)$ of $\mso(2,3)$ are related \cite{Gunaydin:1999jb} (see also \cite{Sezgin:2002rt}) by a similarity transformation via a non-unitary operator that, with the above choice of embedding, corresponds to $e^{L^+_3}$. One can show, in particular, that ($i,j=\{1,2\}$)
\bea & \displaystyle  e^{L^+_3}\star P_3\star e^{-L^+_3}  =  i\left(E-\frac12 L^-_3\right) \ ,\qquad 
e^{L^+_3}\star M_{ij}\star e^{-L^+_3}  =  M_{ij} \ , &\\
& \displaystyle e^{L^+_3}\star M_{0i}\star e^{-L^+_3}  =   iM_{3i}+\frac12 L^-_i \ ,\quad 
e^{L^+_3}\star K_0 \star e^{-L^+_3}  =  \frac12 L^-_3\ ,\quad 
e^{L^+_3}\star K_i \star e^{-L^+_3}  =  \frac{-i}2 L^-_i \ ,  &\eea
and that the highest/lowest-weight states in the two basis are connected via
\be e^{-L^\pm_3}\star e^{\mp 4E}\star \pi(e^{L^\pm_3}) \ = \ e^{-L^\pm_3}\star e^{\mp 4E}\star e^{L^\mp_3} \ \propto \ e^{\pm4iP}  \ . \label{NUtrasf}\ee
%

%%%%%%%%%%%%%%%%%%%%%%%%%%%%%%%%%%%%%%%%%%%%%%%%%%%%%%%%%%%%%%%%%%%%%%%%%
\scss{Regular presentation}\label{Sec:genproj}
%%%%%%%%%%%%%%%%%%%%%%%%%%%%%%%%%%%%%%%%%%%%%%%%%%%%%%%%%%%%%%%%%%%%%%%%%

The enveloping-algebra realization \eq{enhanced2} of the projectors ensures their idempotency, but results in a divergent star product between states with positive and negative $K_{(q)}$-eigenvalue. As shown in \cite{us}, this is due to the divergence arising from the star product between Fock and anti-Fock space elements,
\be e^{-2\eta w}\star e^{-2\eta' w} \ = \ \frac{1}{1+\eta\eta'}\,e^{-2\frac{\eta+\eta'}{1+\eta\eta'} w} \ , \qquad \eta,\eta'  = \pm 1 \ . \label{EstarE} \ee
Therefore, local data of the form  \eq{enhanced2}, while suitable for encoding linear solution to the equations, cannot be used to proceed at higher orders in the perturbative expansion \eq{eq:V^m fact}, since repeated star products of $\Psi'$ would give rise to divergencies -- unless one restricts the expansion \eq{expphi} to elements that only have regular star product among themselves\footnote{This is permitted by reality conditions e.g. for expansions \eq{expphi} only on black-hole states ($\tn_n=0$) and either $\n_{n<0}=0$ or $\n_{n>0}=0$ in ${\cal M}(E,J)$ \cite{us,us2}.}. 

This problem can be solved by adopting a different presentation for the $\cP_n$ that, while equivalent at the linearized level, has the advantage of regularizing the star products (see \cite{us,Iazeolla:2017vng,Aros:2017ror,Aros:2019pgj,DeFilippi:2019jqq}),
\bea {\cal P}_{n}(K_{(q)}) \ = \ 2(-1)^{n-1}\ve_n\,\oint_{C(\ve_n)} \frac{d\eta}{2\pi i}\,\left(\frac{\eta+1}{\eta-1}\right)^{n}\,e^{-4\eta K_{(q)}}\ , \qquad n=\pm1, \pm2,... \ ,\label{enhanced3}\eea
where the contour integral is performed around a small contour $C(\ve_n)$  
encircling $\ve_n:=n/|n|$. Evaluating the residue gives back \eq{enhanced2}. However, evaluating star products before the auxiliary closed-contour integral, it is possible to achieve orthonormality. Indeed, using \eq{EstarE}, and defining $u:=(\eta+\eta')/(1+\eta\eta')$ at fixed $s\in C(\ve)$\footnote{If $\ve=\ve'$ then $u\in C(\ve')$ provided $|\eta'-\ve'|<<1$ and $|\eta-\ve|<<1$ while if $\ve=-\ve'$ then $u\in C(\ve')$ provided $|\eta'-\ve'|<<|\eta-\ve|<<1$.}, one finds
\bea {\cal P}_{n}\star\cP_m &=&  
4(-1)^{m-1}\ve_n\ve_m'\,\oint_{C(\ve)} \frac{d\eta}{2\pi i}\,\frac{(\eta+1)^{n-m-1}}{(\eta-1)^{n-m+1}}\oint_{C(\ve')}\frac{d u}{2\pi i}\,\left(\frac{u+1}{u-1}\right)^{m}\,e^{-4u K_{(q)}}\nn\\
&=&\d_{nm}\cP_m\ .\label{projint}\eea
As for the twisted projectors $\tcP_n$ built from $\cP_n(E)$ and $\cP_n(iP)$, the regular presentation is given by
\bea  \tcP_n \ := \ \cP_n(K_{(q)})\star\k_y \ = \ 4\pi (-)^{n-1}\ve_n\,\oint_{C(\ve_n)} \frac{d\eta}{2\pi i}\,\left(\frac{\eta+1}{\eta-1}\right)^{n}\,\d^2(y-i\eta v\yb)\ ,\label{tPE} \eea
where $v_\a{}^{\bd}=(\s_0)_\a{}^{\bd}$ for $E$ and $v_\a{}^{\bd}=i(\s_3)_\a{}^{\bd}$ for $iP$. With the help of such regular presentations, projectors and twisted projectors indeed form an subalgebra of \eq{projalg}, containing an ideal spanned by the projectors, \emph{viz.}
\bea \cP_n\star \cP_m & = & \d_{nm}\cP_n \ , \label{genproj1}\\[5pt]
\cP_n\star \tcP_m & = & \d_{nm}\tcP_n \ , \label{genproj2}\\[5pt]
\tcP_n\star \cP_m & = & \d_{n,-m}\tcP_n \ , \label{genproj3}\\[5pt]
 \tcP_n\star \tcP_m & = & \d_{n,-m}\cP_n \ .\label{genproj4}\eea

Let us consider first the compact case ${\cal M}(E,J)$. The roles of projectors and twisted projectors, as local data of massless particle and black-hole solutions, respectively, in the expansion \eq{expphi}, are reversed in the expansion of the adjoint local datum $\Psi'=\Phi'\star\k_y=\sum_n(\tn_n\tcP_n+\n_n\cP_n)$. Hence, using \eq{genproj1}-\eq{genproj4} in computing the non-linear corrections \eq{eq:V^m fact}, one concludes that, as expected, the black-hole sector closes on itself (due to \eq{genproj1}), whereas massless particles generate black-hole modes via interactions, starting at second order\footnote{It will be interesting to see if and how these conclusions will be altered in moving to the $G$-gauge via perturbative corrections to the vacuum gauge function.}. It is important to note that, endowed with the above regular presentation, the solution space ${\cal M}(E,J)$ can be elevated to an exact solution via the  expansion \eq{eq:Phi^m fact}-\eq{eq:U^m fact}, as its elements generate an (associative) star-product algebra ${\cal A}({\cal Y}_4)$ which can be further extended with the algebra of the $v_n(z)$ without losing associativity  \cite{us,Iazeolla:2017vng}\footnote{This happens because both the projector and twisted projector basis and the $Z$-dependent coefficients $v_n(z)$ separately self-replicate under star product, preserving associativity  --- with the only proviso that star products have to be performed before  auxiliary integrations. The proof of associativity within our solution space, however, does \emph{not} necessarily rely on the factorization of the Ansatz. As specified in \cite{Iazeolla:2017vng}, the factorized form only speeds up the calculations, but one can equally well first take the star products among $Y$-dependent and $Z$-dependent factors  together and then work with primed fields in normal ordered basis. The associativity within the solution space can still be checked in this way.}.

The same can be repeated for ${\cal M}(iP,iB)$. However, using \eq{enhanced3} for the ${\cP_n(iP)}$, local data for bulk-to-boundary propagators, has some undesirable consequence\footnote{It has been observed that the gauge-invariant quantity ${\rm Tr}(\Psi(x_1)\star\Psi(x_2))$ reproduces the correct boundary two-point functions \cite{Colombo:2012jx,DidenkoSkvortsov,Bonezzi:2017vha}, with their physical divergence at colliding points. For $x_1\to x_2$, by cyclicity the calculation reduces to computing a trace of the star product of a $\cP_n$ and its $\pi$-conjugate $\cP_{-n}$. Hence, in order to keep this divergence one should either not use \eq{enhanced3} but rather the simpler \eq{enhanced2}, which would not enable to use \eq{eq:V^m fact} to dress bulk-to-boundary propagators into a fully non-linear solution, or resort to a different integral presentation of the regular ${\cal M}(iP,iB)$ solution space.}. The regular presentation above was used in \cite{us,us2,Iazeolla:2017vng,review,DeFilippi:2019jqq} to deal with the simple structure of the rotationally-invariant particle and black-hole modes (as well as for solutions of cosmological interest with scalar profiles on $(A)dS$ background in \cite{Aros:2017ror}), reproducing their orthogonality by means of higher-spin invariant quantities\footnote{The simplest such invariant is ${\rm Tr}(\Psi'\star\Psi')$ \cite{Sezgin:2005pv,Iazeolla:2008ix,Colombo:2010fu,FCS,DeFilippi:2019jqq}, which was used to define an inner product on the massless particle modules already in \cite{Iazeolla:2008ix}.}. In fact, the simple small-contour integral presentation reviewed above needs to be modified in order to be extended to the most general non-diagonal operator $P_{\mathbf{n}_L|\mathbf{n}_R}$ (see \cite{Aros:2019pgj} for more comments and \cite{Iazeolla:2008ix,Yin:2019xxs} for more general integral presentations). Moreover, as we have seen, the regular presentation may have to be adapted to the solution space in exam --- for instance, to the compact/non-compact nature of the principal Cartan generator, which is related to the existence of normalizable solution spaces with elements distinguished by discrete/continuous parameters --- in such a way that it is compatible with all the observable features of the solutions.

%dev'essere probabilmente alterata ad una piu generale per estendere a tuttli gli stati (citaz a BTZ e Yihao per roba ancora piu general) . Possibile che reg pres debba essere adattata alle caratteristiche dell spazio di soluz in esame, o forse la somma dev essere organizzata dibversamente (introducendo moduli continui.. ricorda che somma su n e sensata per gener compatti, e la somma su tutti st normalizzab... ma fors eper st noncomp avrei anche un integrale su vettore di lorentz 3D ... dovr essere Tm commutante, non compatto... come lo stato coerente scritto da gunaydin (ma quello non entra gia attraverso gf? direi di si... )) per generatori non compatti. Per iP, dato il significato dei proiettori come local data of Btb, need non regolarizzare per significato fisico singolarita. 

%In order to understand better the difference, it is useful to rewrite \eq  in terms of a gaussian integrand. COndition ... on K amounts to identifying K with one of the Gamma matrices of the $\msp(4;\Comp)$ generators $M_AB = -1/8... $. Let us represent their 2x2 blocks universally as in terms of 2x2 matrices as ... . vark will be (up to signs) \cite{} a sigma ab and v to some sigma a.  Now the star product with ky will pull down the gaussian determinant ... which dictates the singularities of the element. For pi odd, vark = 0 --> delta function. 

%%%%%%%%%%%%%%%%%%%%%%%%%%%%%%%%%%%%%%%%%

\scss{Spacetime dressing of the fibre modules} \label{spacetime}

%%%%%%%%%%%%%%%%%%%%%%%%%%%%%%%%%%%%%%%%%

We shall now reconstruct the spacetime dependence of the solutions in compact and conformal basis by star-multiplying the local data obtained in the previous Section with the gauge function as in \eq{UCfirst} (or, equivalently, \eq{psipr}). We will thereby verify that the fibre elements that we have studied --- and put in correspondence with particle, black-hole modes, etc. purely on the basis of an algebraic state/operator correspondence --- are indeed the ``seed'' of the solutions in Sections \ref{ptandbh} and \ref{sptP}.  The choice of coordinates can be encoded via the choice of vacuum gauge function \eq{Lgf}-\eq{LP}. 

On the basis of the representation theory studied in the previous Section, we expect that solutions belonging to lowest/highest-weight spaces, i.e., composite-massless scalar modes in compact basis and composite-massless scalar bulk-to-boundary propagators and their descendants in conformal basis, will be obtained from applying \eq{LCpiL}  to the regular fibre elements
\be \Phi_{{\rm reg}}^{\prime(1)}(Y)\ = \ \sum_n \tn_n\cP_n  \ ,\ee 
where $\cP_n=\cP_n(E)$ and $\cP_n=\cP_n(iP)$, respectively; while singular solutions, i.e., black-hole modes in compact basis and  singular solutions with vanishing scaling dimension in conformal basis, from the singular elements
\be \Phi_{{\rm sing}}^{\prime(1)}(Y) \ = \ \sum_n \n_n\cP_n\star\k_y   \ee
again with $\cP_n=\cP_n(E)$ and $\cP_n=\cP_n(iP)$, respectively. Therefore, the Weyl zero-forms that we shall examine are
\bea & \Phi_{{\rm reg}}^{(1)}(x,Y)  \ = \ \sum_n \tn_n \,L^{-1}\star \cP_n\star \pi(L) \ = \ \sum_n \tn_n \tcP^L_n\star\k_y \ , &\label{phiregL}\\
& \Phi_{{\rm sing}}^{(1)} (x,Y)  \ = \ \sum_n \n_n\, L^{-1}\star \tcP_n\star \pi(L) \ = \ \sum_n \n_n \cP^L_n\star\k_y \ , \label{phisingL}&
\eea
where we recall the notation for the adjoint rotation of a purely $Y$-dependent Weyl-ordered symbol $f^L(x,Y):=L^{-1}\star f(Y)\star L$ and we are using the property \eq{eq:k fact} that $\pi(f(x,Y))=\k_y\star f(x,Y) \star \k_y$. Recalling that, due to condition \eq{K2}, $K_{(q)}$ is just one specific $AdS_4$ isometry generator (up to a sign,  $K_{(q)\au\bu}$ is one of the  $(\Gamma_{AB})_{\au\bu}$ matrices), the universal form for the two building blocks of \eq{phiregL}-\eq{phisingL} can be written as
\bea & \displaystyle \tcP^L_n\star\k_y \ = \ {\cal O}_{\eta,n} \frac{1}{\det A}\,\exp\left(iyM\yb\right)  \ , &\label{bbphiregL}\\
&\displaystyle \cP^L_n\star\k_y \ = \ {\cal O}_{\eta,n} \frac{1}{\eta\sqrt{\det \vark^L}}\,\exp\left[-\frac1{2\eta}y(\vark^L)^{-1}y+iy(\vark^L)^{-1}v^L\yb-\frac{\eta}{2}\yb(\bar \vark^L)^{-1}\yb\right] \ , \label{bbphisingL}&
\eea
where $(\vark^L)_{\a\b}^{-1}= -\frac{\vark^L_{\a\b}}{\det\vark^L}$, we have introduced the shorthand notation 
\be {\cal O}_{\eta,n} \ = \  2(-1)^{n-1}\ve_n\,\oint_{C(\ve_n)} \frac{d\eta}{2\pi i}\,\left(\frac{\eta+1}{\eta-1}\right)^{n} \ ,\ee
$M_{\a}{}^{\bd}:=(A^{-1}B)_{\a}{}^{\bd}$ and $A_\a{}^\b$ and $B_{\a}{}^{\bd}$ are matrices resulting from the $L$-rotation of the argument of the delta function in \eq{tPE}, $y^L-i\eta v \yb^L=: Ay +B\yb$ (see \eq{YLst}-\eq{ybLP} for the rotated oscillators in stereographic and Poincar\'e coordinates, and \cite{us,us2,Iazeolla:2017vng,Aros:2019pgj,DeFilippi:2019jqq} for more details on each step of the calculation). 

\scsss{Lowest/highest-weight modules}\label{Sec:sptreg}

The expansion of $\Phi_{{\rm reg}}^{(1)}$ in $Y$ only contains equal powers of 
$y_\a$ and $\yb_{\ad}$, \emph{i.e.}, all Weyl tensors of spin $1,2,3,...$ vanish and the coefficients of the expansion correspond to a scalar field --- corresponding to $\Phi_{{\rm reg}}^{(1)}(x,0)={\cal O}_{\eta,n} \frac{1}{\det A}$, which is regular everywhere for particles --- plus all its on-shell non-trivial derivatives. 

\paragraph{Particle modes.} In particular, for the lowest-weight scalars $\mD(1,(0))$  and $\mD(2,(0))$ (corresponding to $n=1,2$) in the particle case ($K_{(q)}=E$), $(\det A)^{-1}$ gives exactly \eq{scalar12}, with $L=L_{\rm stereo}$, or equivalently \eq{lwspherical} with $L=L_{{\rm spherical}}$. The general result reads, in stereographic coordinates,
\bea \Phi_{{\rm reg}}^{(1)}(E^L) \ = \ (1-x^2) \sum_{n\neq 0} \mu_{n}{\cal O}_{\eta,n}
\frac{e^{iy M(x,\eta) \yb}}{1+2i\eta x^0+\eta^2 x^2} \ ,\label{Phipt}\eea
where 
\bea & \displaystyle M_\a{}^{\bd} \ := \ f_1(x,\eta)x_\a{}^{\bd}-if_2(x,\eta)(\s_0)_\a{}^{\bd} \ ,&\\ 
 & \displaystyle f_1 \ := \ \frac{1-2i\eta x_0+\eta^2}{1-2i\eta x_0 + \eta^2 x^2} \ , \qquad f_2 
 \ := \ \eta \,\frac{1-x^2}{1-2i\eta x_0 + \eta^2 x^2}   \label{A-1B} &\ ,\eea
All the other rotationally-invariant (anti-)particle modes can be obtained from \eq{Phipt} by computing the residues of the integrals in ${\cal O}_{\eta,n}$ at varying $n$. We refer to \cite{Iazeolla:2017vng,DeFilippi:2019jqq} for the detailed steps.

 \paragraph{Bulk-to-boundary propagators.}  Adapting the same steps to  projectors with $K_{(q)}=iP$, massless scalar bulk-to-boundary propagators can be obtained (connected to the particle modes via the non-unitary transformation \eq{NUtrasf}). The general solution, e.g., in Poincar\'e coordinates, takes the form
\be \Phi_{{\rm reg}}^{(1)}(iP^L)\ = \ \sum_n \tn_n {\cal O}_{\eta,n}\frac{4z}{[(1+\eta)+(1-\eta)z]^2+(1-\eta)^2\bfx^m\bfx_m}\exp\left(iyM\yb\right)\ ,\label{PhiP}\ee
where
 \bea M_{a\bd} & = & \frac{1}2\frac{z}{z^2+\eta^2\bfx^m\bfx_m}\left\{ \left[(1+\eta)+(1-\eta)\frac{z^2-\bfx^m\bfx_m}{z}\right](\s_3)_{\a\bd} \right.\nn\\
&& \hspace{4cm}+\left.\left[\frac{-\eta(1+\eta)}{z}-2(1-\eta)\right]\bfx^m(\s_m)_{\a\bd}\right\} \ ,\label{MP}\eea
which generalizes the $n=-1$ case of \cite{Giombi:2009wh,Giombi:2010vg,DidenkoSkvortsov} (that can be obtained from \eq{PhiP}-\eq{MP}, up to an overall factor, by simply setting $\eta=-1$) and incorporates it into the general formalism of Section \ref{fibresol}. Indeed the cases $n=-1,-2$ correspond to the scalar bulk-to-boundary propagators \eq{Btb}, and one can also easily check that the cases $n=1,2$ correspond to their counterparts under inversion \eq{antiBtb}, all appearing here with their $\mso(1,2)$-invariant descendants. %\framebox{say here which differential op annihilates them, corresponding to K and T? give their form in appendix? or mayb ein the section on KVF?}. 
As the $n>0$- and $n<0$-sectors are connected via the $\pi$-map, it becomes clear that, in conformal basis, the latter essentially implements a coordinate inversion --- not surprisingly, since here $n$ is the eigenvalue of $D=P_3$, corresponding to the Killing vector that extracts the scaling dimension in the Poincar\'e patch \eq{KVD}.

\paragraph{Wedge modes.} 

Incidentally, even though in this paper we do not attempt to find the non-linear completion of the wedge modes\footnote{What sets this problem slightly outside the scope of the present work is that, apparently, the non-linear corrections in the internal connection require a different type of regular presentation than the  naive one we use here based on an open-contour integral  (see, e.g., \cite{Iazeolla:2008ix}). We hope to complete this analysis in a future work.}, for completeness we present explicitly the calculation showing that the initial datum $\Phi'=\sinh(4E)/4E$ corresponds to the static, rotationally-invariant wedge mode $\phi_{0;(0)}=\xi/\tan\xi$. Indeed, using
\bea \Phi'_{\rm 0,(0)} \ := \ \frac{\sinh(4E)}{4E} \ = \ \frac12 \int_{-1}^1 d\t\,e^{4\t E} \ , \eea
the calculation of the $x$-dependent Weyl zero-forms proceeds in the same way as for the particle sector\footnote{In fact, one immediately notes that the scalar static wedge mode has a twistor-space representative very similar to that of the scalar particle state $\phi_{1;(0)}$, with the difference that the closed-contour integral \eq{enhanced3} is here substituted by an open-contour integral with a continuous parameter $\t$ varying over an interval on the real line. Indeed, both types of twistor-space elements can be subsumed under a common contour-integral presentation \cite{Iazeolla:2008ix}.}, since 
\bea L^{-1}\star e^{4\t E}\star\pi(L) \ = \ \d^2(y^L+i\t \s_0\yb^L) \star \k_y \ = \ \frac{1-x^2}{1+2i\t x_0+\t^2 x^2}\,e^{iy N(x,\t) \yb}\ ,\eea
and 
\bea \Phi_{0,(0)} \ = \ \frac{(1-x^2)}{2} \int_{-1}^1
\frac{e^{iy^\a N_\a{}^{\bd}(x,\t) \yb_{\bd}}}{1+2i\t x_0+\t^2 x^2} \,d\t\ ,\label{Phiw}\eea
where $N(x,\t)=M(x,-\t)$. This is again a scalar Weyl zero-form (containing a scalar field and all its on-shell nontrivial derivatives), and in particular the scalar field is obtained from $\Phi_{0,(0)}|_{y=0=\yb}$ as
\bea  \phi_{0,(0)} \ = \  \frac{1-x^2}{2\sqrt{x^k x_k}}\,\arctan\frac{2\sqrt{x^k x_k}}{1-x^2}\ = \ \frac{\xi}{\tan\xi} \ = \ \frac{\arctan r}{r} \ ,\eea	
$k=1,2,3$. 

\scsss{Singular solutions}\label{Sec:singsol}

Singular solutions are instead obtained from $\Phi_{\rm sing}^{(1)}$, where now the singularity comes from the zeroes  of the determinant of the Killing two-form matrix $\vark^L_{\a\b}(x)$ (see eq. \eq{Gammakv}). That the latter vanishes at certain spacetime points can be traced back to the fact that the original, unrotated $K_{(q)\underline{\a\b}}$ matrix, for solutions with principal Cartan generators $E$ and $iP$, is purely off-diagonal, hence $\vark^L_{\a\b}$ vanishes at the unfolding point\footnote{This is of course only true for families of solutions with $\pi$-odd principal Cartan generators. For $\pi$-even ones the Killing two-form is non-degenerate at the unfolding point, even though, according to the compact or non-compact nature of the principal Cartan generator, it may still degenerate on some other spacetime surface \cite{us,Aros:2019pgj}.}. The (selfdual part of the) Killing two-form $\vark^L_{\a\b}(x)$ can be written in canonical form on the basis of its eigenspinors $(u^{+}_{(K_{(q)})\a} , u^{-}_{(K_{(q)})\a})$, $u^{+\a}_{(K_{(q)})} u^{-}_{(K_{(q)})\a}=1$ \cite{us}, as
\be \vark^L_{\a\b} \ \propto \ \sqrt{\det \vark^L}{\cal D}^{(K_{(q)})}_{\a\b}(x) \ , \qquad {\cal D}^{(K_{(q)})}_{\a\b} :=2 u^{+}_{(K_{(q)})(\a} u^{-}_{(K_{(q)})\b)} \ ,  \ee
(analogously for the anti-selfdual components). The $Y$-expansion of $\Phi_{\rm sing}^{(1)}$ contains a tower of spin-$s$ Weyl tensors and all their derivatives for all $s=0,1,2,....$.

\paragraph{Black-hole modes.}  For twisted projectors based on $E$, $\det\vark^L=-r^2$, in spherical coordinates. Thus, the Weyl tensor generating function extracted from \eq{phisingL} and \eq{bbphisingL} is 
\be  \left.\Phi^{(1)}_{\rm sing}(E^L) \right|_{\bar y=0} \ = \
\frac{1}{ir}\sum_{n\neq 0}\n_n{\cal O}_{\eta,n} \frac{1}{\eta}\,\exp \left(\ft{1}{2\eta r}\,
y^\a{\cal D}^{(E)}_{\a\b}y^\b\right)\ ,\label{o3Weylnm}\ee
(analogously  for the anti-selfdual components) leading to type-D spherically-symmetric spin-$s$ Weyl tensors of the form
\be C^{(1)}_{{\textrm{bh}},n,\a(2s)}\ \sim \ \frac{i^{n-1}\mu_n}{r^{s+1}}\,(u_{(E)}^+
 u_{(E)}^- )^s_{\a(2s)} \ ,\label{nsWeylspher}  \ee
coinciding with \eq{WeylD}, from each twisted projector $\tcP_n$. Note that twisted projectors of energy level $n$ odd (even) contribute electric (magnetic) Weyl tensors. It will be interesting to elaborate further on the nature of the mass-like parameter that turns on each individual spin-$s$ Weyl tensor, a linear combination of the deformation parameters $\n_n$, coming from the sum over $n$ in \eq{o3Weylnm}, which is related to asymptotic charges \cite{Didenko:2015pjo} --- a task that we postpone to a future publication. At any rate, as expected, $\Phi^{(1)}_{\rm sing}(E^L)$ contains the AdS-Schwarzschild Weyl tensor (for $s=2$)  (see \cite{Didenko:2009td} and references therein) accompanied by its counterparts for all integer spins, and each individual Weyl tensor blows up at $r=0$. See \cite{us,us2,Iazeolla:2017vng,DeFilippi:2019jqq} for extended and more detailed results. 

\paragraph{Singular solutions with vanishing scaling dimension.}  Finally, singular solutions with vanishing scaling dimension like \eq{0confweight} are obtained from \eq{phisingL} and \eq{bbphisingL}, where in this case $\det\vark^L=\frac{\bfx^m\bfx_m}{z^2}$, in Poincar\'e coordinates. In the Poincar\'e patch, denoting $\bfx^2:=\bfx^m\bfx_m$, \eq{phisingL} specialized to $iP$ as principal Cartan generator reads
\bea  \Phi^{(1)}_{\rm sing}(iP^L)  & = & \sum_n\n_n {\cal O}_{\eta,n} \frac{iz}{\eta\sqrt{\bfx^2}} \nn\\ 
 &\times&\exp\left[-\frac{iz}{2\bfx^2}\left(\frac1{\eta}\bfx^m y\sigma_{m3}y+\eta \bfx^m \yb\bar \sigma_{m3}\yb\right)+i\left(y\s_3\yb-\frac{z}{\bfx^2}\,\bfx^m y\s_m\yb\right)\right] \label{PhiLconf}\eea
and, by bringing the Killing two-form on canonical basis \cite{us}, it is possible to extract a tower of type-D Weyl tensor for every $n$ of the form
\be C^{(1)}_{\D=0,n,\a(2s)}\ \sim \ \nu_n\left(\frac{z}{\sqrt{\bfx^2} }\right)^{s+1}\,(u_{(iP)}^+
 u_{(iP)}^- )^s_{\a(2s)} \ ,\label{Weylconf}  \ee
all having vanishing scaling dimension, a ``Wick-rotated'' counterpart of the static solutions \eq{nsWeylspher}. As anticipated, these solutions correspond to factorized combinations $z^{s+1}\phi_{m_1\ldots m_s}(\bfx)$ where $\phi_{m_1\ldots m_s}(\bfx)$ solves the boundary wave equation, being of the form $\frac{1}{(\sqrt{\bfx^2})^{1+2s}}\bfx_{\{m_1}\ldots \bfx_{m_s\}}$, where $\{...\}$ denotes traceless projection\footnote{This index structure is a consequence of the fact that the  spin-$s$ Weyl tensor obtained from \eq{PhiLconf} is of the form $C_{\a(2s)}\propto\frac{1}{(\sqrt{\det\vark^L})^{1+2s}}(\vark^L)^s_{\a(2s)}$, and the indices of the Killing two-form are carried by $\s_{m3}$, thereby reducing the indices of the first row of the Weyl tensors' Young diagram to be purely $\mso(1,2)$ indices.}. This is a direct consequence of the fact that d'Alembertian operator on Weyl tensors $C_{a(s),b(s)}(z,\bfx)$ splits into a piece that purely acts on the $z$-dependence plus another that only acts on the $\bfx$-dependence, $z^2\partial^m\partial_m$ (as in \eq{KGP} for the scalar case), and that the factor $z^{s+1}$ solves the equation on its own. In order to have total scaling dimension $\D=0$, $\varphi(\bfx)$ must therefore solve the boundary wave equation and have scaling (mass) dimension $s+1$, leading to the form above. $\phi_{m_1\ldots m_s}(\bfx)$ with $s\geq 1$ are also descendants, via $\partial_m$, of the $s=0$ solution $\frac{1}{\sqrt{\bfx^2}}$  which essentially gives the massless scalar propagator in the flat $(1+2)$-dimensional boundary. 

%It will be interesting to investigate the most appropriate regular presentation, compatible with all the expected physical features of such field configurations, that enables to elevate these linearized solution spaces to solutions of the full theory. We defer this issue, together with a more detailed and systematic analysis of these representations, to a future paper.

\paragraph{Resolution of curvature singularities.} The construction of solutions via unfolding and their embedding into higher-spin gravity provides some interesting insight on representation-theoretic properties of bulk solution spaces and on their boundary CFT duals.  In fact, as we have seen, reformulating the Klein-Gordon and the Bargmann-Wigner equations as covariant constancy conditions on the twisted-adjoint zero-form, besides packaging all solutions together in a higher-spin covariant fashion, also implies a correspondence between spacetime and the fibre-space behaviour of the solutions. The arranging of solutions into $\mso(2,3)$-modules is reproduced in the $Y$-fibre in terms of the action of the rigid $AdS_4$ isometry generators acting on the fibre-space counterparts $\Phi'(Y)$ of the solutions. This has the advantage of offering a purely algebraic setup for investigating certain properties of the solution space, such as their decomposability: for instance, in such setup it was possible to recognize the indecomposable structure of the regular twisted-adjoint module, revealing that, e.g., particle states can be generated from (in general non-normalizable) wedge states \cite{Iazeolla:2008ix}. Moreover, the properties of the fibre elements under separate left and right action of the algebra store information about the nature of the boundary CFT duals of the corresponding bulk solutions: in the most familiar example, fibre elements encoding massless (anti-)particle modes are operators on the (anti-)singleton Hilbert space. Hence, as shown explicitly in  \cite{Iazeolla:2008ix}, they can be reflected into specific states belonging to the tensor product of two singleton, in a different version of the Flato-Fronsdal theorem. However, as we have reviewed here, one can now apply the same procedure to the fibre elements encoding singular solutions, such as black-hole modes, which suggests that their CFT duals should be obtained from the tensor product of singleton and anti-singleton irreps \cite{Iazeolla:2017vng}. And in principle similar considerations can be repeated for any slicing one chooses of $\mso(2,3)$-modules.

The spacetime-fibre mapping that the unfolded procedure encodes becomes particularly interesting for the singular branch of solutions. While the regular solutions are encoded in regular (yet non-polynomial) elements on ${\cal Y}_4$, the fibre counterparts of singular solutions are distributions, delta functions in $Y$ and derivatives. Indeed, it is easy to see from the form of $\Phi^{(1)}_{\rm sing}(x,Y)$ that, approaching the regions where $\det\vark^L(x)$ vanishes, the twisted adjoint field \eq{bbphisingL} for singular solutions resembles a delta sequence: away from the surface $\det\vark^L(x)=0$, the Weyl zero-form is a smooth Gaussian function of the oscillators, while it approaches a Dirac delta function on ${\cal Y}_4$ in the $\det\vark^L(x)\to 0$ limit. Indeed, rearranging the exponent in \eq{bbphisingL} in terms of modified oscillators $\widetilde{y}_\a:= y_a-i\eta (v^L\yb)_a$ \cite{us,Aros:2019pgj,DeFilippi:2019jqq}, and with $\widetilde y^\pm=u^{+\a}_{(K_{(q)})}\widetilde y_\a $, each term in the sum \eq{phisingL} is proportional to
 \be {\cal O}_{n,\eta} \frac{1}{\sqrt{\det \vark^L}}\,\exp\left(i\frac{\widetilde y^+\widetilde y^-}{\sqrt{\det\vark^L}}\right)  \quad \xrightarrow[\det\vark^L\to 0]{} \quad {\cal O}_{n,\eta} 2\pi\delta^2(\hat y) \ ,  \ee
where $\hat y_\a := \widetilde y_\a|_{\det\vark^L=0}=y_\a-i\eta (v\yb)_\a$. So unfolding trades curvature singularities for a distributional behaviour in $Y$. However, unlike the delta function on a commutative space, a delta function in the non-commutative $Y$-space, thought of as a symbol for an element of a star product algebra, is smooth. In other words, the singularities of a tower of Weyl tensors can be handled better once mapped to the fibre, as the resulting distributions have good star-product properties and can be considered elements of an associative algebra\footnote{In fact, one may further speculate that a delta function of the oscillators  could be considered to some extent equivalent to a bounded function (which would give an even stronger meaning to the notion of resolution of curvature singularities) in the sense that, on a non-commutative space, a change in the ordering prescription can turn a delta function into a smooth symbol (for instance, an exponential in normal ordering on $Y$ \cite{us}). Changes of ordering are formally part of the gauge transformation that leave the classical observables of the Vasiliev system invariant (with important subtleties that are currently being studied \cite{Vasiliev:2015mka,DeFilippi:2019jqq}). So in such view, resolution of curvature singularities in higher-spin gravity would amount to saying that the latter are simply an artifact of the ordering choice for the infinite-dimensional symmetry algebra governing the Vasiliev system.}, see \cite{us,Iazeolla:2017vng,Aros:2017ror,Aros:2019pgj}. In particular this means that, at a curvature singularity, the Weyl 0-form master field and associated higher-spin invariant quantities are still well-defined, even though the coefficients of their expansion in $Y$ are not.

Summarizing, the examples here treated suggest that we can describe Vasiliev's higher-spin gravity as a theory in which higher-spin geometries are properly described only via master fields. At generic points on the base manifold, they are real-analytic in fibre coordinates,  and the coefficients of their power series expansion in $Y$ are bounded component fields which,  in weak-field regions, satisfy Fronsdal's equations. At special surfaces they can approach non-analytic functions (and distributions) in $Y$ but remain well-defined as star product algebra elements. Only their interpretation in terms of component Fronsdal fields breaks down.

%%%%%%%%%%%%%%%%%%%%%%%%%%%%%%%%%%%%%%%%%%%%%%%

\scs{Conclusions and perspectives}

%%%%%%%%%%%%%%%%%%%%%%%%%%%%%%%%%%%%%%%%%%%%%%%

We have reviewed some aspects of Vasiliev's higher-spin gravity, focussing in particular on perturbative schemes and on the encoding of the propagating degrees of freedom into fibre elements, and extended the investigation of this sort of spacetime/fibre duality to AdS irreps in conformal basis. 

As for perturbative schemes, we have reviewed an approach proposed in \cite{DeFilippi:2019jqq} in which solutions are built by means of a convenient choice of homotopy-contraction operator and then subjected to asymptotically anti-de Sitter boundary conditions by adjusting a gauge function and the integration constants of the system order by order. More precisely, the perturbative corrections to the bulk master fields involve star-product interactions that may affect their leading order in the asymptotic expansion, in which case corrections to the local data stored in the integration constants, as well as to the gauge function, will be required. It is important to note that, by affecting the integration constants, such procedure affects the observables of the theory, in particular the on-shell action (i.e., the second Chern class on ${\cal Z}_4$, see \cite{DeFilippi:2019jqq} and references therein for details). Classical observables, which encode truly physical information (such as boundary correlation functions), are indeed the right testing ground for the correctness of this as well as other perturbative approaches. To do so, it will first of all be important to push this boundary condition proposal to second order in the interactions.

Then, we have reviewed and further investigated the construction of solutions to linearized field equations for massless fields in AdS via unfolding, by exhibiting the parallels between the familiar spacetime analysis and the corresponding fibre algebraic construction. We first reviewed the fibre representatives of solutions built by solving the Klein-Gordon equation in AdS in global spherical coordinates, that is, slicing the $\mso(2,3)$ modules in compact basis. Regular and normalizable solutions correspond to scalar particle modes, and we singled out some solutions belonging to the singular branch that, together with their higher-spin counterparts, form spherically-symmetric higher-spin black hole solutions, solving both the linear and fully non-linear Vasiliev equations. Fibre representatives of such singular elements are delta functions in $Y$ oscillators --- in fact, being able to handle singular functions of $Y$ and to extract corresponding gauge field generating functions was one of the reasons why the perturbative scheme here reviewed was elaborated in the first place. The construction of solution spaces in the fibre was here extended to modules in conformal basis, corresponding to the spacetime analysis in Poincar\'e patch coordinates. Fibre representatives of the bulk-to-boundary propagators and their descendants, as well as their counterparts under inversion have been exhibited, as well as the non-unitary transformation that connects them to particle states in compact basis. Finally, we briefly examined certain singular solution with vanishing scaling dimension, related to boundary Green's functions. We hope to return to a more detailed and systematic investigation of this solution space in a future work.

\vspace{1.5cm}

\paragraph{Acknowledgments}

It is a pleasure to thank the Organizers of CORFU2019, in particular Harold Steinacker and George Zoupanos, for arranging a stimulating and very pleasant workshop.  
A large part of the results reported on in this paper have been obtained in collaboration with D. De Filippi and P. Sundell, whom I would like to thank for many valuable and helpful discussions. I am also grateful to M. Bianchi, N. Boulanger,  V.E. Didenko, A.V. Korybut, N.G. Misuna, J. Raeymaekers, E. Sezgin, C. Sleight, E.D. Skvortsov, H.C. Steinacker, M. Taronna, M.A. Vasiliev, Y. Yin for stimulating discussions.

\vspace{1cm}

\begin{appendix}

%%%%%%%%%%%%%%%%%%%%%%%%%%%%%%%%%%%%%%%%%%%%%%%%%%%%%%%%%%%%%%%%%%%%%%%%%%

\scs{Spinor conventions and $AdS_4$ Background}\label{App:conv}

%%%%%%%%%%%%%%%%%%%%%%%%%%%%%%%%%%%%%%%%%%%%%%%%%%%%%%%%%%%%%%%%%%%%%%%%%%

We use the conventions of \cite{Iazeolla:2008ix} in which $SO(2,3)$ generators $M_{AB}$ with $A,B=0,1,2,3,0'$ obey
\be [M_{AB},M_{CD}]\ =\ 4i\y_{[C|[B}M_{A]|D]}\ ,\qquad
(M_{AB})^\dagger\ =\ M_{AB}\ ,\label{sogena}\ee
which can be decomposed using $\eta_{AB}~=~(\eta_{ab};-1)$ with $a,b=0,1,2,3$ as
\be [M_{ab},M_{cd}]_\star\ =\ 4i\y_{[c|[b}M_{a]|d]}\ ,\qquad
[M_{ab},P_c]_\star\ =\ 2i\y_{c[b}P_{a]}\ ,\qquad [P_a,P_b]_\star\ =\
i\lambda^2 M_{ab}\ ,\label{sogenb}\ee
where $M_{ab}$ generate the Lorentz subalgebra $\mso(1,3)$, and $P_a=\l M_{0'a}$ with $\l$ being the inverse $AdS_4$ radius related to the cosmological constant via $\L=-3\l^2$. We set $\l=1$ in the following, as we do in the body of the paper.

It is possible to decompose further under the maximal compact subalgebra $\mso(2)\oplus\mso(3)$, generated by the compact $AdS_4$ energy generator $E=P_0=\l M_{0'0}$ and the spatial rotation generators $M_{rs}$ with $r,s=1,2,3$. The remaining generators then arrange into energy-raising and energy-lowering combinations identified with
\bea L^\pm_r&=& M_{0r}\mp iM_{0'r}\ =\ M_{0r}\mp iP_r\
,\label{Lplusminus}\eea
leading to the following $E$-graded decomposition of the
commutation rules \eq{sogena}:
\bea [L^-_r,L^+_s] & = & 2iM_{rs}+2\d_{rs}E \ , \label{ll}\\[5pt]
[M_{rs},M_{tu}]&=& 4i\d_{[t|[s}M_{r]|u]} \ , \label{mm}\\[5pt]
[E,L^{\pm}_r] & = & \pm L^{\pm}_r\ ,\label{el}\\[5pt] [M_{rs},L^\pm_t]&=&
2i\d_{t[s}L^\pm_{r]}\ .\label{ml}\eea
The generators $(E,M_{rs},L^\pm_r)$ are also referred to as generators of the \emph{compact basis}, or \emph{compact split} of $\mso(2,3)$.

$\mso(2,3)$ also admits a three-grading with respect to the non-compact subalgebra $\mso(1,1)\oplus\mso(1,2)$, which is most often used in the AdS/CFT context, as it highlights the Lorentz symmetry algebra of the three-dimensional boundary. The $\mso(1,1)$ is generated by a non-compact generator that can be identified with any spacelike transvection, e.g. $P_3$. With this convention, we can identify the (boundary) dilatation generator $D=P_3$,  the (boundary) Lorentz generators $M_{mn}$, $m,n=1,2,3$, and $D$-raising and $D$-lowering combinations
\be T_m \ = \ M_{m3}-P_m \ ,\qquad K_m \ = \ M_{m3}+P_m \ ,\ee
where $T_m$ are boundary (commuting) translations, satisfying
\bea & [K_m,T_n] \ = \ 2i(\eta_{mn}D-M_{mn}) \ , &\label{kt}\\[5pt]
&[M_{mn},M_{pq}] \ =  \ 4i\eta_{[p|[n}M_{m]|q]} \ , &\label{mlml}\\[5pt]
& [D,T_m] \ = \ iT_m\ ,\qquad [D,K_m] \ = \ -iK_m \ ,&\label{dt}\\[5pt] 
& [M_{mn},T_p] \ = \ 2i\eta_{p[n}T_{m]}\ , \qquad [M_{mn},K_p] \ = \ 2i\eta_{p[n}K_{m]}\ .&\label{mt}\eea
The generators $(D,M_{mn},T_m,K_m)$ are also referred to as generators of the \emph{conformal basis}, of \emph{conformal split} of $\mso(2,3)$ \cite{Gunaydin:1999jb,Sezgin:2002rt}. 

In terms of the oscillators $Y_{\underline\a}=(y_\a,\yb_{\ad})$, the realization of the generators of $\mso(2,3)$ is taken to be
\be M_{AB}~=~ -\ft18  (\C_{AB})_{\underline{\a\b}}\,Y^{\underline\a}\star Y^{\underline\b}\ ,\label{MAB}\ee
 \be
 M_{ab}\ =\ -\frac18 \left[~ (\s_{ab})^{\a\b}y_\a\star y_\b+
 (\sb_{ab})^{\ad\bd}\bar y_{\ad}\star \yb_{\bd}~\right]\ ,\qquad P_{a}\ =\
 \frac{1}4 (\s_a)^{\a\bd}y_\a \star \yb_{\bd}\ ,\label{mab}
 \ee
using Dirac matrices obeying $(\C_A)_{\underline\a}{}^{\underline\b}(\C_B C)_{\underline{\b\c}}=
\eta_{AB}C_{\underline{\a\c}}+(\C_{AB} C)_{\underline{\a\c}}$, 
\begin{equation}
\left( \Gamma ^{0'a}\right) _{\underline{\alpha }}^{\ \ \underline{\beta }
}\equiv \left( \Gamma ^{a}\right) _{\underline{\alpha }}^{\ \ \underline{\beta }
}=\left( 
\begin{array}{cc}
0 & \left( \sigma ^{a}\right) _{\alpha }^{\ \ \dot{\beta}} \\ 
\left( \bar{\sigma}^{a}\right) _{\dot{\alpha}}^{\ \ \beta } & 0%
\end{array}
\right) \text{ ,}
\end{equation}
and
\begin{equation}
\left( \Gamma _{ab}\right) _{\underline{\alpha \beta }}=\left( 
\begin{array}{cc}
\left( \sigma _{ab}\right) _{\alpha \beta } & 0 \\ 
0 & \left( \bar{\sigma}_{ab}\right) _{\dot{\alpha}\dot{\beta}}%
\end{array}
\right) \text{ .}
\end{equation}
and van der Waerden symbols obeying
 \be
  (\s^{a})_{\a}{}^{\ad}(\sb^{b})_{\ad}{}^{\b}~=~ \y^{ab}\d_{\a}^{\b}\
 +\ (\s^{ab})_{\a}{}^{\b} \ ,\qquad
 (\sb^{a})_{\ad}{}^{\a}(\s^{b})_{\a}{}^{\bd}~=~\y^{ab}\d^{\bd}_{\ad}\
 +\ (\sb^{ab})_{\ad}{}^{\bd} \ ,\label{so4a}\ee\be
 \ft12 \e_{abcd}(\s^{cd})_{\a\b}~=~ i (\s_{ab})_{\a\b}\ ,\qquad \ft12
 \e_{abcd}(\sb^{cd})_{\ad\bd}~=~ -i (\sb_{ab})_{\ad\bd}\ ,\label{so4b}
\ee
\be ((\s^a)_{\a\bd})^\dagger~=~
(\sb^a)_{\ad\b} ~=~ (\s^a)_{\b\ad} \ , \qquad ((\s^{ab})_{\a\b})^\dagger\ =\ (\sb^{ab})_{\ad\bd} \ .\ee
and raising and lowering spinor indices according to the
conventions $A^\a=\epsilon^{\a\b}A_\b$ and $A_\a=A^\b\epsilon_{\b\a}$ where
\be \e^{\a\b}\e_{\c\d} \ = \ 2 \d^{\a\b}_{\c\d} \ , \qquad
\e^{\a\b}\e_{\a\c} \ = \ \d^\b_\c \ ,\qquad (\e_{\a\b})^\dagger \ = \ \e_{\ad\bd} \ .\ee
In order to avoid cluttering the expression with many spinor indices, in the paper we also use the matrix notations 
\bea & A^{\underline{\a}} B_{\underline{\a}} \ = :\  AB \ = \ ab+\bar a \bar b \ := \ a^\a b_\a + \bar a^{\ad}\bar b_{\ad} \ ,&\\
& aMb \ := \ a^\a M_{\a}{}^\b b_\b \ , \qquad aN\bar{b} \ := \ a^\a N_{\a}{}^{\bd} \bar{b}_{\bd} \ . &
\eea
The $\mso(2,3)$-valued connection
 \be
  \O~:=~-i \left(\frac12 \omega^{ab} M_{ab}+e^a P_a\right) ~:=~ \frac1{2i}
 \left(\frac12 \omega^{\a\b}~y_\a \star y_\b
 +  e^{\a\dot\b}~y_\a \star {\bar y}_{\dot\b}+\frac12 \bar{\omega}^{\dot\a\dot\b}~{\bar y}_{\dot\a}\star {\bar y}_{\dot\b}\right)\
 ,\label{Omega}
 \ee
  \be
 \o^{\a\b}~=~ -\ft14(\s_{ab})^{\a\b}~\o^{ab}\ , \qquad \omega_{ab}~=~\ft12\left( (\s_{ab})^{\a\b} \o_{\a\b}+(\bar\s_{ab})^{\ad\bd} \bar\o_{\ad\bd}\right)\ ,\ee
 \be e^{\a\dot\a}~=~ \ft{1}2(\s_{a})^{\a \dot\a}~e^{a}\ , \qquad e_a~=~ - (\s_a)^{\a\ad} e_{\a\ad}\ ,\label{convert}\ee
and field strength
\bea {\cal R} & := & d\O+\O\star \O~:=~-i \left(\frac12 {\cal R}^{ab}M_{ab}+{\cal R}^a P_a\right) \nn\\[5pt]
& :=&  \frac1{2i}
 \left(\frac12 {\cal R}^{\a\b}~y_\a \star y_\b
 +  {\cal R}^{\a\dot\b}~y_\a \star {\bar y}_{\dot\b}+\frac12 \bar{\cal R}^{\dot\a\dot\b}~{\bar y}_{\dot\a}\star {\bar y}_{\dot\b}\right)\
 ,\label{calRdef}\eea
\be
 {\cal R}^{\a\b}\ =\ -\ft14(\s_{ab})^{\a\b}~{\cal R}^{ab}\ ,
 \qquad {\cal R}_{ab}~=~\ft12\left( (\s_{ab})^{\a\b} {\cal R}_{\a\b}+(\bar\s_{ab})^{\ad\bd} \bar{\cal R}_{\ad\bd}\right)\ ,\ee
 \be
 {\cal R}^{\a\dot\a}\ =\ \ft{1}2(\s_{a})^{\a \dot\a}~{\cal R}^{a}\ ,
 \qquad {\cal R}_a~=~ - (\s_a)^{\a\ad} {\cal R}_{\a\ad}\ .\ee
In these conventions, it follows that
 \be
 {\cal R}_{\a\b}~=~ d\o_{\a\b} -\o_{\a}^{\c}\o_{\c\b}-
 e_{\a}^{\cd}\bar e_{\cd\b}\ ,\qquad
 {\cal R}_{\a\dot\b}~=~  de_{\a\bd}+ \o_{\a\c}\wedge
 e^{\c}{}_{\bd}+\bar{\o}_{\bd\dd}\wedge e_{\a}{}^{\dd}\
 ,\ee\be
 {\cal R}^{ab}~=~ R_{ab}+
 e^a\wedge e^b\ ,\qquad R_{ab}~:=~d\o^{ab}+\o^a{}_c\wedge\o^{cb}\ ,\ee\be
 {\cal R}^a~=~ T^a ~:=~d e^a+\o^a{}_b\wedge e^b\ ,
 \label{curvcomp} \ee
where $R_{ab}:=\frac12 e^c e^d R_{cd,ab}$ and $T_a:=e^b e^c T^a_{bc}$ are the Riemann and torsion two-forms.
The metric $g_{\mu\nu}:=e^a_\mu e^b_{\nu}\eta_{ab}$. The $AdS_4$ vacuum solution $\O_{(0)}=e_{(0)}+\o_{(0)}$ obeying $d\O_{(0)}+\O_{(0)}\star\O_{(0)}=0$, with Riemann tensor $ R_{(0)\m\n,\r\s}=
 - \left( g_{(0)\mu\rho} g_{(0)\nu\sigma}-
  g_{(0)\nu\rho} g_{(0)\mu\sigma} \right)$ and vanishing torsion, can be expressed as $\O_{(0)}=L^{-1}\star dL$ where the gauge function $L\in SO(2,3)/SO(1,3)$. 
  
 The stereographic coordinates $x^\mu$ of Eq. \eq{stereo}, are related to the coordinates $X^A$ of the five-dimensional embedding space with metric
$ds^2  =dX^A dX^B\eta_{AB}$,
in which $AdS_4$ is embedded as the hyperboloid
$X^A X^B \eta_{AB}=  -1 $,
as
\bea x^a \ = \ \frac{X^a}{1+X^{0'}} \ , \ \ \qquad X^a \ = \ \frac{2x^a}{1- x^2}\ , \quad X^{0'} \ = \ \frac{1+x^2}{1-x^2}\ .\label{A.15}\eea
%
%while the extra time coordinate $X_{0'}$ can be solved for from \eq{emb}.
$AdS_4$ can be covered by two sets of stereographic coordinates, $x^a_{(i)}$, $i=N,S$, related by the inversion $x^a_N = -x^a_S/( x_S)^2$ in the overlap region $ (x_N)^2,  (x_S)^2  <  0$, and the transition function $T_N^S=(L_N)^{-1}\star L_S\in SO(1,3)$. The inversion $x^a \rightarrow -x^a/( x)^2$ leaves the metric invariant, maps the future and past time-like cones into themselves and exchanges the two space-like regions $0< x^2< 1$ and $ x^2 > 1$ while leaving the boundary $ x^2 =1$ fixed. It follows that the single cover of $AdS_4$ is formally covered by taking $x^a\in \Real^{1,3}$. 

The global spherical coordinates $(t,r,\theta,\varphi)$, in which the metric reads
\bea  ds^2 \ = \ -(1+r^2)dt^2+\frac{dr^2}{1+ r^2}+r^2(d\theta^2+\sin^2\theta d\varphi^2) \ ,\label{metricglob}\eea
are especially adapted to the compact split, and are related to the embedding coordinates by
\bea & X^0 \ = \ \sqrt{1+r^2}\sin t \ , \qquad X^{0'} \ = \ \sqrt{1+r^2}\cos t \ , & \nn\\[5pt]
& X^1 \ = \ r\sin\theta\cos\varphi \ , \quad  X^2 \ = \ r\sin\theta\sin\varphi \ , \quad X^3 \ = \ r\cos\theta \ ,& \label{AdSspherical}\eea
providing a one-to-one map if $t\in [0,2\pi)$, $r\in[0,\infty)$, $\theta\in[0,\pi]$ and $\varphi\in[0,2\pi)$, defining the single cover of $AdS_4$. 

We also use the global coordinates $(t,\xi,\theta,\varphi)$, where $\xi\in[0,\pi/2]$ is related to the radial coordinate $r$ of the spherical coordinates as $\xi=\arctan r$, %the embedding coordinates by
%
%\bea &\displaystyle X_0 \ = \ -\frac{\sin t}{\cos \xi} \ , \qquad X_{0'} \ = \ -\frac{\cos t}{\cos \xi} \ , & \nn\\[5pt]
%& \displaystyle X_1 \ = \ \tan\xi \sin\theta\cos\phi \ , \quad  X_2 \ = \ \tan\xi\sin\theta\sin\phi \ , \quad X_3 \ = \ \tan\xi\cos\theta \ ,& \label{globxi}\eea
%
in which the metric reads 
\be  ds^2 \ = \ \frac{1}{\cos^2\xi}\left[-dt^2+d\xi^2+\sin^2\xi(d\theta^2+\sin^2\theta d\varphi^2)\right]  \ .\label{globxi}\ee

The Poincar\'e patch coordinates $(z,\bfx^m)$, $z\in(0,+\infty)$, $\bfx^m\in \Real$, $m=0,1,2$ adapted to the conformal split, cover a region of $AdS_4$ in which the metric is conformal to half of a flat Minkowski spacetime,
\be ds^2 \ = \ \frac{dz^2 +d\bfx^m d\bfx_m}{z^2} \ . \label{Poincx}\ee
Their relation to the embedding coordinates is given by
\be  X^m \ = \ \frac{\bfx^m}{z} \ , \qquad X^3 \ = \ \frac{1-\bfx^m\bfx_m-z^2}{2z} \ , \qquad \displaystyle X^{0'} \ = \ \frac{1+\bfx^m\bfx_m+z^2}{2z}  \ .\label{Pemb} \ee

In the computation of the $x$-dependent master fields we made use of \eq{YLrotn}. The matrix representative of $L_{\rm stereo}$ \eq{Lgf} is
\be
(L_{\rm stereo})_{\underline{\a}}{}^{\underline{\b}} ~=~ \frac{1}{h}\left(\ba{cc} \d_\a{}^\b & x_\a{}^{\bd} \\[5pt] \bar x_{\ad}{}^{\b} & \d_{\ad}{}^{\bd}\ea\right)\label{3.20}\ ,\ee
giving the rotated oscillators 
\be y^L_\a \ = \ \frac{1}{h} (y_\a+ x_\a{}^{\bd}\yb_{\bd}) \ , \qquad \yb^L_{\ad} \ = \  \frac{1}{h} (\yb_{\ad}+ \bar x_{\ad}{}^{\b}y_{\b}) \ .\label{YLst}\ee
For Poincar\'e coordinates, that is, $L$-rotating by means of \eq{LP}, one obtains
\bea & \displaystyle y^L_\a \ = \ \frac{1}{2\sqrt{z}}\left[(1+z)y_\a-\bfx^m(\s_{m3}y)_\a+(1-z)(\s_3\yb)_\a-\bfx^m(\s_{m}\yb)_\a
)\right] \ ,&\label{yLP}\\ 
&  \displaystyle \yb^L_{\ad} \ = \ \frac{1}{2\sqrt{z}}\left[(1+z)\yb_{\ad}-\bfx^m(\bar \s_{m3}\yb)_{\ad}+(1-z)(\bar \s_3y)_{\ad}-\bfx^m(\bar \s_{m}y)_{\ad}
)\right] \ .& \label{ybLP}\eea

\end{appendix}


\begin{thebibliography}{99}


%\cite{Gross:1988ue}
\bibitem{Gross}
D.~J.~Gross,
``High-Energy Symmetries of String Theory,''
Phys. Rev. Lett. \textbf{60} (1988), 1229
doi:10.1103/PhysRevLett.60.1229
%297 citations counted in INSPIRE as of 29 Apr 2020

%\cite{Sundborg:2000wp}
\bibitem{sundborg}
  B.~Sundborg,
  ``Stringy gravity, interacting tensionless strings and massless higher spins,''
  Nucl.\ Phys.\ Proc.\ Suppl.\  {\bf 102} (2001) 113
  doi:10.1016/S0920-5632(01)01545-6
  [hep-th/0103247].
  %%CITATION = doi:10.1016/S0920-5632(01)01545-6;%%
  %242 citations counted in INSPIRE as of 25 Mar 2017
  
  %\cite{Engquist:2005yt}
\bibitem{perjohan}
  J.~Engquist and P.~Sundell,
  ``Brane partons and singleton strings,''
  Nucl.\ Phys.\ B {\bf 752} (2006) 206
  [hep-th/0508124].
  %%CITATION = HEP-TH/0508124;%%
  
  %\cite{Chang:2012kt} 
  \bibitem{minwalla} 
  C.~M.~Chang, S.~Minwalla, T.~Sharma and X.~Yin, 
  ``ABJ Triality: from Higher Spin Fields to Strings,'' J.\ Phys.\ A {\bf 46} (2013) 214009 
  doi:10.1088/1751-8113/46/21/214009 [arXiv:1207.4485 [hep-th]]. 
  
   %\cite{Gaberdiel:2014cha}
\bibitem{Gaberdiel:2014cha}
  M.~R.~Gaberdiel and R.~Gopakumar,
  ``Higher Spins \& Strings,''
  JHEP {\bf 1411} (2014) 044
  doi:10.1007/JHEP11(2014)044
  [arXiv:1406.6103 [hep-th]].
  %%CITATION = doi:10.1007/JHEP11(2014)044;%%
  %69 citations counted in INSPIRE as of 25 Mar 2017
  
    
  %\cite{Gaberdiel:2015wpo}
\bibitem{Gaberdiel:2015wpo}
  M.~R.~Gaberdiel and R.~Gopakumar,
  ``String Theory as a Higher Spin Theory,''
  JHEP {\bf 1609} (2016) 085
  doi:10.1007/JHEP09(2016)085
  [arXiv:1512.07237 [hep-th]].
  %%CITATION = doi:10.1007/JHEP09(2016)085;%%
  %21 citations counted in INSPIRE as of 25 Mar 2017
  

%\cite{Sezgin:2002rt}
\bibitem{Sezgin:2002rt}
  E.~Sezgin and P.~Sundell,
  ``Massless higher spins and holography,''
  Nucl.\ Phys.\  B {\bf 644} (2002) 303
  [Erratum-ibid.\  B {\bf 660} (2003) 403]
  [arXiv:hep-th/0205131].
  %%CITATION = NUPHA,B644,303;%%
 
  %\cite{Klebanov:2002ja}
\bibitem{Klebanov:2002ja}
  I.~R.~Klebanov and A.~M.~Polyakov,
  ``AdS dual of the critical O(N) vector model,''
  Phys.\ Lett.\  B {\bf 550} (2002) 213
  [arXiv:hep-th/0210114].
  %%CITATION = PHLTA,B550,213;%%
   

%\cite{Douglas:2010rc}
\bibitem{Douglas:2010rc}
  M.~R.~Douglas, L.~Mazzucato and S.~S.~Razamat,
  ``Holographic dual of free field theory,''
  Phys.\ Rev.\  D {\bf 83} (2011) 071701
  [arXiv:1011.4926 [hep-th]].
  %%CITATION = PHRVA,D83,071701;%%

\bibitem{Gaberdiel:2010pz}
  M.~R.~Gaberdiel and R.~Gopakumar,
  ``An $AdS_3$ Dual for Minimal Model CFTs,''
  Phys.\ Rev.\ D {\bf 83} (2011) 066007
  doi:10.1103/PhysRevD.83.066007
  [arXiv:1011.2986 [hep-th]].
  %CITATION = doi:10.1103/PhysRevD.83.066007;%%
  %276 citations counted in INSPIRE as of 21 Jan 2017
  %\cite{Gaberdiel:2011wb}

%\cite{Eberhardt:2019ywk}
\bibitem{Eberhardt}
L.~Eberhardt, M.~R.~Gaberdiel and R.~Gopakumar,
%``Deriving the AdS$_{3}$/CFT$_{2}$ correspondence,''
JHEP \textbf{02} (2020), 136
doi:10.1007/JHEP02(2020)136
[arXiv:1911.00378 [hep-th]].
%9 citations counted in INSPIRE as of 29 Apr 2020

%\cite{vasiliev}
\bibitem{vasiliev}
M.~A.~Vasiliev,
``Consistent equations for
interacting gauge fields of all spins in $3+1$ dimensions,'' Phys.
Lett. {\bf B243} (1990) 378.
%\cite{Vasiliev:1988xc}

\bibitem{Vasiliev:1990vu}
  M.~A.~Vasiliev,
  ``Properties of equations of motion of interacting gauge fields of all spins
  in (3+1)-dimensions,''
  Class.\ Quant.\ Grav.\  {\bf 8} (1991) 1387.
  %%CITATION = CQGRD,8,1387;%%
  %\cite{more}

\bibitem{more}
  M.~A.~Vasiliev,
  ``More on equations of motion for interacting massless fields of all spins in
  (3+1)-dimensions,''
  Phys.\ Lett.\  B {\bf 285} (1992) 225.
  %%CITATION = PHLTA,B285,225;%%
  
  %\cite{Vasiliev:2003ev}
\bibitem{Vasiliev:2003ev}
M.~Vasiliev,
``Nonlinear equations for symmetric massless higher spin fields in (A)dS(d),''
Phys. Lett. B \textbf{567} (2003), 139-151
doi:10.1016/S0370-2693(03)00872-4
[arXiv:hep-th/0304049 [hep-th]].
%494 citations counted in INSPIRE as of 27 Apr 2020

%\cite{Vasiliev:1999ba}
\bibitem{Vasiliev:1999ba}
M.~A.~Vasiliev,
``Higher spin gauge theories: Star product and AdS space,''
%doi:10.1142/9789812793850_0030
[arXiv:hep-th/9910096 [hep-th]].
%450 citations counted in INSPIRE as of 29 Apr 2020


%\cite{Bekaert:2005vh}
\bibitem{Bekaert:2005vh}
X.~Bekaert, S.~Cnockaert, C.~Iazeolla and M.~Vasiliev,
``Nonlinear higher spin theories in various dimensions,''
[arXiv:hep-th/0503128 [hep-th]].
%442 citations counted in INSPIRE as of 29 Apr 2020

  %\cite{Iazeolla:2008bp}
\bibitem{Iazeolla:2008bp}
C.~Iazeolla,
``On the Algebraic Structure of Higher-Spin Field Equations and New Exact Solutions,''
[arXiv:0807.0406 [hep-th]].
%37 citations counted in INSPIRE as of 29 Apr 2020

%\cite{Didenko:2014dwa}
\bibitem{Didenko:2014dwa}
  V.~E.~Didenko and E.~D.~Skvortsov,
  ``Elements of Vasiliev theory,''
  arXiv:1401.2975 [hep-th].
  %%CITATION = ARXIV:1401.2975;%%
  %65 citations counted in INSPIRE as of 21 Jan 2017
  
  %\cite{Vasiliev:2017cae}
\bibitem{Vasiliev:2017cae}
M.~Vasiliev,
``On the Local Frame in Nonlinear Higher-Spin Equations,''
JHEP \textbf{01} (2018), 062
doi:10.1007/JHEP01(2018)062
[arXiv:1707.03735 [hep-th]].
  
  %\cite{Gelfond:2018vmi}
\bibitem{Gelfond:2018vmi}
O.~Gelfond and M.~Vasiliev,
``Homotopy Operators and Locality Theorems in Higher-Spin Equations,''
Phys.\ Lett.\ B \textbf{786} (2018), 180-188
doi:10.1016/j.physletb.2018.09.038
[arXiv:1805.11941 [hep-th]].
%7 citations counted in INSPIRE as of 13 Apr 2020
  
  %\cite{Didenko:2018fgx}
\bibitem{Didenko:2018fgx}
V.~Didenko, O.~Gelfond, A.~Korybut and M.~Vasiliev,
``Homotopy Properties and Lower-Order Vertices in Higher-Spin Equations,''
J.\ Phys.\ A \textbf{51} (2018) no.46, 465202
doi:10.1088/1751-8121/aae5e1
[arXiv:1807.00001 [hep-th]].
%14 citations counted in INSPIRE as of 12 Apr 2020
  
  %\cite{Didenko:2019xzz}
\bibitem{Didenko:2019xzz}
V.~Didenko, O.~Gelfond, A.~Korybut and M.~Vasiliev,
``Limiting Shifted Homotopy in Higher-Spin Theory and Spin-Locality,''
JHEP \textbf{12} (2019), 086
doi:10.1007/JHEP12(2019)086
[arXiv:1909.04876 [hep-th]].
%4 citations counted in INSPIRE as of 12 Apr 2020

%\cite{Gelfond:2019tac}
\bibitem{Gelfond:2019tac}
O.~Gelfond and M.~Vasiliev,
``Spin-Locality of Higher-Spin Theories and Star-Product Functional Classes,''
JHEP \textbf{03} (2020), 002
doi:10.1007/JHEP03(2020)002
[arXiv:1910.00487 [hep-th]].

  %\cite{Boulanger:2015ova}
\bibitem{Boulanger:2015ova}
  N.~Boulanger, P.~Kessel, E.~D.~Skvortsov and M.~Taronna,
  ``Higher spin interactions in four-dimensions: Vasiliev versus Fronsdal,''
  J.\ Phys.\ A {\bf 49} (2016) no.9,  095402
  doi:10.1088/1751-8113/49/9/095402
  [arXiv:1508.04139 [hep-th]].
  %%CITATION = doi:10.1088/1751-8113/49/9/095402;%%
  %20 citations counted in INSPIRE as of 20 Jan 2017
  
  
\bibitem{Bekaert:2015tva}
  X.~Bekaert, J.~Erdmenger, D.~Ponomarev and C.~Sleight,
  ``Quartic AdS Interactions in Higher-Spin Gravity from Conformal Field Theory,''
  JHEP {\bf 1511} (2015) 149
  doi:10.1007/JHEP11(2015)149
  [arXiv:1508.04292 [hep-th]].
  %%CITATION = doi:10.1007/JHEP11(2015)149;%%
  %36 citations counted in INSPIRE as of 09 Feb 2017
  %\cite{Sleight:2016dba}
  
  %\cite{Skvortsov:2015lja}
\bibitem{Skvortsov:2015lja}
  E.~D.~Skvortsov and M.~Taronna,
  ``On Locality, Holography and Unfolding,''
  JHEP {\bf 1511} (2015) 044
  doi:10.1007/JHEP11(2015)044
  [arXiv:1508.04764 [hep-th]].
  %%CITATION = doi:10.1007/JHEP11(2015)044;%%
  %16 citations counted in INSPIRE as of 20 Jan 2017


%\cite{Sleight:2017pcz}
\bibitem{Sleight:2017pcz}
C.~Sleight and M.~Taronna,
``Higher-Spin Gauge Theories and Bulk Locality,''
Phys.\ Rev.\ Lett.\  \textbf{121} (2018) no.17, 171604
doi:10.1103/PhysRevLett.121.171604
[arXiv:1704.07859 [hep-th]].
%44 citations counted in INSPIRE as of 13 Apr 2020

 %\cite{Vasiliev:2012vf}
\bibitem{Vasiliev:2012vf}
  M.~A.~Vasiliev,
  ``Holography, Unfolding and Higher-Spin Theory,''
  J.\ Phys.\ A {\bf 46} (2013) 214013
  doi:10.1088/1751-8113/46/21/214013
  [arXiv:1203.5554 [hep-th]].
  %%CITATION = doi:10.1088/1751-8113/46/21/214013;%%
  %82 citations counted in INSPIRE as of 21 Jan 2017 
  
  %\cite{Iazeolla:2008ix}
\bibitem{Iazeolla:2008ix}
  C.~Iazeolla and P.~Sundell,
  ``A Fiber Approach to Harmonic Analysis of Unfolded Higher-Spin Field
  Equations,''
  JHEP {\bf 0810} (2008) 022
  [arXiv:0806.1942 [hep-th]].
  %%CITATION = JHEPA,0810,022;%%
  
  %\cite{Iazeolla:2011cb}
\bibitem{us}
  C.~Iazeolla and P.~Sundell,
  ``Families of exact solutions to Vasiliev's 4D equations with spherical, cylindrical and biaxial symmetry,''
  JHEP {\bf 1112} (2011) 084
  [arXiv:1107.1217 [hep-th]].
  %%CITATION = ARXIV:1107.1217;%%

  %\cite{Iazeolla:2012nf}
\bibitem{us2}
  C.~Iazeolla and P.~Sundell,
  ``Biaxially symmetric solutions to 4D higher spin gravity,''
  J.\ Phys.\ A {\bf 46} (2013) 214004
  [arXiv:1208.4077 [hep-th]].
  
  %\cite{Iazeolla:2015tca}
\bibitem{Iazeolla:2015tca}
  C.~Iazeolla and J.~Raeymaekers,
  ``On big crunch solutions in Prokushkin-Vasiliev theory,''
  JHEP {\bf 1601} (2016) 177
  doi:10.1007/JHEP01(2016)177
  [arXiv:1510.08835 [hep-th]].
  %%CITATION = doi:10.1007/JHEP01(2016)177;%%
  %2 citations counted in INSPIRE as of 18 Jan 2017

%\cite{Iazeolla:2017vng}
\bibitem{Iazeolla:2017vng}
  C.~Iazeolla and P.~Sundell,
  ``4D Higher Spin Black Holes with Nonlinear Scalar Fluctuations,''
  JHEP {\bf 1710} (2017) 130
  doi:10.1007/JHEP10(2017)130
  [arXiv:1705.06713 [hep-th]].
  %%CITATION = doi:10.1007/JHEP10(2017)130;%%
  
%\cite{Iazeolla:2017dxc}
\bibitem{review}
  C.~Iazeolla, E.~Sezgin and P.~Sundell,
  ``On Exact Solutions and Perturbative Schemes in Higher Spin Theory,''
  Universe {\bf 4} (2018) no.1,  5
  doi:10.3390/universe4010005
  [arXiv:1711.03550 [hep-th]].
  %%CITATION = doi:10.3390/universe4010005;%%
  %8 citations counted in INSPIRE as of 22 Mar 2020
  
   \bibitem{Aros:2017ror}
R.~Aros, C.~Iazeolla, J.~Nore\~{n}a, E.~Sezgin, P.~Sundell and Y.~Yin,
``FRW and domain walls in higher spin gravity,''
JHEP \textbf{03} (2018), 153
doi:10.1007/JHEP03(2018)153
[arXiv:1712.02401 [hep-th]].
%11 citations counted in INSPIRE as of 13 Apr 2020
  
  
  %\cite{Aros:2019pgj}
\bibitem{Aros:2019pgj}
  R.~Aros, C.~Iazeolla, P.~Sundell and Y.~Yin,
  ``Higher spin fluctuations on spinless 4D BTZ black hole,''
  JHEP {\bf 1908} (2019) 171
  doi:10.1007/JHEP08(2019)171
  [arXiv:1903.01399 [hep-th]].
  %%CITATION = doi:10.1007/JHEP08(2019)171;%%
  %2 citations counted in INSPIRE as of 22 Mar 2020
  
 
  
  %\cite{DeFilippi:2019jqq}
\bibitem{DeFilippi:2019jqq}
  D.~De Filippi, C.~Iazeolla and P.~Sundell,
  ``Fronsdal fields from gauge functions in Vasiliev's higher spin gravity,''
  JHEP {\bf 1910} (2019) 215
  doi:10.1007/JHEP10(2019)215
  [arXiv:1905.06325 [hep-th]].
  %%CITATION = doi:10.1007/JHEP10(2019)215;%%
  %4 citations counted in INSPIRE as of 22 Mar 2020



%\cite{Didenko:2009td}
\bibitem{Didenko:2009td}
  V.~E.~Didenko and M.~A.~Vasiliev,
  ``Static BPS black hole in 4d higher spin gauge theory,''
  Phys.\ Lett.\  B {\bf 682} (2009) 305
  [arXiv:0906.3898 [hep-th]].
  %%CITATION = PHLTA,B682,305;%%
  
  %\cite{Giombi:2009wh}
\bibitem{Giombi:2009wh}
  S.~Giombi and X.~Yin,
  ``Higher Spin Gauge Theory and Holography: The Three-Point Functions,''
  JHEP {\bf 1009} (2010) 115
  [arXiv:0912.3462 [hep-th]].
  %%CITATION = JHEPA,1009,115;%%


  
  %\cite{Giombi:2010vg}
\bibitem{Giombi:2010vg}
  S.~Giombi and X.~Yin,
  ``Higher Spins in AdS and Twistorial Holography,''
  JHEP {\bf 1104} (2011) 086
  [arXiv:1004.3736 [hep-th]].
  %%CITATION = JHEPA,1104,086;%%
  %\cite{Giombi:2011ya}
  

%\cite{Didenko:2012tv}
\bibitem{DidenkoSkvortsov}
  V.~E.~Didenko and E.~D.~Skvortsov,
  ``Exact higher-spin symmetry in CFT: all correlators in unbroken Vasiliev theory,''
  JHEP {\bf 1304} (2013) 158
  doi:10.1007/JHEP04(2013)158
  [arXiv:1210.7963 [hep-th]].
  %%CITATION = doi:10.1007/JHEP04(2013)158;%%
  %59 citations counted in INSPIRE as of 18 Apr 2017
  
  %\cite{Sezgin:2003pt}
\bibitem{Sezgin:2003pt}
  E.~Sezgin and P.~Sundell,
  ``Holography in 4D (super) higher spin theories and a test via cubic scalar
  couplings,''
 JHEP {\bf 0507} (2005) 044
  [arXiv:hep-th/0305040].
  %%CITATION = JHEPA,0507,044;%%
  
  %\cite{berezin}
\bibitem{berezin}
  F.~A.~Berezin and M.~A.~Shubin,
  ``The Schr\"{o}dinger Equation,''
  Moscow University Press, (Moscow, 1983)
  
  %\cite{Bolotin:1999fa}
\bibitem{Bolotin:1999fa}
K.~I.~Bolotin and M.~A.~Vasiliev,
``Star-product and massless free
field dynamics in AdS(4)'', Phys.\ Lett.\ B {\bf 479} (2000) 421
[arXiv:hep-th/0001031].

%\cite{Sezgin:2005pv}
\bibitem{Sezgin:2005pv}
  E.~Sezgin and P.~Sundell,
  ``An exact solution of 4D higher spin gauge theory,''
  Nucl.\ Phys.\  B {\bf 762}, 1 (2007)
  [arXiv:hep-th/0508158].
  

  %\cite{Rahman:2015pzl}
\bibitem{Rahman:2015pzl}
  R.~Rahman and M.~Taronna,
  ``From Higher Spins to Strings: A Primer,''
  arXiv:1512.07932 [hep-th].
  %%CITATION = ARXIV:1512.07932;%%
  %13 citations counted in INSPIRE as of 02 Jun 2017

%\cite{Didenko:2008va}
\bibitem{Didenko:2008va}
  V.~E.~Didenko, A.~S.~Matveev and M.~A.~Vasiliev,
  ``Unfolded Description of AdS(4) Kerr Black Hole,''
  Phys.\ Lett.\ B {\bf 665} (2008) 284
  doi:10.1016/j.physletb.2008.05.067
  [arXiv:0801.2213 [gr-qc]].
  %%CITATION = doi:10.1016/j.physletb.2008.05.067;%%
  %18 citations counted in INSPIRE as of 18 Apr 2017
  
  %\cite{Didenko:2009tc}
\bibitem{Didenko:2009tc}
  V.~E.~Didenko, A.~S.~Matveev and M.~A.~Vasiliev,
  ``Unfolded Dynamics and Parameter Flow of Generic AdS(4) Black Hole,''
  arXiv:0901.2172 [hep-th].
  %%CITATION = ARXIV:0901.2172;%%
  %11 citations counted in INSPIRE as of 20 Aug 2017
  
%\cite{Fradkin:1986ka}
\bibitem{Fradkin:1986ka}
E.~Fradkin and M.~A.~Vasiliev,
``Candidate to the Role of Higher Spin Symmetry,''
Annals Phys. \textbf{177} (1987), 63
doi:10.1016/S0003-4916(87)80025-8
%223 citations counted in INSPIRE as of 29 Apr 2020

%\cite{Vasiliev:1986qx}
\bibitem{Vasiliev:1986qx}
  M.~A.~Vasiliev,
  ``Extended Higher Spin Superalgebras and Their Realizations in Terms of Quantum Operators,''
  Fortsch.\ Phys.\  {\bf 36} (1988) 33.
  %%CITATION = FPYKA,36,33;%%
  %126 citations counted in INSPIRE as of 09 Sep 2017
  
  %\cite{Konstein:1989ij}
\bibitem{Konstein:1989ij}
S.~Konstein and M.~A.~Vasiliev,
``Extended Higher Spin Superalgebras and Their Massless Representations,''
Nucl. Phys. B \textbf{331} (1990), 475-499
doi:10.1016/0550-3213(90)90216-Z
%108 citations counted in INSPIRE as of 29 Apr 2020
 
  
%\cite{Sezgin:2002ru}
\bibitem{Sezgin:2002ru}
  E.~Sezgin and P.~Sundell,
 ``Analysis of higher spin field equations in four dimensions,''
  JHEP {\bf 0207} (2002) 055 [arXiv:hep-th/0205132].
  

%\cite{Didenko:2015cwv}
\bibitem{Didenko:2015cwv}
  V.~E.~Didenko, N.~G.~Misuna and M.~A.~Vasiliev,
  ``Perturbative analysis in higher-spin theories,''
  JHEP {\bf 1607} (2016) 146
  doi:10.1007/JHEP07(2016)146
  [arXiv:1512.04405 [hep-th]].
  %%CITATION = doi:10.1007/JHEP07(2016)146;%%
  %7 citations counted in INSPIRE as of 02 Sep 2017
  
  \bibitem{Vasiliev:1988sa}
  M.~A.~Vasiliev,
  ``Consistent equations for interacting massless fields of all spins in the first order in curvatures,''
  Annals Phys.\  {\bf 190}, 59 (1989).
  %%CITATION = APNYA,190,59;%%
   %\cite{Vasiliev:1990vu}


%\cite{Vasiliev:1990bu}
\bibitem{Vasiliev:1990bu}
  M.~A.~Vasiliev,
  ``Algebraic aspects of the higher spin problem,''
  Phys.\ Lett.\  B {\bf 257} (1991) 111.
  %%CITATION = PHLTA,B257,111;%%
  
  %\cite{Iazeolla:2007wt}
\bibitem{Iazeolla:2007wt}
  C.~Iazeolla, E.~Sezgin and P.~Sundell,
  ``Real Forms of Complex Higher Spin Field Equations and New Exact
  Solutions,''
  Nucl.\ Phys.\  B {\bf 791} (2008) 231
  [arXiv:0706.2983 [hep-th]].
  %%CITATION = NUPHA,B791,231;%%

%\cite{Sundell:2016mxc}
\bibitem{Sundell:2016mxc}
  P.~Sundell and Y.~Yin,
  ``New classes of bi-axially symmetric solutions to four-dimensional Vasiliev higher spin gravity,''
  JHEP {\bf 1701} (2017) 043
  doi:10.1007/JHEP01(2017)043
  [arXiv:1610.03449 [hep-th]].
  %%CITATION = doi:10.1007/JHEP01(2017)043;%%
  %1 citations counted in INSPIRE as of 16 Feb 2017  
  
  %\cite{Prokushkin:1998bq}
\bibitem{Prokushkin:1998bq}
  S.~F.~Prokushkin and M.~A.~Vasiliev,
  ``Higher-spin gauge interactions for massive matter fields in 3D AdS
  spacetime,''
  Nucl.\ Phys.\  B {\bf 545} (1999) 385 [arXiv:hep-th/9806236].
  
  %\cite{Kraus:2012uf}
\bibitem{Kraus:2012uf}
  P.~Kraus and E.~Perlmutter,
  ``Probing higher spin black holes,''
  JHEP {\bf 1302} (2013) 096
  doi:10.1007/JHEP02(2013)096
  [arXiv:1209.4937 [hep-th]].
  %%CITATION = doi:10.1007/JHEP02(2013)096;%%
  %38 citations counted in INSPIRE as of 28 Jan 2017

 %\cite{Sezgin:2011hq}
\bibitem{Sezgin:2011hq}
E.~Sezgin and P.~Sundell,
``Geometry and Observables in Vasiliev's Higher Spin Gravity,''
JHEP \textbf{07} (2012), 121
doi:10.1007/JHEP07(2012)121
[arXiv:1103.2360 [hep-th]].
%43 citations counted in INSPIRE as of 27 Apr 2020

\bibitem{FCS}
  N.~Boulanger, E.~Sezgin and P.~Sundell,
  ``4D Higher Spin Gravity with Dynamical Two-Form as a Frobenius-Chern-Simons Gauge Theory,''
  arXiv:1505.04957 [hep-th].
  %%CITATION = ARXIV:1505.04957;%%
  %14 citations counted in INSPIRE as of 16 May 2017
  
  \bibitem{Bonezzi:2016ttk}
  R.~Bonezzi, N.~Boulanger, E.~Sezgin and P.~Sundell,
  ``Frobenius–Chern–Simons gauge theory,''
  J.\ Phys.\ A {\bf 50} (2017) no.5,  055401
  doi:10.1088/1751-8121/50/5/055401
  [arXiv:1607.00726 [hep-th]].
  %%CITATION = doi:10.1088/1751-8121/50/5/055401;%%
  %2 citations counted in INSPIRE as of 16 May 2017

  
  %\cite{Boulanger:2008up}
\bibitem{Boulanger:2008up}
  N.~Boulanger, C.~Iazeolla and P.~Sundell,
  ``Unfolding Mixed-Symmetry Fields in AdS and the BMV Conjecture: I. General Formalism,''
  JHEP {\bf 0907} (2009) 013
  doi:10.1088/1126-6708/2009/07/013
  [arXiv:0812.3615 [hep-th]].
  %%CITATION = doi:10.1088/1126-6708/2009/07/013;%%
  %77 citations counted in INSPIRE as of 29 Oct 2017
  
  %\cite{Boulanger:2008kw}
\bibitem{Boulanger:2008kw}
  N.~Boulanger, C.~Iazeolla and P.~Sundell,
  ``Unfolding Mixed-Symmetry Fields in AdS and the BMV Conjecture. II. Oscillator Realization,''
  JHEP {\bf 0907} (2009) 014
  doi:10.1088/1126-6708/2009/07/014
  [arXiv:0812.4438 [hep-th]].
  %%CITATION = doi:10.1088/1126-6708/2009/07/014;%%
  %61 citations counted in INSPIRE as of 29 Oct 2017
  
 
     
  %\cite{Colombo:2010fu}
\bibitem{Colombo:2010fu}
  N.~Colombo and P.~Sundell,
  ``Twistor space observables and quasi-amplitudes in 4D higher spin gravity,''
  JHEP {\bf 1111} (2011) 042
  doi:10.1007/JHEP11(2011)042
  [arXiv:1012.0813 [hep-th]].
  %%CITATION = doi:10.1007/JHEP11(2011)042;%%
  %21 citations counted in INSPIRE as of 18 Apr 2017
  
  \bibitem{Colombo:2012jx}
  N.~Colombo and P.~Sundell,
  ``Higher Spin Gravity Amplitudes From Zero-form Charges,''
  arXiv:1208.3880 [hep-th].
  %%CITATION = ARXIV:1208.3880;%%
  %43 citations counted in INSPIRE as of 14 Apr 2017
  
  %\cite{Bonezzi:2017vha}
\bibitem{Bonezzi:2017vha}
R.~Bonezzi, N.~Boulanger, D.~De Filippi and P.~Sundell,
``Noncommutative Wilson lines in higher-spin theory and correlation functions of conserved currents for free conformal fields,''
J. Phys. A \textbf{50} (2017) no.47, 475401
doi:10.1088/1751-8121/aa8efa
[arXiv:1705.03928 [hep-th]].
%21 citations counted in INSPIRE as of 28 Apr 2020
  


  %{Flato:1978qz}
\bibitem{Flato:1978qz}
  M.~Flato and C.~Fronsdal,
  ``One Massless Particle Equals Two Dirac Singletons: Elementary Particles In
  A Curved Space. 6,''
  Lett.\ Math.\ Phys.\  {\bf 2} (1978) 421.
  
   
  
   
  %\cite{Lunin:2002qf}
\bibitem{Lunin:2002qf}
  O.~Lunin and S.~D.~Mathur,
  ``Statistical interpretation of Bekenstein entropy for systems with a stretched horizon,''
  Phys.\ Rev.\ Lett.\  {\bf 88} (2002) 211303
  doi:10.1103/PhysRevLett.88.211303
  [hep-th/0202072].
  %%CITATION = doi:10.1103/PhysRevLett.88.211303;%%
  %150 citations counted in INSPIRE as of 04 May 2017

%\cite{Mathur:2002ie}
\bibitem{Mathur:2002ie}
  S.~D.~Mathur,
  ``A Proposal to resolve the black hole information paradox,''
  Int.\ J.\ Mod.\ Phys.\ D {\bf 11} (2002) 1537
  doi:10.1142/S0218271802002852
  [hep-th/0205192].
  %%CITATION = doi:10.1142/S0218271802002852;%%
  %19 citations counted in INSPIRE as of 04 May 2017
  
 %\cite{Mathur:2005zp}
\bibitem{Mathur:2005zp}
  S.~D.~Mathur,
  ``The Fuzzball proposal for black holes: An Elementary review,''
  Fortsch.\ Phys.\  {\bf 53} (2005) 793
  doi:10.1002/prop.200410203
  [hep-th/0502050].
  %%CITATION = doi:10.1002/prop.200410203;%%
  %448 citations counted in INSPIRE as of 04 May 2017
  
  %\cite{Skenderis:2008qn}
\bibitem{Skenderis}
  K.~Skenderis and M.~Taylor,
  ``The fuzzball proposal for black holes,''
  Phys.\ Rept.\  {\bf 467} (2008) 117
  doi:10.1016/j.physrep.2008.08.001
  [arXiv:0804.0552 [hep-th]].
  %%CITATION = doi:10.1016/j.physrep.2008.08.001;%%
  %187 citations counted in INSPIRE as of 04 May 2017
 

%\cite{Avis:1977yn}
\bibitem{Avis:1977yn}
  S.~J.~Avis, C.~J.~Isham and D.~Storey,
  ``Quantum Field Theory in anti-De Sitter Space-Time,''
  Phys.\ Rev.\ D {\bf 18} (1978) 3565.
  doi:10.1103/PhysRevD.18.3565
  %%CITATION = doi:10.1103/PhysRevD.18.3565;%%
  %344 citations counted in INSPIRE as of 11 Aug 2017
  
  %\cite{Breitenlohner:1982jf}
\bibitem{Breitenlohner:1982jf}
  P.~Breitenlohner and D.~Z.~Freedman,
  ``Stability In Gauged Extended Supergravity,''
  Annals Phys.\ {\bf 144} (1982) 249.
  %%CITATION = APNYA,144,249;%%
  
  %\cite{Mezincescu:1984ev}
\bibitem{Mezincescu:1984ev}
  L.~Mezincescu and P.~K.~Townsend,
  ``Stability at a Local Maximum in Higher Dimensional Anti-de Sitter Space and Applications to Supergravity,''
  Annals Phys.\  {\bf 160} (1985) 406.
  doi:10.1016/0003-4916(85)90150-2
  %%CITATION = doi:10.1016/0003-4916(85)90150-2;%%
  %160 citations counted in INSPIRE as of 11 Aug 2017
  
  %\cite{Balasubramanian:1998sn}
\bibitem{Balasubramanian:1998sn}
  V.~Balasubramanian, P.~Kraus and A.~E.~Lawrence,
  ``Bulk versus boundary dynamics in anti-de Sitter space-time,''
  Phys.\ Rev.\ D {\bf 59} (1999) 046003
  doi:10.1103/PhysRevD.59.046003
  [hep-th/9805171].
  %%CITATION = doi:10.1103/PhysRevD.59.046003;%%
  %416 citations counted in INSPIRE as of 11 Aug 2017
  
  %\cite{Son:2002sd}
\bibitem{SonStarinets}
D.~T.~Son and A.~O.~Starinets,
``Minkowski space correlators in AdS / CFT correspondence: Recipe and applications,''
JHEP \textbf{09} (2002), 042
doi:10.1088/1126-6708/2002/09/042
[arXiv:hep-th/0205051 [hep-th]].
%976 citations counted in INSPIRE as of 23 Apr 2020

%\cite{Nastase:2015wjb}
\bibitem{Nastase:2015wjb}
H.~Nastase,
``Introduction to the ADS/CFT Correspondence,'' Cambridge University Press, 2015
%6 citations counted in INSPIRE as of 23 Apr 2020
  
    
  %\cite{Vasiliev:2015wma}
%\bibitem{Vasiliev:2015wma}
%  M.~A.~Vasiliev,
 % ``Star-Product Functions in Higher-Spin Theory and Locality,''
 % JHEP {\bf 1506} (2015) 031
 % doi:10.1007/JHEP06(2015)031
%  [arXiv:1502.02271 [hep-th]].
  %%CITATION = doi:10.1007/JHEP06(2015)031;%%
  %15 citations counted in INSPIRE as of 20 Jan 2017
  
  %\cite{Vasiliev:2016xui}
%\bibitem{Vasiliev:2016xui}
 % M.~A.~Vasiliev,
  %``Current Interactions, Locality and Holography from the 0-Form Sector of Nonlinear Higher-Spin Equations,''
  %arXiv:1605.02662 [hep-th].
  %%CITATION = ARXIV:1605.02662;%%
  %1 citations counted in INSPIRE as of 19 Jul 2016
  
 % \bibitem{Gelfond:2017wrh}
 % O.~A.~Gelfond and M.~A.~Vasiliev,
 % ``Current Interactions from the One-Form Sector of Nonlinear Higher-Spin Equations,''
%  arXiv:1706.03718 [hep-th].
  %%CITATION = ARXIV:1706.03718;%%
  %7 citations counted in INSPIRE as of 21 Oct 2017})
  
  
  %\cite{Sezgin:2017jgm}
%\bibitem{Sezgin:2017jgm}
 % E.~Sezgin, E.~D.~Skvortsov and Y.~Zhu,
 % ``Chern-Simons Matter Theories and Higher Spin Gravity,''
 % arXiv:1705.03197 [hep-th].
  %%CITATION = ARXIV:1705.03197;%%
  %1 citations counted in INSPIRE as of 21 May 2017
  
  %\cite{Didenko:2017lsn}
%\bibitem{Didenko:2017lsn}
 % V.~E.~Didenko and M.~A.~Vasiliev,
 % ``Test of the local form of higher-spin equations via AdS/CFT,''
 % arXiv:1705.03440 [hep-th].
  %%CITATION = ARXIV:1705.03440;%%
  %1 citations counted in INSPIRE as of 16 May 2017
  
  
  %\cite{Taronna:2016xrm}
%\bibitem{Taronna:2016xrm}
 % M.~Taronna,
  %``A Note on Field Redefinitions and Higher-Spin Equations,''
 % EPJ Web Conf.\  {\bf 125} (2016) 05025
 % doi:10.1051/epjconf/201612505025, 10.1088/1751-8121/aa55f0, 10.1051/epjconf/201612505025
  %[arXiv:1607.04718 [hep-th]].
  %%CITATION = doi:10.1051/epjconf/201612505025, 10.1088/1751-8121/aa55f0, 10.1051/epjconf/201612505025;%%
  %4 citations counted in INSPIRE as of 24 Feb 2017
  
  
  

%\cite{Vasiliev:2015mka}
\bibitem{Vasiliev:2015mka}
  M.~A.~Vasiliev,
  ``Invariant Functionals in Higher-Spin Theory,''
  Nucl.\ Phys.\ B {\bf 916} (2017) 219
  doi:10.1016/j.nuclphysb.2017.01.001
  [arXiv:1504.07289 [hep-th]].
  %%CITATION = doi:10.1016/j.nuclphysb.2017.01.001;%%
  %25 citations counted in INSPIRE as of 24 Oct 2017
        
  
%\cite{Raeymaekers:2016mmm}
\bibitem{Raeymaekers:2016mmm}
J.~Raeymaekers,
``On matter coupled to the higher spin square,''
J. Phys. A \textbf{49} (2016) no.35, 355402
doi:10.1088/1751-8113/49/35/355402
[arXiv:1603.07845 [hep-th]].
%4 citations counted in INSPIRE as of 18 Apr 2020

%\cite{Kessel:2018zqm}
\bibitem{Kessel:2018zqm}
P.~Kessel and J.~Raeymaekers,
``Simple unfolded equations for massive higher spins in AdS$_{3}$,''
JHEP \textbf{08} (2018), 076
doi:10.1007/JHEP08(2018)076
[arXiv:1805.07279 [hep-th]].
%2 citations counted in INSPIRE as of 18 Apr 2020

%\cite{Raeymaekers:2019dkc}
\bibitem{Raeymaekers:2019dkc}
J.~Raeymaekers,
``On tensionless string field theory in AdS$_3$,''
JHEP \textbf{07} (2019), 019
doi:10.1007/JHEP07(2019)019
[arXiv:1903.09647 [hep-th]].
%1 citations counted in INSPIRE as of 18 Apr 2020

%\cite{Gunaydin:1999jb}
\bibitem{Gunaydin:1999jb}
M.~Gunaydin,
``AdS / CFT dualities and the unitary representations of noncompact groups and supergroups: Wigner versus Dirac,''
Turk. J. Phys.
[arXiv:hep-th/0005168 [hep-th]].
%18 citations counted in INSPIRE as of 18 Apr 2020

%\cite{Yin:2019xxs}
\bibitem{Yin:2019xxs}
Y.~Yin,
``Higher-spin initial data in twistor space with complex stargenvalues,''
[arXiv:1909.12097 [hep-th]].
%0 citations counted in INSPIRE as of 20 Apr 2020

%\cite{Basile:2018dzi}
\bibitem{Basile:2018dzi}
T.~Basile, X.~Bekaert and E.~Joung,
``Twisted Flato-Fronsdal Theorem for Higher-Spin Algebras,''
JHEP \textbf{07} (2018), 009
doi:10.1007/JHEP07(2018)009
[arXiv:1802.03232 [hep-th]].
%5 citations counted in INSPIRE as of 22 Apr 2020

%\cite{Didenko:2015pjo}
\bibitem{Didenko:2015pjo}
V.~Didenko, N.~Misuna and M.~Vasiliev,
``Charges in nonlinear higher-spin theory,''
JHEP \textbf{03} (2017), 164
doi:10.1007/JHEP03(2017)164
[arXiv:1512.07626 [hep-th]].
%14 citations counted in INSPIRE as of 29 Apr 2020

\end{thebibliography}
\end{document}